\documentclass[11pt,a4paper]{JHEP3}
\usepackage{amsmath, amssymb}
\usepackage{euscript}

\def\be{\begin{eqnarray}}
 \def\ee{\end{eqnarray}}
\def\0{\nonumber}
\newcommand\EE{\EuScript{E}}
\newcommand\EU{\EuScript{U}}
\usepackage{amsfonts}
\usepackage{verbatim}

\newcommand\z{\zeta}

\newcommand\zh{\theta}

\def\k{\kappa}

\def\Ct{{\cal C}}

\def\sin{{\rm sin}}

\def\sinh{{\rm sinh}}
\def\cosh{{\rm cosh}}

\def\k{\kappa}

\def\bra#1{\langle #1 |}
\def\ket#1{|#1 \rangle}
\def\0{\nonumber}

\def \today{\ifcase\month\or
January\or February\or March\or April \or May\or June\or
July\or August\or September\or October\or November\or December\fi
\space\number\day,\space\number\year}

\preprint{SISSA/44/2009/EP\\ ULB-TH/09-24\\  \tt hep-th/0908.0055\\}

\title{Ghost story. II. The midpoint ghost vertex}

\author{ L.Bonora\\
International School for Advanced Studies (SISSA/ISAS)\\
Via Beirut 2--4, 34014 Trieste, Italy, and INFN, Sezione di
Trieste\\
E-mail:   \email{bonora@sissa.it},}

\author{C.Maccaferri\\
International Solvay Institutes, Physique Th\'eorique et Math\'ematique,\\
ULB C.P. 231, Universit\'e Libre de Bruxelles, \\
B-1050, B-1050 Brussels, Belgium\\
E-mail: \email{cmaccafe@ulb.ac.be},}

\author{R.J.Scherer Santos\\
Centro Brasileiro de Pesquisas Fisicas (CBPF-MCT)-LAFEX\\
R. Dr. Xavier Sigaud, 150 - Urca - Rio de Janeiro - Brasil - 22290-180\\
E-mail: \email{scherer@cbpf.br},}

\author{D.D.Tolla\\
Department of Physics and University College,
Sungkyunkwan University,
Suwon 440-746, South Korea\\
E-mail:  \email{ddtolla@skku.edu}}

\abstract{We construct the ghost number 9 three strings vertex for OSFT in
the natural
normal ordering. We find two versions, one with a ghost insertion at $z=i$ and
a twist--conjugate one with insertion at $z=-i$. For this reason we call them
midpoint vertices. We show that the relevant Neumann
matrices commute among themselves and with the matrix $G$ representing the
operator $K_1$. We analyze the spectrum of the latter and find that beside
a continuous spectrum there is a (so far ignored) discrete one. We are
able to write spectral formulas for all the Neumann matrices involved and
clarify the important role of the integration contour over the continuous spectrum.
We then pass to examine the (ghost) wedge states. We compute the discrete and
continuous eigenvalues of the corresponding Neumann matrices and show that
they satisfy the appropriate recursion relations. Using these results we
show that the formulas for our vertices correctly define the star
product in that, starting from the data of two ghost number 0 wedge states, they allow us
to reconstruct a ghost number 3 state which is the expected wedge state
with the ghost insertion at the midpoint, according to the star recursion relation.}

\keywords{String Field Theory, Ghost Wedge States}

\begin{document}


\section{Introduction}

Wedge states are associated to an integer $n$ and are defined in the
abstract by the $*$--multiplication rule
\be
|n\rangle \star |m\rangle = |n+m-1\rangle\label{wedgedef}
\ee
They may have different `embodiments', \cite{Rastelli:2000iu, Schnabl:2002gg}.
They are surface states and, as such, may be realized as
 squeezed states in the oscillator formalism or as
exponentials of the operator ${\cal L}_0 +
{\cal L}_0^{\dagger}$ applied to the vacuum; other representations are also
possible. Our purpose, in
the series of papers started with \cite{BMST}, is to find the
correspondence between {\it the different realizations of the ghost part}
of the wedge states.

We recall that in \cite{BMST} we were concerned with proving the
equation
\be
e^{ - \frac {n-2}2\left({\cal L}^{(g)}_0 +
{\cal L}_0^{(g)\dagger}\right)}|0\rangle
= {\cal N}_n\, e^{ c^\dagger S_n b^\dagger}|0\rangle = |n\rangle
\label{ghostwedge}
\ee
where $|n\rangle$ are the ghost wedge states in the oscillator
formalism, which is a crucial ingredient of the analytic solution
of SFT found in \cite{Schnabl05} (see also
\cite{Okawa1, Fuchs0, Ellwood, RZ06, ORZ, Erler:2009uj, Erler:2007xt,
Aref'eva:2008ad, Okawa2, Okawa3, Schnabl:2007az, KORZ, Fuchs3, Kwon:2008ap, Lee:2007ns, Kiermaier:2007ki, Kiermaier:2007vu, Erler:2007rh,
 Kroyter:2009zj, Kroyter:2009bg, Ellwood:2009zf, Kiermaier:2008qu, Ellwood:2008jh}
and \cite{Fuchs4} for an updating on  recent progress).
It is known that the LHS of this equation can also be written
as a squeezed state whose defining matrix is that of a surface state (with ghost insertion at 0 in the UHP). In \cite{BMST} (also
referred to henceforth as I) we dealt mostly
with it from the oscillator point of view.
We showed that it can be cast into the
midterm form in (\ref{ghostwedge}) and we diagonalized the matrix
$S_n$ in a {\it continuous basis of eigenvectors}. Then we proved that,
{\it if} we are allowed to star--multiply the squeezed states representing
the ghost wedge states $|n\rangle$ the same way we do for the matter
wedge states and diagonalize the corresponding matrices, the eigenvalue
we obtain satisfies the wedge states recursion relation. This was based on
the expectation that all the (twisted) Neumann matrices entering the game could
be diagonalized in the same basis (this is what happens for the matrices
of the matter sector). In the course of the continuation of
this research  we realized that the expectation of I was a bit too optimistic
and had to readjust our line of arguments. The main reason for this
is that the spectral theory of the Neumann matrices that characterize
the ghost sector of the three strings vertex and wedge states is significantly different
from ordinary spectral theory of real symmetric matrices, which are the
basic example of Neumann
matrices in the matter sector. Once the eigenvalues and eigenvectors of the latter
matrices are given, their reconstruction via the spectral formula is unique.
In the ghost case instead the spectral formula is not uniquely determined but depends
on the integration contour over the continuous spectrum: this is one of the
basic results of our analysis. It should be clarified that such spectral formulas
we obtain for the ghost Neumann matrices are not derived on the basis of general
theorems in operator theory, which to our best knowledge do not exist in the
literature, but on a heuristic basis. We thinks we have checked them beyond any doubt
both numerically and using consistency with other methods.

In this paper we introduce a three--strings
vertex for the ghost part in order to be able to explicitly perform
the star product in (\ref{wedgedef}), up to a midpoint ghost insertion. Moreover we complete the spectral
analysis of the ghost bases by computing the relevant discrete bases of eigenvectors
(which were missing in \cite{BMST,BST}). Finally we show
that the states in the LHS of (\ref{ghostwedge}) do satisfy
the recursion relations for the wedge states (the RHS), although
not in the form expected in I. We show in fact that only the eigenvalues
of the relevant matrices satisfy the appropriate recursion relations.
Based on this, we can reconstruct, in the sense mentioned above, Neumann
matrices which represent ghost number 3 states and show that the latter
are surface states with a midpoint insertion, representing, at $gh$=3, the expected wedge states. So, it is true
that in the ghost number 3 sector things work much as in the matter sector.
However the same is not true for the ghost number 0 sector. In fact
what remains to be done
is reconstructing from the ghost number 3 the ghost number 0 wedge states
we started from (eq.(\ref{ghostwedge})). This would close the circle and fully justify
our claim about the consistency of our three strings ghost vertex
and the correctness of (\ref{ghostwedge}). This rather non--trivial task will
be carried out in another paper \cite{BMT3}, referred to as III.

{\bf Notation}. Any infinite matrix we meet in this paper is either square short
or long legged, or lame. In this regard we will often use a compact
notation: a subscript $_s$ will represent an integer label $n$ running
from $2$ to $\infty$, while a subscript $_l$ will represent a label
running from $-1$ to $+\infty$. So $Y_{ss}, Y_{ll}$ will denote
square short and long legged matrices, respectively; $Y_{sl}, Y_{ls}$ will
denote short--long and long--short lame matrices, respectively.
With the same meaning we will say that a matrix is $(ll),(ss),(sl)$ or
$(ls)$. The $(ss)$ part of a matrix $M$ will be referred to as the bulk of $M$.
In a similar way we will denote by $V_s$ and $V_l$ a short and long
infinite vector, to which the above matrices naturally apply.
Moreover, while $n,m$ represent generic matrix indices, at times we will
use $N,M$ to represent `long' indices, i.e. $N,M\geq -1$. In this case
$n,m$ will represent short indices, i.e. $n,m\geq 2$.

We will also use the symbol $\Ct$ to represent the twist matrix,
$\Ct_{n,m}=(-1)^n \delta_{n,m}$. Given any matrix $M$, we generically represent
the twisted matrix $\Ct M$ by $\tilde M$. Finally we use the symbol
$gh$ to denote the ghost number.

\subsection{A summary of the results}

Since the paper is rather long and elaborate we would like to start
with an outline of it and a summary of the main results.

To start with we first
recall the basic anti--commutator for the $b,c$ ghost oscillators
and $bpz$ transformation properties
\be
[c_n,b_m]_+=\delta_{n+m,0}, \quad \quad bpz(c_n) = -(-1)^n c_{-n},
\quad\quad bpz(b_n) = (-1)^n b_{-n},\quad \quad bpz(|0\rangle)=\langle 0|
\0\ee
where $|0\rangle$ is the SL(2,R)--invariant vacuum.
Next we define the state $|\hat 0\rangle = c_{-1}c_0c_1|0\rangle$ and the
tensor product of states
\be
_{123}\langle \hat\omega|= {}_1\langle \hat 0|_2\langle \hat 0 |_3\langle \hat 0|
\label{omegahat}
\ee
carrying total $gh=9$, and
\be
|\omega\rangle_{123} &=& | 0\rangle_1| 0\rangle_2| 0\rangle_3
\label{omega}
\ee
carrying total $gh$=0. They satisfy
$_{123}\langle \hat\omega|\omega\rangle_{123}=1$. Finally
we write down the general form of the three strings vertices we will find below
(section 2). The first two are
\be
\langle \hat V_{(\pm i)3}|=\,\hat{\cal K}_{(\pm i)}\,{}_{123}\langle \hat \omega|
e^{\hat E_{(\pm i)}}, \quad\quad
\hat E_{(\pm i)}=-\sum_{r,s=1}^3\sum_{n,m}^{\infty}c_n^{(r)}
\hat V_{(\pm i)nm}^{rs}b_m^{(s)}\label{V3ghpmi}
\ee
where
\be
\hat V_{(i)nm}^{rs}
= \oint\frac{dz}{2\pi i}\oint\frac{dw}{2\pi i}\frac{1}{z^{n-1}}
\frac{1}{w^{m+2}} \left( \frac {\left(\frac d{dz}  f_r(z)\right)^2}
{\frac d{dw} f_s(w)}\, \frac  1{f_r(z)-f_s(w)}\left(\frac { f_s(w)}{f_r(z)}\right)^3-
\frac {\delta^{rs}}{z-w} \right)\label{Vnmrsi}
\ee
and
\be
\hat V_{(-i)nm}^{rs}
= \oint\frac{dz}{2\pi i}\oint\frac{dw}{2\pi i}\frac{1}{z^{n-1}}
\frac{1}{w^{m+2}} \left( \frac {\left(\frac d{dz}  f_r(z)\right)^2}
{\frac d{dw} f_s(w)}\, \frac  1{f_r(z)-f_s(w)} -
\frac {\delta^{rs}}{z-w} \right)\label{Vnmrsmi}
\ee
The labels $(\pm i)$ refer to the ghost insertion at the string midpoint $i$
and image point $-i$, respectively (see below). These Neumann matrices are complex.

We will also use a third auxiliary vertex (a sort of average of the previous two)
whose Neumann matrices are real
\be
\langle \hat V_{3}|=\,\hat{\cal K}\,{}_{123}\langle \hat \omega|
e^{\hat E}, \quad\quad
\hat E=-\sum_{r,s=1}^3\sum_{n,m}^{\infty}c_n^{(r)} \hat V_{nm}^{rs}b_m^{(s)}
\label{V3gh}
\ee
where
\be
\hat V_{nm}^{rs}
=\frac 12 \oint\frac{dz}{2\pi i}\oint\frac{dw}{2\pi i}\frac{1}{z^{n-1}}
\frac{1}{w^{m+2}} \left( \frac {\left(\frac d{dz} \ln f_r(z)\right)^2}
{\frac d{dw} \ln f_s(w)}\, \frac {f_r(z)+f_s(w)}{f_r(z)-f_s(w)}-
\frac {\delta^{rs}}{z-w} \right)\label{Vnmrs}
\ee
All these vertices satisfy cyclicity
\be
\hat V_{nm}^{rs}=\,\hat V_{nm}^{r+1,s+1},\quad\quad
\hat V_{(\pm i)nm}^{rs}=\,\hat V_{(\pm i)nm}^{r+1,s+1}\label{cyclgh}
\ee
The third vertex satisfies twist--covariance
\be
\hat V_{nm}^{rs}=(-1)^{n+m}\hat V_{nm}^{sr}\label{twistgh}
\ee
while the first two are twist conjugate
\be
\hat V_{(-i)nm}^{rs}=(-1)^{n+m}\hat V_{(i)nm}^{sr}\label{twistghpmi}
\ee
The latter are BRST invariant. Dual vertices can also be defined.
We will show that the twisted Neumann matrices of each vertex commute.

The constants ${\cal K}$ and ${\cal K}_{(\pm i)}$ turn out to be 1.

In the previous formulas
\be
f_r(z_r)=\alpha^{2-r} f(z_r) \, ,\, r=1,2,3\label{fi}
\ee
where
\be
f(z)=\Big{(} \frac{1+iz}{1-iz}\Big{)} ^{\frac{2}{3}}\label{f}
\ee
Here $\alpha=e^{\frac{2\pi i}{3}}$.

In section 3 we will show that the twisted Neumann matrices of all the vertices
just introduced commute with the matrix $G$, which represents the operator
$K_1=L_1+L_1^\dagger$. This allows us to diagonalize the matrices that commute
with $G$ on the basis of its eigenvectors. In section 4 we explicitly compute the
bases corresponding to the discrete spectrum of $G$, while the continuous spectrum
had already been computed in \cite{Belov1,Belov2,BeLove}. We also write down
the spectral formula for $G$ and notice that it depends on the contour one takes
in order to integrate over the continuous eigenvalue $\k$: only in a certain range of
$\Im(\k)$ do we correctly reproduce $G$. We also give (partial) reconstruction formulas
for the matrices $A,B,C,D$ of I.

In section 5 we write down spectral formulas for the (twisted) Neumann matrices
of the above constructed vertices. We show that the integration contour over the
continuous spectrum plays a fundamental role. In fact different vertices have the
same spectral formulas but differ by the integration contour and can be obtained from
one another by changing it.

The main purpose of section 6 is to extract information about the eigenvalues
of the Neumann coefficients of the ghost number 0 wedge states from solving
the KP equation, \cite{KPot} , as was done in I. The main difference with I is that we do not
use commutativity of the matrices $A,B,C,D$ but solve the equation for their
eigenvalues. In such a way we are able to prove that both the
continuous and discrete eigenvalues of the wedge states satisfy the appropriate
recursion relation. With these results at hand, in section 7 we pass to the
task of reconstructing the twisted Neumann matrices of the ghost number 3 wedge states.
Once again the integration path over the continuous spectrum plays a crucial role
and allows us to pass from one possible representation to another of
these states. It is clear that in so doing we are assuming that the eigenvalues
are common to all the representations of a given wedge state both with ghost number 3
and with ghost number 0. This assumption turns out to be correct but will be fully
justified only in paper III.

Finally, five appendices contains auxiliary material, calculations and complements.

\section{The three strings vertex}

In order to construct the ghost three string vertex in the oscillator
formalism (for previous literature, see \cite{Samu,CST,GJ1,GJ2,Ohta}; problems related
to the present paper are treated in \cite{RSZ3,GRSZ2,MM,Oku2,Furu,Kishimoto,FKM,Fuchs1})
we have to face a number of problems which are not met
in the matter sector. The first is normal ordering. Let us recall
that one can envisage two main types of normal orderings, which we have
called in \cite{BMST} the {\it natural} and {\it conventional
normal ordering}. The former is the obvious normal ordering
required when the vacuum is $|0\rangle$, the latter is instead
requested by the vacuum state $c_1|0\rangle$ (of course, in
principle, one could consider other possibilities). A second problem
is generated by the ghost insertions, which are a priori free.
It is clear that the three strings vertex will depend to some extent both
on the normal ordering and the ghost insertion. Finally the vertices
must be BRST invariant.

To start with, in this paper we will use the natural normal ordering.
This is at variance with the existing ghost three strings vertex
\cite{Samu,CST,GJ1,GJ2,Ohta}, which is based
on the conventional normal ordering. This innovation is required by the new
non--perturbative analytic solution of SFT found by Schnabl, \cite{Schnabl05},
where ghost number 0 wedge states are used,
for which the old ghost vertex is ineffective, see for instance \cite{BST}.

Using the definitions (\ref{V3ghpmi},\ref{V3gh}), our aim now is
to explicitly compute $\hat V_{(\pm i)nm}^{rs},\hat V_{nm}^{rs}$.
The method is well--known: one expresses the propagator
$\ll c(z) b(w)\gg$ (see Appendix A) in two different ways, first as a
CFT correlator and then in terms of $\hat V_3$ and equates the two
expressions after mapping them to the disk via the maps (\ref{fi}).
However this recipe leaves several uncertainties.

First we have to insert the three $c$ zero modes. We can either, for instance, insert
three separate fields $c(z_i)$, (\ref{leclair}), or use
$Y(z)= \frac 12 \partial^2 c(z) \partial c(z) c(z)$. In order to pair
ghost number 3 and ghost number 0 states so that they preserve their
conformal properties, we should use a ghost number 3 primary field insertion
with
vanishing conformal weight. This implies the use of $Y$, which has this property.
Even so there remain many possibilities.
Let us make the obvious remark that, given the vacuum $|0>$, there are
many ways to define a conjugate vacuum $|0^c>$ such that
$<0^c|0>=<0|0^c>=1$. The simplest example is given by
$|0^c>=Y(0)|0>=c_{-1}c_0c_1|0>$.
However this is not the only possible choice since
$\partial_z <0|Y(z)|0>=0.$
So, in principle, any choice of $|0^c>=Y(z)|0>$ is a good conjugate
vacuum and we can choose the insertion point as we like.

The above can be understood in terms of $Q_B$ cohomology.
Remembering that $$\{Q, c(z)\}=c\partial c(z),$$ we have
$$\{Q, Y(z)\}=0, \quad\quad {\rm and}\quad\quad
\partial Y(z)=Q(...).$$
This means that the point where one inserts
$Y$ is irrelevant when the other string fields in the game are in
the kernel of $Q$. This is in particular true for surface states in
critical dimension. For any surface state $\Sigma$ we have
$$\partial_z\langle \Sigma|Y(z)|\Sigma\rangle=0$$ if
$Q|\Sigma\rangle=0$.

In defining the vertex, however, the place where $Y$ is inserted matters.
This is because the $*$--product treats the midpoint as special (it is the
only point which is common to the three interacting strings). So, out of the
infinite places where we could insert $Y$, we make
the most symmetric choice of inserting $Y$ at the midpoint. Since $Y$ is a
weight zero primary, this will not cause the typical divergences
of midpoint insertions. The vertex we are constructing is thus meant to
perform the $*$--product (which is ghost number preserving) and then to add
a $Y$ midpoint insertion to the result. Calling $\bra{\hat V_3}$
such a vertex, we can define it symbolically as
\be
\bra{\hat
V_3}\ket{\psi_1}\ket{\psi_2}=\bra{\psi_1*\psi_2}{\Big(}Y(i)\, ,\,Y(-i){\Big)}
\ee
In other words, calling $\bra {V_3}$ the usual three--strings vertex without
insertions, we can write
\be
\bra{\hat V_3}=\bra {V_3}{\Big(}Y(i)\, ,\,Y(-i){\Big)}={\Big(}\bra
{\hat V_{(i)}}\,,\,\bra{\hat V_{(-i)}}{\Big)}\label{V3couple}
\ee
meaning that, as we will see, we need both insertions in order to
correctly represent the star product (this is just the  doubling trick).

In the natural normal ordering it is impossible to represent $\bra {V_3}$
in a squeezed state form which is cyclic in the string indices.
That is not a problem, in principle, but it would give rise to very
complicated Neumann coefficients matrices. On the other hand
the midpoint inserted vertex $\bra{\hat V_3}$ is expressed in terms
of two cyclic squeezed states: each of them can actually be
used independently of each another. Computations with  $\bra{\hat
V_{(i)}}$ will be related to $\bra{\hat V_{(-i)}}$ by
twist--conjugation. The price we have to pay for this choice
in the vertex is that, when we midpoint--multiply 2 $gh=0$ states,  we get
a $gh=3$ result. Going back to $gh=0$  will be the subject of III.

\subsection{Three zero modes insertion}

We start by inserting the operator $Y$ at the point $t$ (for simplicity we understand
the dependence on $t$ in the vertex, until further notice).
We use the correlator (\ref{leclair}) in Appendix A
and compare
\be
\langle f_j \circ Y(t) \, f_r \circ c^{(r)}(z) \, f_s \circ b^{(s)}(w) \rangle
\label{corrcb}
\ee
with
\be
\langle \hat V_{3}|R(c^{(r)}(z)\, b^{(s)}(w)) |\omega\rangle_{123}
 \label{V3cb}
\ee
where $R$ denotes radial ordering. If :: denotes the natural normal
ordering, we have for instance (see Appendix A)
\be
R(c(z)\,b(w)) = \sum_{n,k} :c_n\, b_k: \,
z^{-n+1} w^{-k-2} +\frac 1{z-w}\label{radial}
\ee
This should be inserted inside (\ref{V3cb}). Let us refer to
the last term in (\ref{radial}) as the {\it ordering term}.

We first compute the $\hat{\cal K}$ constant. By making use
of $\langle 0|Y(t)|0\rangle=1$ for any $t$, we have
\be
&&\langle \hat V_{3}|\omega\rangle_{123}=\hat {\cal K}
= \langle f_j \circ Y(t)\rangle =1\label{calK}
\ee
for any $j$.
Now
\be
&&\langle \hat V_{3}|R (c^{(r)}(z)\, b^{(s)}(w))
|\omega\rangle_{123}\0\\
&&= \langle \hat V_{3}|\sum_{n,k} :c^{(r)}_{n}\, b^{(s)}_{k}: \, z^{-n+1} w^{-k-2}
+\frac {\delta^{rs}}{z-w}|\omega\rangle_{123} \0\\
&&
=  -\hat V^{sr}_{kn} \, z^{n+1} w^{k-2} + \frac {\delta^{rs}}
{z-w}\label{cbcorr1}
\ee
On the other hand, from direct computation,
\be
&& \langle f_r\circ c(z) \, f_s \circ b(w)\, f_j\circ Y(t)\rangle\0\\
&=& \frac {(f_s'(w))^2}{f_r'(z)} \, \frac 1{f_r(z)-f_s(w)}
\left(\frac{ f_j(t) -f_r(z)}
{f_j(t) -f_s(w)}\right)^3\label{cbcorr2}
\ee
Comparing the last two equations and using (\ref{calK}) we get
\be
\hat V_{kn}^{sr} &=& -\oint \frac {dz}{2\pi i}  \oint \frac {dw}{2\pi i}
\frac 1{z^{n+2}} \frac 1{w^{k-1}}\cdot \label{cbcorr3}\\
&&\cdot \left(
\frac {(f_s'(w))^2}{f_r'(z)} \, \frac 1{f_r(z)-f_s(w)} \
 \left(\frac {f_j(t) -f_r(z)}
 {f_j(t) -f_s(w)}\right)^3-\frac {\delta^{rs}}
{z-w}\right)\0
\ee
After obvious changes of indices and variables we end up with
\be
\hat V_{nm}^{rs} &
=&\oint\frac{dz}{2\pi i}\oint\frac{dw}{2\pi i}\frac{1}{z^{n-1}}
\frac{1}{w^{m+2}}\label{cbcorr4}\\
&&\cdot \left(\frac{(f'_r(z))^2}{(f'_s(w))}\,
\frac{1}{f_r(z)-f_s(w)}\left(
\frac{f_s(w)-f_j(t)}{f_r(z)-f_j(t)}\right)^3- \frac {\delta^{rs}}
{z-w}\right)\0
\ee
After some elementary algebra, using
$f'(z)=\frac{4i}{3}\frac{1}{1+z^2}f(z)$, one finds
\be
\hat V_{nm}^{rs}=\frac{1}{3}(E_{nm}+\bar{\alpha}^{r-s} U_{nm}+
\alpha^{r-s}\bar{U}_{nm})\label{decompgh}
\ee
where
\be
E_{nm}&=&\oint\frac{dz}{2\pi i}\oint\frac{dw}{2\pi i}\frac{1}{z^{n+1}}
\frac{1}{w^{m+1}}\left[
\Big{(}\frac{1}{1+zw}-\frac{w}{w-z}\Big{)}\right.\label{E}\\
&&\left.~~~~~~ ~~~~~~~~~~~~~~~~~~~~~
\cdot (1-p_t(z,w))-\frac {z^2}w \frac 1{z-w}\right] \0\\
U_{nm}&=&\oint\frac{dz}{2\pi i}\oint\frac{dw}{2\pi
i}\frac{1}{z^{n+1}}\frac{1}{w^{m+1}}\left[\frac{f(z)}{f(w)}
\Big{(}\frac{1}{1+zw}-\frac{w}{w-z}\Big{)}\right.\label{U}\\
&&\left. ~~~~~~~~~~~~~~~~~~~~~~~~~~~
\cdot (1-p_t(z,w))- \frac {z^2}w \frac 1{z-w} \right]\0\\
\bar{U}_{nm}&=&\oint\frac{dz}{2\pi i}\oint\frac{dw}{2\pi
i}\frac{1}{z^{n+1}}\frac{1}{w^{m+1}}\left[\frac{f(w)}{f(z)}
\Big{(}\frac{1}{1+zw}
-\frac{w}{w-z}\Big{)}\right.\label{Ubar}\\
&&\left. ~~~~~~~~~~~~~~~~~~~~~~~~~~~
\cdot (1-p_t(z,w))-\frac {z^2}w \frac 1{z-w} \right]\nonumber
\ee
for $t=\pm i$. In the above equations
\be
p_t(z,w)= \frac {t(w-z)(1+wz)} {w(t-z)(1+tz)}\label{qzw}
\ee
This function enjoys the properties
\be
p_t\left(-\frac 1z,w\right)= p_t(z,w),\quad  p_t\left(z,-\frac 1w\right)=
p_t(z,w),\quad p_t(z,z)=0, \quad p_t(0,z)=1 \label{qprop}
\ee
which will be of great importance later on.

It is immediate to check cyclicity (with $t=\pm i$)
\be
\hat V_{nm}^{rs}=\hat V_{nm}^{r+1,s+1},\0
\ee
Moreover we have the twist covariance property
\be
\hat V_{(i)nm}^{rs}=(-1)^{n+m}\hat V_{(-i)nm}^{sr}\0
\ee
that is the vertex with $Y$ insertion at $i$ is twist conjugate to the one
with insertion at $-i$. This is due, in particular, to the property
\be
p_i(-z,-w)= p_{-i}(z,w)\label{pip-i}
\ee

So we have a couple of twist--conjugate
vertices. Due to the considerations at the beginning of this section, these two
vertices are BRST invariant. As we will see in the sequel, they have the properties
we need, therefore we stick to them even though they are complex.
They are the two vertices defined by formulas (\ref{Vnmrsi},\ref{Vnmrsmi}).
The corresponding $E,U,\bar U$ are the ones defined by
eqs.(\ref{E},\ref{U},\ref{Ubar}), with $t=i$ and $-i$, respectively, in
$p_t$.

\subsubsection{The midpoint Neumann coefficients}

In conclusion, our midpoint vertices are defined as
\be
\hat V_{(\pm i)nm}^{rs}=\frac{1}{3}(E_{(\pm i)nm}+\bar{\alpha}^{r-s} U_{(\pm i)nm}+
\alpha^{r-s}\bar{U}_{(\pm i)nm})\label{decompghpmi}
\ee
in terms of the quantities
\be
E_{(\pm i)}= \EE_{(\pm i)}+Z, \quad\quad U_{(\pm i)} = \EU_{(\pm i)}+Z,\quad\quad
\bar U_{(\pm i)} =\bar\EU_{(\pm i)}+Z \label{EU+Z}
\ee
where
\be
\EE_{(\pm i)nm}&=&\oint\frac{dz}{2\pi i}\oint\frac{dw}{2\pi i}\frac{1}{z^{n+1}}
\frac{1}{w^{m+1}} \Big{(}\frac{1}{1+zw}-\frac{w}{w-z}\Big{)}
\,(1-p_{\pm i}(z,w))   \label{Enew}\\
 \EU_{(\pm i)nm}&=&\oint\frac{dz}{2\pi i}\oint\frac{dw}{2\pi
i}\frac{1}{z^{n+1}}\frac{1}{w^{m+1}} \frac{f(z)}{f(w)}
\Big{(}\frac{1}{1+zw}-\frac{w}{w-z}\Big{)}
\, (1-p_{\pm i}(z,w))  \label{Unew}\\
{\bar\EU}_{(\pm i)nm} &=&\oint\frac{dz}{2\pi i}\oint\frac{dw}{2\pi
i}\frac{1}{z^{n+1}}\frac{1}{w^{m+1}} \frac{f(w)}{f(z)}
\Big{(}\frac{1}{1+zw}
-\frac{w}{w-z}\Big{)} \,(1-p_{\pm i}(z,w)) \label{Ubarnew}
\ee
with the ordering term
\be
Z_{nm}= \oint\frac{dz}{2\pi i}\oint\frac{dw}{2\pi i}\frac{1}{z^{n+1}}
\frac{1}{w^{m+1}}  \left(-\frac {z^2}w \frac 1{z-w}\right)  \label{Znm}
\ee

\subsection{The average (real) vertex}

In addition to these two vertices we will construct a third one
which is twist invariant and real, although it does not evidently
respect BRST invariance.

One way to get a twist invariance vertex
is to average between the two above, that is to make the replacement
\be
\left(
\frac{f_s(w)-f_j(t)}{f_r(z)-f_j(t)}\right)^3 \longrightarrow
\frac 12\left( \left(
\frac{f_s(w)-f_j(t)}{f_r(z)-f_j(t)}\right)^3+ \left(
\frac{f_s(w)-f_j(\bar t)}{f_r(z)-f_j(\bar t)}\right)^3
 \right)\label{images}
\ee
in the above definitions, mimicking the method of images. We stress that
here we refer to the average of the vertex exponents.

This leads to (\ref{decompgh0},\ref{E},\ref{U},\ref{Ubar}) with
$p_t$ replaced by
\be
p_0(z,w) \equiv \frac 12 (p_i(z,w)+p_{-i}(z,w)) =
\frac{(z-w)(1+zw)(z^2-1)}{w(1+z^2)^2}\label{p0}
\ee
However this choice produces a singularity in the product $\EU^2$
(see subsection 2.4), a singularity which is due to the double pole of $p_0(z,w)$
at $z=i$ and $z=-i$. The definition of the twist--invariant midpoint vertex
requires a not a priori obvious
modification, which is as follows. We replace $p_t$ with $p_0$ in $\EE$,
with $p_i$ in $\EU$ and with $p_{-i}$ in $\bar \EU$,
(\ref{E},\ref{U},\ref{Ubar}) respectively.
We notice that, beside the properties (\ref{qprop}), one has
\be
p_0(-z,-w)=p_0(z,w)\label{parity}
\ee
As is easily verified, this property guarantees twist--invariance of
the Neumann matrices. We will denote the corresponding Neumann matrices
simply by $\hat V^{rs}_{nm}$.

In summary the average (regularized) vertex is defined in terms of
$U_{(i)},\bar U_{(-i)}$ and
\be
E=\EE+Z,\quad\quad  \EE= \frac 12 \Big{(}\EE_{(i)}+\EE_{(-i)}\Big{)}\label{avE}
\ee
as follows
\be
\hat V_{nm}^{rs}=\frac{1}{3}(E_{nm}+\bar{\alpha}^{r-s} U_{(i)nm}+
\alpha^{r-s}\bar{U}_{(-i)nm})\label{decompgh0}
\ee
Now one can easily show that the Neumann
matrices $\hat V^{rs}_{nm}$ can be written in the compact form (\ref{Vnmrs})
(apart from the ordering term).

\subsection{Two remarks}

The matrices $\hat V_{nm}^{rs},\hat V_{(\pm i)nm}^{rs} $ are all $sl$.
However, when $r=s$, it is always possible to add to them an upper left $3\times 3$
matrix  $z$, where $z_{ij}= \delta_{i+j,0}$, with $-1\leq i,j\leq 1$.
The addition of the matrix $z$ to $\hat V^{rr}$ does not change the vertex
provided we understand that the expression of the vertex is normal ordered, since,
in the definition (\ref{V3gh}), the vertex is applied to the vacuum $\langle \hat 0|$.
In fact we have more:
\be
\langle \hat 0|:e^{c_i \, \tau_{ij}\, b_j + c_n V_{nM}b_M}: \,=
 \langle \hat 0|e^{c_n V_{nM}b_M} \0
\ee
for any matrix $\tau_{ij}$. This ambiguity is allowed by the formalism and actually
it turns out to be very useful. This remark will be crucial in the sequel.

Another remark concerning the just defined vertices is the following.
While the expressions $\EE,\EU$ and $\bar \EU$ are ambiguous, due to the
presence of the factor $1/(z-w)$, in (\ref{EU+Z})
any ambiguity has disappeared. This is evident for $E$,
but is true also for $U$ and $\bar U$. For instance
\be
\frac {f(z)}{f(w)} \frac{w}{z-w} &=&\frac{f(w) +
(z-w) f'(w) + 1/2 (z-w)^2 f''(w)+\ldots}
{f(w)}\, \frac{w}{z-w} \0\\
&=& \frac{w}{z-w} + w f'(w) +\frac 12 (z-w)w f''(w)+\ldots\0
\ee
Of course only the first term in the RHS is ambiguous when inserted in the
double contour integral (\ref{Unew}), but it is cancelled by the ordering
term.
Therefore all the double integrals above are unambiguous. But if we
evaluate separately (as it will happen) $\EU$ and $Z$, for instance, we have
to be careful to use the same prescriptions, because each separate term
is ambiguous.

Finally we record the twist properties
\be
\Ct E_{(i)}= E_{(-i)}\Ct ,\quad\quad \Ct U_{(\pm i)}= \bar U_{(\mp i)}\Ct\label{twist}
\ee

\subsection{Fundamental properties of the Neumann coefficients}

In this subsection we will analytically prove certain fundamental relations
for the matrices (\ref{Enew},\ref{Unew},\ref{Ubarnew}) and (\ref{Znm}),
following the methods of \cite{tope}.
We remark that the analytic proof in this case is essential, because
the numerical analysis, while confirming the analytic results, is
hindered by the poor convergence properties of the product matrices.

\subsubsection{$\EU_{(p)} \EU_{(p')}$ }

Our first aim is to evaluate the product $(\EU_{(p)}\EU_{(p')})_{nm}$ where
$p$ and $p'$ stand for either $i$ or $-i$ and denote generically
the dependence on $p_{\pm i}$.
Since this result is specially important we present the calculation in
full detail as a model for many others that occur in the paper.

Let us consider the product $\sum_{k=-1}^\infty \EU_{(p)nk}\,\EU_{(p')_{km}}$.
In the first $\EU$ we use the integration variables $z$ and $\z$ and in the second $\zh$
and $w$. We assume $|z|<|\z|$ and $|\zh|>|w|$. This means that we have first
to integrate in $\z$ and $\zh$ and then in $z$ and $w$. This prescription is
arbitrary. We have to be careful to use the same prescription when
computing the other pieces of $U_{(p)} U_{(p')}$.

We use for $\EU$ the definition above, (\ref{Unew}), and perform the
intermediate summation in $\sum_{k=-1}^\infty \EU_{(p)nk}\,\EU_{(p')km}$:
\be
\sum_{k=-1}^\infty \frac 1{(\z\zh)^{k+1}}= \frac {\z\zh}{\z\zh-1} \label{ksum}
\ee
This is true if $|\z\zh|>1$. If the latter condition holds we have
\be
\sum_{k=-1}^\infty \EU_{(p)nk}\,\EU_{(p')km}&=& \oint\frac{dz}{2\pi i}\frac{1}{z^{n+1}}
\oint\frac{d\z}{2\pi i}\oint\frac{d\zh}{2\pi i} \oint\frac{dw}{2\pi i}
\frac{1}{w^{m+1}}\frac {\z\zh}{\z\zh-1} \label{U21new}\\
&&\cdot \frac{f(z)}{f(\z)} \Big{(}\frac{1}{1+z\z} -\frac{\z}{\z-z}\Big{)}
  \Big{(}1- p(z,\zeta) \Big{)} \cdot\0\\
&&\cdot \frac{f(\zh)}{f(w)} \Big{(}\frac{1}{1+\zh w}
-\frac{w}{w-\zh}\Big{)}\,
  \Big{(}1- p'(\zh,w) \Big{)}= *\0
\ee
We notice that $p(z,\zeta)$ has a double pole in $z=i$ and a simple pole in $\z=0$.
$p'(\zh,w)$ has a double pole in $\zh=\pm i$ and a simple pole in $w=0$.
In order to avoid the pole at $\zh=i$ it is
more convenient to integrate first with respect to $\z$.
In the integrand there are poles in $\z=z,-\frac 1z, \frac 1{\zh}$. In
order to guarantee $|\z\zh>1|$ we have to take $|\z| > \frac 1{|\zh|}$.
Therefore the integration contour in $\z$ will include
the poles in $z, \frac 1{\zh}$, but excludes $-\frac 1z$.

So we take $|\z| > 1$ and $|\zh|< 1$. Notice that we have
\be
|z|<|\z|,\quad \quad |\zh|>|w|,\quad\quad |\z|>\frac 1{|\zh|},
\quad \quad \frac 1{|z|}> |\z|> \frac 1{|\zh|},\quad {\rm i.e.}\quad
|\zh|>|z|\label{ineq1}
\ee
To comply with the condition $|\z| > 1$ we deform the $\z$ contour
while keeping the $\zh$ contour fixed. In doing so we have to be
careful to avoid possible singularities in $\z$. The latter are poles
at $\z = z,-\frac 1z, \frac 1{\zh}$ and branch cuts at $\z = \pm i$,
due to the $f(\z)$ factor. One can deform the $\z$ contour in such a
way as to keep the pole at $-\frac 1{z}$ external to the contour,
since the $z$ contour is as small as we wish around the origin. But,
of course, one cannot avoid the branch points at $\z = \pm i$. To make
sense of
the operation we introduce a regulator $K>1$ and modify the integrand
by modifying $f(\z)$
\be
f(\z) \to f_K(\z) =
\left(\frac {K +i \z}{K- i\z}\right)^{\frac 23}\0
\ee
We will take $K$ as large as needed and eventually move back to $K=1$.

Under these circumstances we can safely perform the summation over $k$,
make the replacement (\ref{ksum}) in the integral and get (\ref{U21new}).
Now we can integrate over $\z$.
The integration contour only surrounds $z,\frac 1{\zh}$. So (\ref{U21new}) becomes
\be
\{\z= \frac 1\zh\}\quad\quad *&=& \oint\frac{dz}{2\pi i}\frac{1}{z^{n+1}}
\oint\frac{d\zh}{2\pi i} \oint\frac{dw}{2\pi i}
\frac{1}{w^{m+1}}\,\left[\frac 1\zh\,\frac{f(z)}{f_K(\frac 1\zh)}
\Big{(} \frac{\zh}{\zh+z}-\frac{1}{1-z\zh} \Big{)}\right.\cdot\0\\
&&\cdot   \Big{(}1- p(z,-\zh) \Big{)}\,
\frac{f(\zh)}{f(w)} \Big{(} \frac1{1+\zh w} -   \frac{w}{ w-\zh}
\Big{)}\cdot\0\\
&&\cdot   \Big{(}1- p'(\zh,w) \Big{)}\label{U22new}\\
\{\z=z\}\,\,\quad\quad  &-& \frac {\zh z^2}{\zh z-1} \, \frac {f(\zh)}{f(w)} \,
\Big{(}\frac{1}{1+\zh w} -\frac{w}{w-\zh}\Big{)}\0\\
&&\left.\cdot   \Big{(}1- p'(\zh,w) \Big{)}\right]=**\0
\ee
where, on the left, in curly brackets we denote the pole that gives rise
to the contribution in the body of the formula.

Next we wish to integrate with respect to $\zh$. There are poles
at $\zh= -z, \frac 1z, w, -\frac 1w$ and possibly at $\zh=\pm i$,  and branch cuts
starting and ending at $\zh =\pm i$ and at $\zh = \pm \frac iK$
(no poles at $\zh=0,\infty$ !).
The singularities trapped
within the $\zh$ contour of integration are the poles at $\zh =-z,w$.
Since above we had
$K>|\zh|>\frac 1{|\z|}$, it follows that $|\zh|>\frac 1K$. Therefore
also the branch points at $\zh = \pm \frac iK$ of $f_K(1/\zh)$ are trapped
inside the $\zh$ contour and we have to compute the relevant contribution
to the integral.  Let us call this cut ${\mathfrak  c}_{1/K}$
and let us fix it to be the semicircle of radius $1/K$ at the LHS of
the imaginary axis; the contour that surrounds it excluding all the other
singularities will be denoted $C_{1/K}$. The other cut, due to $f(\zh)$,
with branch points at $\zh=\pm i$, will be denoted ${\mathfrak  c}_1$;
the contour that surrounds it (another semicircle of radius 1)
excluding all the other singularities will be denoted $C_1$. The
forthcoming argument requires that we split the branch point at
$\zh =i$ from the pole at the same point coming from $p'(\zh,w)$.
Therefore  we will introduce a regulator in $p'(\zh,w)$ to move away
this singularity and return eventually to the initial condition.
This regulator is simply to help keeping the branch point and
the pole of  $p'(\zh,w)$ at $\zh=i$ distinct.
The role of the poles of $p'(\zh,w)$ at $\zh=i$ will be analyzed further on.

Evaluating (\ref{U22new}) we get
\be
\{\zh=w\} \quad\quad ** &=&  \oint\frac{dz}{2\pi i}\frac{1}{z^{n+1}}
\oint\frac{dw}{2\pi i} \frac{1}{w^{m+1}}\,\left[ - \frac{f(z)}{f_K(\frac 1w)}
\right.\0\\
&& \cdot \left(\frac 1{1-zw} -\frac w{z+w}\right) \,
 \Big{(}1- p(z,-w)\Big{)} \0\\
&& + \frac {z^2 w^2}{1-zw}\label{U23new}\\
\{\zh=-z\}\quad \quad\quad &+& \frac{f(z)}{f_K(-\frac 1z)}\, \left(\frac 1{1-zw} -\frac w{z+w}\right)
 \Big{(}1- p'(-z,w) \Big{)}\ \,
\left.\frac{f(-z)}{f(w)}\right]\0\\
 &+& \oint_{C_{1/K}} d\zh (\ldots) \quad\quad\0
\ee
where the last term refers to the integral along the contour $C_{1/K}$.
We have used
\be
\frac{zw}{zw-1} - \frac z{z+w} = \frac w{z+w}- \frac{1}{1-zw}\0
\ee

The problem now is to evaluate the integral around
the cut. Fortunately this can be reduced to an evaluation of contributions
from poles. To see this, we first recall the properties of $f(z)$.
It is easy to see that
\be
f(1/z) =\gamma f(-z) \quad{\rm and}\quad f(-z)= 1/f(z)\label{fprop}
\ee
This comes from
\be
f(\frac 1z)=\Big{(} \frac{1+\frac iz}{1-\frac iz}\Big{)} ^{\frac{2}{3}}
= \Big{(}- \frac{1-iz}{1+iz}\Big{)} ^{\frac{2}{3}}\0
\ee
Above $\gamma$ is either $1,\alpha$ or $\bar \alpha$, depending on what Riemann
sheet we choose. However, denoting with an arrow the effect of a transformation
$\z\rightarrow -\frac 1{\z}$ we get
\be
f(-\frac 1{\z})=\gamma f(-\z) \rightarrow \gamma f(\frac 1{\z}) = \gamma^2 f(-\z)\0
\ee
On the other hand
\be
f(\frac 1{\z}) \rightarrow f(-\z)\0
\ee
Thus $\gamma^2=1$, which implies $\gamma=1$. We remark that this result comes
from requiring that the entry of $f$ takes values on a Riemann sphere.
The value of $\gamma$, however, does nott really matter provided we choose always
the same sheet.

Therefore, in the limit $K\to 1$, the factor $f_K(1/\zh)/f(\zh)$
tends, up to the $\gamma$ factor, to $(f(-\zh))^2$. As a consequence,
in the same limit, the integral
of $(\ldots)$ around the ${\mathfrak  c}_{1/K}$ cut is the same as the integral around
the ${\mathfrak  c}_1$ cut. To be more explicit in (\ref{U23new}) we have
\be
\oint_{C_{1/K}}\frac{d\zh}{2\pi i} \frac {f(\zh)}{f_K(\frac 1\zh)}\ldots
=\frac 1\gamma \oint_{C_{1/K}}\frac{d\zh}{2\pi i}  {f_K(\zh)f(\zh)}\ldots\0
\ee
In this expression the relevant cut is ${\mathfrak  c}_{1/K}$.
On the other hand
\be
\oint_{C_{1}}\frac{d\zh}{2\pi i} \frac {f(\zh)}{f_K(\frac 1\zh)}\ldots
=\frac 1\gamma \oint_{C_{1}}\frac{d\zh}{2\pi i} {f_K(\zh)f(\zh)}\ldots\0
\ee
In this expression the relevant cut is ${\mathfrak  c}_{1}$.
It is evident that in the limit $K\to 1$ the two expressions become one
and the same.

At this point it is convenient to take,
instead of the integral around one contour, the half sum of the
integral around both. But using a well-known argument,
the integral around both cuts equals minus the integral around all the other
singularities in the complex $\zh$--plane.
I.e. the overall contour integral around the cuts equal the
negative of the integral of $(\ldots)$ about all the remaining singularities
in the complex $\zh$--plane, {\it which are poles}
at $\zh = -z,w,1/z,-1/w,\pm i$.

Returning to (\ref{U22new}), the integral over $C_{1/K}$ involves only the
first part of  (\ref{U22new}) the one containing $f_K$, because the second
part does not
contain any trapped contour. As for the possible double poles at $\zh=\pm i$, they can
at the worst be simple because the double pole of $p_{\pm i}(\zh,w)$ are partly compensated
by the zero of $ \frac \zh{z+\zh}-\frac 1{1-z\zh}$.
Evaluating the residues at the poles we get
\be
 \oint_{C_{1/K}} d\zh \ldots&=& -\frac 12 \left\{\frac{}{} \right.\0\\
\{\zh=w\}\quad \quad\quad &-&  \oint\frac{dz}{2\pi i}\frac{1}{z^{n+1}}
\oint\frac{dw}{2\pi i} \frac{1}{w^{m+1}}\,\left[ - \frac{f(z)}{f(\frac 1w)}
\left(\frac 1{1-zw} -\frac w{z+w}\right) \right.\0\\
&& \cdot
  \Big{(}1- p(z,-w) \Big{)}\0 \\
\{\zh=-z\}\quad\quad\quad &+& \frac{f(z)}{f_K(-\frac 1z)}\, \left(\frac 1{1-zw} -
\frac w{z+w}\right) \Big{(}1- p'(-z,w) \Big{)} \,
\frac{f(-z)}{f(w)}\0\\
\{\zh=-\frac 1w\}\quad\quad\quad&-&\frac{f(z)}{f_K(-w)} \left(\frac 1{1-zw} -\frac w{z+w}\right)\,
\frac{f(-\frac 1w)}{f(w)}
 \Big{(}1- p(z,-w) \Big{)} \0\\
\{\zh =\frac 1z\}\quad\quad\quad&+& \frac {f(z)}{f_K(z)}
\frac {f(\frac 1z)}{f(w)} \left(\frac 1{1-zw}
-\frac w{z+w}\right) \Big{(}1- p'(-z,w) \Big{)}\0\\
\{\zh=\pm i\}\quad\quad\quad&+&  \ldots {\Big]}{\Big\}}\label{U24new}
\ee
where ellipses represent possible contributions of poles at $\zh=\pm i$.
The term $\zh=-z,\frac 1z$ cancel exactly the term
$\zh=-z$ of (\ref{U23new}) and the term $\zh=w,-\frac 1w$ cancel
the term $\zh=w$ in (\ref{U23new}). The $f$ factors in each of them become either
\be
\gamma{f(z)f(w)},\quad\quad {\rm or} \quad\quad \frac {\gamma}{f(z)f(w)} \0
\ee

In order to evaluate the contributions of the poles at $\zh=\pm i$
we have to distinguish various cases. If $p=p_i, p'=p_{i}$,
then the contribution of the pole
at $\zh= i$ does not appear because, 1) the double pole is partly compensated
by the zero of $ \frac \zh{z+\zh}-\frac 1{1-z\zh}$, and thus is a simple pole;
2) the residue
of this simple pole vanishes due to the factor $f(\zh)^2$, which vanishes
as $(\zh -i)^{\frac 43}$ when $\zh\to i$.

Let us consider next the case $p=p_i, p'=p_{-i}$. The double pole of $p_{-i}$ at
$\zh =-i$ is compensated by the zeroes of $ \frac \zh{z+\zh}-\frac 1{1-z\zh}$
and $1-p_i(z,-\zh)$. Therefore the pole disappears.

In the case $p=p_{-i}, p'=p_{i}$ we have a double pole at $\zh =i$, which is partly
compensated by the zero of $ \frac \zh{z+\zh}-\frac 1{1-z\zh}$. The remaining simple pole
has a zero residue due to $f(\zh)^2$, as in the case $p=p_i, p'=p_{i}$.

Thus in all three cases just considered, the ellipses at the
end of (\ref{U24new}) correspond to a vanishing contribution.

In the case $p=p_{-i}, p'=p_{-i}$, we have a double pole at $\zh =-i$, which is
partly compensated by the usual zero of $ \frac \zh{z+\zh}-\frac 1{1-z\zh}$. But
the residue of the simple pole is divergent due to $f(\zh)$. Therefore, in this
case we have a divergent result.

Finally we can write
\be
&&(\EU_{(i)}\,\EU_{(i)})_{nm}=(\EU_{(i)}\,\EU_{(-i)})_{nm}=(\EU_{(-i)}\,\EU_{(i)})_{nm}\0\\
&& =\oint\frac{dz}{2\pi i}\frac{1}{z^{n+1}}
 \oint\frac{dw}{2\pi i} \frac{1}{w^{m+1}} \frac {z^2 w^2}{1-zw}=
\left\{ \begin{matrix}\delta_{nm},\quad\quad & n,m\geq 2\\
                            0,\quad\quad & -1\leq n\,{\rm or}\,m \leq 1
\end{matrix} \right.\label{U2=1new}
\ee
The ${ss}$ matrix in the RHS of this equation will be denoted by ${\bf 1}_{ss}$.

\noindent On the other hand, $\EU_{(-i)}\EU_{(-i)}$ is singular.\\

\noindent After twist conjugation we get also
\be
\bar\EU_{(-i)}\,\bar\EU_{(-i)}=\bar\EU_{(i)}\,\bar\EU_{(-i)}=
\bar\EU_{(-i)}\,\bar\EU_{(i)}= {\bf 1}_{ss}
\label{barU2=1}
\ee
while $\bar\EU_{(i)}\,\bar\EU_{(i)}$ is singular.

On the basis of the previous discussion one can better understand the
origin of the singularity, mentioned in subsection 2.2,
which arises if $p_i$ is simply replaced by $p_0$. The latter contains both
$p_i$ and $p_{-i}$ and we have seen above that when two $p_{-i}$
simultaneously enter into the game we cannot avoid a singularity.

\subsection{Fundamental properties of $V^{rs}$}

The calculations relevant to the fundamental properties of the Neumann coefficients
for the real vertex $V^{rs}$ are completed in Appendix B. To summarize
the results obtained, after
incorporating those of Appendix B, in a compact form, we will use
the 3x3 matrix $z$ introduced in sec. 2.3, and introduce the $\infty \times 3$
matrix $u$, $u_{n,i}=U_{n,i}$ ($n\geq 2, -1 \geq i\geq 1$), as well as its twist
conjugate $\bar u$, and an analogous matrix $e$, $e_{n,i}=E_{n,i}$ . Then
\be
U^2 = (\EU +Z)(\EU+Z) = \EU^2+ \EU Z = {\bf 1}_{ss} - u\,z,\label{Usquare}
\ee
\be
\bar U^2 = (\bar\EU +Z)(\bar\EU+Z) = {\bf 1}_{ss}- \bar u\, z\label{Ubarsquare}
\ee
and
\be
E^2 = (\EE +Z)(\EE+Z) = \EE^2+ \EE Z= {\bf 1}_{ss} - e\,z\label{Esquare}
\ee

Likewise we have
\be
E\,U = \Ct\, U,\quad\quad
U\,E= U\,\Ct- \Ct \,e\,z- u\,z\label{EUUE}
\ee
i.e., after twisting and combining,
\be
E\,\bar U = \Ct\,\bar U,\quad\quad
\bar U\,E=\bar U\,\Ct+ \,e\,z-\bar u\,z\,\hat c\label{EbarUbarUE}
\ee
where $\hat c$ is $\Ct$ reduced to the first $3\times 3$ block.

It was noted in sec. 2.3 that we could add the $3\times 3$ matrix $z$ to
$E,U$ and $\bar U$ without changing the three string vertex. We use this freedom
to redefine the vertex Neumann matrices. This simple move will dramatically
simplify everything.

Let us set
\be
E' = E+ z, \quad\quad U' = U+ z, \quad\quad \bar U' =\bar U+ z\label{E'U'}
\ee
and let us compute $U'^2$:
\be
U'^{2}= U^2+ U\, z+ z^2 = {\bf 1}_{ss}- u\,z+ u\,z + 1_{3x3}
= 1\label{U'2}
\ee
where, now, 1 is the identity matrix in the full range $-1 \leq n,m<\infty$.
In the last derivation we have used the fact that $z\, U=0$.
Similarly we can prove that
\be
\bar U'^2=1, \quad \quad E'^2=1\label{E'2}
\ee

Moreover, using again the results of subsection 2.6:
\be
&&E'U'= \Ct U + 1_{3x3} + e\,z \label{E'U'1}\\
&&U'E' = U\Ct - \Ct \,e\,z +  1_{3x3} \label{U'E'}
\ee
Twist--conjugating the second equation we get
\be
\bar U'E' = \Ct U  + e\,z +  1_{3x3} \label{barU'E'}
\ee
Therefore
\be
E'U' = \bar U' E'\label{E'U'barU'E'}
\ee
Twist--conjugating this
\be
E'\bar U' = U' E'\label{E'barU'U'E'}
\ee
We can now define two types of $X$ matrices, $X_E^{rs}= E' \hat V^{'rs}$ and
$X'^{rs}= \Ct \hat V^{'rs}$ ($\hat V^{'rs}=\hat V^{rs}+z\delta_{rs}$),
\be
X_E^{rs} =  \frac{1}{3}(1+\bar{\alpha}^{r-s}E' U'+
\alpha^{r-s}E'\bar{U}')\label{X_E}
\ee
or
\be
X'^{rs} =  \frac{1}{3}(\Ct E'+\bar{\alpha}^{r-s}\Ct U'+
\alpha^{r-s}\Ct\bar{U}')\label{X'rs}
\ee
The ghost three string vertex Neumann matrices $\hat V^{rs}$, obtained from
$X'^{rs}$ dropping the $z$ matrix,  are
those defined in section 1.1, eq.(\ref{Vnmrs}).

Our first aim is to prove that
\be
X_E^{rs}X_E^{r's'}= X_E^{r's'}X_E^{rs}\label{XEXEcomm}
\ee
and
\be
X^{'rs}X^{'r's'}= X^{'r's'}X^{'rs}\label{X'X'comm}
\ee
for any $r,s,r',s'$. For conciseness we write $\alpha^{s-r}=\beta$ and
$\alpha^{s'-r'}=\beta'$. To start with, using (\ref{twist}) we get
\be
X^{'rs}X^{'r's'}&=& \frac 19 \Big( E'E' + \beta' E'U'+ \bar \beta' E'\bar U' +
\beta \bar U' E' + \beta\beta' \bar U'U'\0\\
&+&\beta \bar \beta' \bar U^{'2} +\bar \beta U'E' +\bar \beta\beta' U^{'2}+
\bar \beta \bar \beta'
U'\bar U'\Big)\label{XrsXr's'}
\ee
Similarly
\be
X^{'r's'}X^{'rs}&=& \frac 19 \Big( E'E' + \beta E'U'+ \bar \beta E'\bar U' +
\beta' \bar U' E' + \beta'\beta \bar U'U'\0\\
&+&\beta' \bar \beta \bar U^{'2} +\bar \beta' U'E' +\bar \beta'\beta U^{'2}+
\bar \beta' \bar \beta
U'\bar U'\Big)\label{Xr's'Xrs}
\ee
The necessary conditions for (\ref{X'X'comm}) to hold are
\be
E'U'=\bar U' E', \quad\quad E'\bar U'= U'E', \quad \quad U^{'2} =\bar U^{'2}
\label{commcond}
\ee
This is certainly true on the basis of the previous results. In the same way
one can prove (\ref{XEXEcomm}). Moreover it is not hard to show
\be
&&X_E+X_E^++ X_E^-=1\0\\
&&X_E^+X_E^-= X_E^2-X_E\label{XEfundid}\\
&&X_E^2+(X_E^+)^2+(X_E^-)^2=1\0\\
&&(X_E^+)^3+(X_E^-)^3=1+2X_E^3-3X_E^2\0
\ee
The analogous relations for the $X'$ matrices are not as simple.
Unfortunately the $X_E$'s are not the matrices that are
going to appear in the star product of two string states (see below).
In the star product the relevant matrices are the $X'$'s.
We have
\be
X'-X_E= \Ct E'-1= -1_{3\times 3}+ \hat c z + \hat c  e
\equiv {\mathfrak E}\label{calE}
\ee
while $X_E^\pm= X'^\pm$.
It is easy to see that ${\mathfrak E}$ has only two nontrivial
columns, precisely the only nonvanishing entries are
${\mathfrak E}_{-1,2n+1}= {\mathfrak E}_{1,2n+1}=- (-1)^n (2n+1)$,
$n=-1,0,1,\ldots$. Moreover ${\mathfrak E}$ commutes with all
$X_E$'s and $X'$'s matrices, and ${\mathfrak E}^2=-2{\mathfrak E}$,
${\mathfrak E} X'=X' {\mathfrak E}= -{\mathfrak E}$. Finally one can prove
\be
&&X'+X'^++ X'^-=1+{\mathfrak E}\0\\
&&X'^+X'^-= X'^2-X'+ {\mathfrak E}\label{X'fundid}\\
&&X'^2+(X'^+)^2+(X'^-)^2=1\0\\
&&(X'^+)^3+(X'^-)^3=1+2X'^3-3X'^2-2{\mathfrak E}\0
\ee

\subsection{Fundamental properties of $V_{(i)}^{rs}$ and $V_{(-i)}^{rs}$ }

We proceed in analogy to the previous case. Although the procedure is the same
many important details are different and we are forced to repeat the derivations.
In the following we consider only $V_{(i)}^{rs}$, because everything concerning
 $V_{(-i)}^{rs}$ can be obtained by twist--conjugation (${\cal C} V_{(i)}^{rs}
{\cal C}= V_{(-i)}^{sr}$).
Many calculations relevant for the fundamental properties of the Neumann coefficients
for the vertex $V_{(i)}^{rs}$ can be found in Appendix B. As before we will
summarize the results in a compact form by means of
the 3x3 matrix $z$  and the $\infty \times 3$
matrix $u_{(i)}$, $u_{(i)n,i}=U_{(i)n,i}$ ($n\geq 2, -1 \geq i\geq 1$),
as well as its twist conjugate $\bar u_{(i)}$ and the analogous
matrix $e_{(i)}$, $e_{(i)n,i}=E_{(i)n,i}$ . Then
\be
U_{(-i)} U_{(i)}= (\EU_{(-i)} +Z)(\EU_{(i)}+Z) = \EU_{(-i)} \EU_{(i)}+ \EU_{(-i)} Z =
{\bf 1}_{ss}
- u_{(-i)}\,z,\label{Uisquare}
\ee
\be
\bar U_{(-i)} \bar U_{(i)}= (\bar\EU_{(-i)} +Z)(\bar\EU_{(i)}+Z) =
{\bf 1}_{ss}- \bar u_{(-i)}\, z\label{Uibarsquare}
\ee
and
\be
E_{(-i)} E_{(i)} = (\EE_{(-i)} +Z)(\EE_{(i)}+Z) =
{\bf 1}_{ss} - e_{(-i)}\,z\label{Eisquare}
\ee

Likewise we have
\be
E_{(-i)}\,U_{(i)}  &=& \Ct\, U_{(i)} ,\quad\quad
U_{(-i)}\,E_{(i)}= U_{(-i)}\,\Ct+ \,e_{(-i)}\,z- u_{(-i)}\,z\label{EiUi}\\
E_{(-i)}\,\bar U_{(i)}&=& \Ct\,\bar U_{(-i)},\quad\quad
\bar U_{(-i)}\,E_{(i)}=\bar U_{(-i)}\,\Ct+ \,e_{(-i)}\,z-
\bar u_{(-i)}\,z\,\hat c\label{EibarUbarUE}
\ee
where $\hat c$ is $\Ct$ reduced to the first $3\times 3$ block.

Now, as noted before, we are allowed to add the $3\times 3$ matrix $z$ to
the $E,U$ and $\bar U$ matrices without changing the three string vertex. We use this freedom
to redefine the vertex Neumann matrices. Again, this will simplify everything.

\noindent Let us set
\be
E'_{(\pm i)} = E_{(\pm i)}+ z, \quad\quad U'_{(\pm i)} = U_{(\pm i)}+ z,
\quad\quad \bar U'_{(\pm i)} =\bar U_{(\pm i)}+ z\label{E'iU'i}
\ee
and let us compute $U'_{(- i)}U'_{(i)}$:
\be
U'_{(- i)}U'_{( i)}= U_{(- i)}  U_{(-i)}+ U_{(i)}\, z+ z^2
= {\bf 1}_{ss}- u_{(-i)}\,z+ u_{(-i)}\,z + 1_{3x3}
= 1\label{U'i2}
\ee
where, now, 1 is the identity matrix in the full range $-1 \leq n,m<\infty$.
In the last derivation we have used the fact that $z\, U_{(\pm i)}=0$.
Similarly we can prove that
\be
\bar U'_{(- i)}\bar U'_{(i)}=1,
\quad \quad E'_{(- i)}E'_{(i)}=1=E'_{(i)}E'_{(-i)}\label{Ei'2}
\ee
Moreover
\be
E'_{(-i)}\,U'_{(i)}  &=&  \bar U'_{(-i)}E'_{(i)} ,\quad\quad
E'_{(-i)}\bar U'_{(i)}=U'_{(-i)}\,E'_{(i)} \label{E'iU'iUi}
\ee
As before we can now define two types of $X$ matrices, $X_{(i)E}^{rs}= E' \hat V_{(i)}^{'rs}$ and
$X_{(i)}^{'rs}= \Ct \hat V_{(i)}^{'rs}$
($\hat V_{(i)}^{'rs}=\hat V_{(i)}^{rs}+z\delta_{rs}$):
\be
X_{(i)E}^{rs} =  \frac{1}{3}(1+\bar{\alpha}^{r-s}E_{(-i)}' U_{(i)}'+
\alpha^{r-s}E_{(-i)}'\bar{U}_{(i)}')\label{XiE}
\ee
or
\be
X_{(i)}^{'rs} =  \frac{1}{3}(\Ct E_{(i)}'+\bar{\alpha}^{r-s}\Ct U_{(i)}'+
\alpha^{r-s}\Ct\bar{U}_{(i)}')\label{Xi'rs}
\ee
The ghost three string vertex Neumann matrices $\hat V^{rs}_{(i)}$, obtained
from $X_{(i)}'^{rs}$ dropping the $z$ matrix, are
those defined in section 1.1, eq.(\ref{Vnmrsi}).

\noindent Using the previous results and the methods of the previous subsection
we can prove that
\be
X_{(i)E}^{rs}X_{(i)E}^{r's'}= X_{(i)E}^{r's'}X_{(i)E}^{rs}\label{XiEXiEcomm}
\ee
and
\be
X_{(i)}^{'rs}X_{(i)}^{'r's'}= X_{(i)}^{'r's'}X_{(i)}^{'rs}\label{Xi'Xi'comm}
\ee
for any $r,s,r',s'$. In addition the $X_{(i)E}$ matrices commute
with the $X_{(i)}'$ ones.

\noindent Moreover it is not hard to show that
\be
&&X_{(i)E}+X_{(i)E}^++ X_{(i)E}^-=1\0\\
&&X_{(i)E}^+X_{(i)E}^-= X_{(i)E}^2-X_{(i)E}\label{XiEifundid}\\
&&X_{(i)E}^2+(X_{(i)E}^+)^2+(X_{(i)E}^-)^2=1\0\\
&&(X_{(i)E}^+)^3+(X_{(i)E}^-)^3=1+2X_{(i)E}^3-3X_{(i)E}^2\0
\ee
The analogous relations for the $X_{(i)}'$ matrices are not as simple.
Unfortunately in the star product the relevant matrices
are the $X_{(i)}'$'s. We have
\be
X_{(i)}'-X_{(i)E}=-1_{3\times 3}+ \hat c z + \hat c  e
\equiv {\mathfrak E}\label{calE'}
\ee
while $X_{(i)E}^\pm= X_{(i)}'^\pm$. Moreover ${\mathfrak E}$ commutes with all
$X_{(i)E}$'s and $X_{(i)}'$'s matrices, and ${\mathfrak E}^2=-2{\mathfrak E}$,
${\mathfrak E} X_{(i)}'=X_{(i)}' {\mathfrak E}= -{\mathfrak E}$. Finally one can prove
\be
&&X_{(i)}'+X_{(i)}'^++ X_{(i)}'^-=1+{\mathfrak E}\0\\
&&X_{(i)}'^+X_{(i)}'^-= X_{(i)}'^2-X_{(i)}'+ {\mathfrak E}\label{Xi'fundid}\\
&&X_{(i)}'^2+(X_{(i)}'^+)^2+(X_{(i)}'^-)^2=1\0\\
&&(X_{(i)}'^+)^3+(X_{(i)}'^-)^3=1+2X_{(i)}'^3-3X_{(i)}'^2-2{\mathfrak E}\0
\ee

\section{Commutators with $K_1$}

The fact that the twisted Neumann matrices of the three strings vertices
introduced in the previous section commute opens the way to their
diagonalization. The basis we will use is formed by the eigenvectors of the
matrix $G$, which represents together with $H^T$ the operator $K_1$.
The operator $K_1$ plays a fundamental auxiliary role in SFT and in
particular in relation with the three ghost strings vertex.
Let us recall the relevant definitions from I.
\be
K_1= \sum_{p,q\geq -1} c_p^\dagger\, G_{pq} \,b_q + \sum_{p,q\geq 2}
b_{p}^\dagger\,H_{ p q}\,
c_{q}- 3c_2\,b_{-1}\label{K1gh}
\ee
where
\be
&&G_{pq}= (p-1) \delta_{p+1,q} + (p+1)\delta_{p-1,q},\0\\
&&H_{ p q}= ( p+2) \delta_{ p+1, q}
+ ( p-2)\delta_{ p-1, q}\label{GF}
\ee
Therefore $G$ is a square long--legged matrix and $H$ a square
short--legged one. In the common overlap
we have $G=H^T$. Since we are able to completely solve the spectral
problem for $G$, the latter will be a constant reference
in the forthcoming developments.

The generating functions for $G$ and $H$ are
\be
G(z,w)&=& \frac {z-2w+3zw^2}{zw(1-zw)^2}\label{Gzw}\\
H(z,w)&=& \frac{wz^2(3+w^2-2zw)}{(1-zw)^2}\label{Hzw}
\ee
As matrices they have the block structure
\be
G= \left(\begin{matrix}g_0 &|&0\\
                --&{}&--\\
                {\hat g} &|&\hat G
\end{matrix}\right), \quad\quad H^T= \left(\begin{matrix}0 &|&0\\
                --&{}&--\\
                0 &|&\hat G
\end{matrix}\right),
\label{AB}
\ee
where $g_0$ is a $3\times 3$ matrix and $\hat g$ represents a $\infty \times 3$ matrix
with only one entry (the one in position (1,2)) different from zero.

\subsection{$G$ and the $V^{rs}$ vertex}

Let us consider $G$ from now on. We want to prove that it commutes with $\Ct U$.
Since $G$ anticommutes with $\Ct$, this is equivalent to compute the
anticommutator of $G$ with $U_{(i)}$, which can be easily done both numerically and
analytically. Here is the analytic proof
\be
( \EU_{(i)}\, G)_{nm} &=&\oint\frac{dz}{2\pi i}\frac{1}{z^{n+1}}
\oint\frac{d\z}{2\pi i}\oint\frac{d\zh}{2\pi i} \oint\frac{dw}{2\pi i}
\frac{1}{w^{m+1}}\frac {\z\zh}{\z\zh-1} \label{EUG1}\\
&&\cdot \frac{f(z)}{f(\z)} \Big{(}\frac{1}{1+z\z} -\frac{\z}{\z-z}\Big{)}
  \Big{(}1- p_i(z,\zeta) \Big{)} \frac {\zh -2w+3\zh w^2}{w\zh (1-w\zh)^2}=*\0
\ee
Integrating wrt to $\zh$ (pole at $\zh=\frac 1\z$)
\be
*&=& \oint\frac{dz}{2\pi i}\frac{1}{z^{n+1}}
\oint\frac{d\z}{2\pi i} \oint\frac{dw}{2\pi i}
\frac{1}{w^{m+1}}\frac {f(z)}{f(\z)}
\Big{(}\frac{1}{1+z\z} -\frac{\z}{\z-z}\Big{)} \label{EUG2}\\
&& \cdot \Big{(}1- p_i(z,\zeta) \Big{)}\frac {\z}{w}
\frac {1 -2w\z+3 w^2}{(\z-w)^2} \0\\
&=& \oint\frac{dz}{2\pi i}\frac{1}{z^{n+1}}
 \oint\frac{dw}{2\pi i}
\frac{1}{w^{m+1}}\left[ - \frac {z^2}w \,\frac {1 -2w z+3 w^2}{(z-w)^2}\right.\0\\
&&+ \left.\frac d{d\z}\left( \frac {f(z)}{f(\z)}
\Big{(}\frac{1}{1+z\z} -\frac{\z}{\z-z}\Big{)}
\Big{(}1- p_i(z,\zeta) \Big{)}\frac {\z}{w}
(1 -2w\z+3 w^2) \right)_{\z=w}\right] \0
\ee
The last two terms come from poles at $\z=z$ and $\z=w$, respectively.

Let us add the ordering term and repeat the same procedure.
\be
(Z\, G)_{nm} &=&\oint\frac{dz}{2\pi i}\frac{1}{z^{n+1}}
\oint\frac{d\z}{2\pi i}\oint\frac{d\zh}{2\pi i} \oint\frac{dw}{2\pi i}
\frac{1}{w^{m+1}}\frac {\z\zh}{\z\zh-1} \label{ZG}\\
&&\cdot \left(-\frac {z^2}{\z} \,\frac 1{z-\z}\right)\,
 \frac {\zh -2w+3\zh w^2}{w\zh (1-w\zh)^2}\0\\
&=& \oint\frac{dz}{2\pi i}\frac{1}{z^{n+1}}
\oint\frac{d\z}{2\pi i}  \oint\frac{dw}{2\pi i}\frac{1}{w^{m+1}}
\left(-\frac {z^2} {z-\z} \,  \frac {1 -2w\z+3 w^2}{w (\z-w)^2}\right)\0\\
&=&\oint\frac{dz}{2\pi i}\frac{1}{z^{n+1}}
\oint\frac{dw}{2\pi i}\frac{1}{w^{m+1}}
\left[\frac {z^2}w \frac  {1 -2w z+3 w^2}{ (z-w)^2}
-\frac d{d\z} \left(  \frac {z^2}w \frac {1 -2w\z+3 w^2}{z-\z}\right)_{\z=w}\right]\0
\ee
The first piece in the RHS of (\ref{EUG2}) cancels exactly
the first piece in the RHS of (\ref{ZG}), so we have only to
evaluate the sum of the two remaining derivatives wrt $\z$.

Next let us compute $G U$. We start with
\be
(G\,\EU_{(i)})_{nm} &=&\oint\frac{dz}{2\pi i}\frac{1}{z^{n+1}}
\oint\frac{d\z}{2\pi i}\oint\frac{d\zh}{2\pi i} \oint\frac{dw}{2\pi i}
\frac{1}{w^{m+1}}\frac {\z\zh}{\z\zh-1} \label{GEU1}\\
&&\cdot\frac {z-2\z+3z\z^2}{z\z(1-z\z)^2} \frac {f(\zh)}{f(w)}
\Big{(}\frac{1}{1+w\zh} -\frac{w}{w-\zh}\Big{)}\Big{(}1- p_i(\zh,w) \Big{)}=*\0
\ee
Integrating over $\z$ we obtain
\be
*&=&\oint\frac{dz}{2\pi i}\frac{1}{z^{n+1}}
\oint\frac{d\zh}{2\pi i} \oint\frac{dw}{2\pi i}
\frac{1}{w^{m+1}}\, \frac {z\zh^2-2\zh+3z}{z(\zh-z)^2} \label{GEU2}\\
&& \cdot\frac {f(\zh)}{f(w)}
\Big{(}\frac{1}{1+w\zh} -\frac{w}{w-\zh}\Big{)}\Big{(}1- p_i(\zh,w) \Big{)}\0\\
&=&\oint\frac{dz}{2\pi i}\frac{1}{z^{n+1}}
\oint\frac{dw}{2\pi i}\frac{1}{w^{m+1}}\, \left[ \frac wz \,
\frac {zw^2-2w+3z}{(w-z)^2}\right.\0\\
&&+\left. \frac d{d\zh} \left( \frac {f(\zh)}{f(w)}
\Big{(}\frac{1}{1+w\zh} -\frac{w}{w-\zh}\Big{)}\Big{(}1- p_i(\zh,w) \Big{)}
\,\frac {z\zh^2-2\zh+3z}{z}\right)_{\zh=z}\right]\0
\ee
The last two terms come from poles at $\zh=w$ and $\zh= z$, respectively.

The ordering term gives,
\be
(G\, Z)_{nm} &=&\oint\frac{dz}{2\pi i}\frac{1}{z^{n+1}}
\oint\frac{d\z}{2\pi i} \oint\frac{d\zh}{2\pi i} \oint\frac{dw}{2\pi i}
\frac{1}{w^{m+1}}\frac {\z\zh}{\z\zh-1} \label{GZ}\\
&&\cdot \frac {z-2\z+3z\z^2}{z\z(1-z\z)^2} \, \frac {\zh^2}{w} \frac {-1}{\zh-w}\0\\
&=&-\oint\frac{dz}{2\pi i}\frac{1}{z^{n+1}}\oint\frac{d\zh}{2\pi i}
\oint\frac{dw}{2\pi i}\frac{1}{w^{m+1}}\, \frac {z\zh^2-2\zh+3z}{z(\zh-z)^2}
\,  \frac {\zh^2}{w} \frac 1{\zh-w}\0\\
&=&-\oint\frac{dz}{2\pi i}\frac{1}{z^{n+1}}
\oint\frac{dw}{2\pi i}\frac{1}{w^{m+1}}\,
\left[ \frac wz \,\frac {zw^2-2w+3z}{(w-z)^2} + \frac d{d\zh}
\left(\frac {z\zh^2-2\zh+3z}{zw}
\,  \frac {\zh^2}{\zh-w}\right)_{\zh=z} \right]\0
\ee
In $G\,U_{(i)}$ the first piece in the RHS of (\ref{GEU2}) cancels the first piece
in the RHS of (\ref{GZ}). What remains is a derivative wrt to $\z$ in
(\ref{EUG2},\ref{ZG}) and wrt to $\zh$ in (\ref{GEU2},\ref{ZG}).
The derivative in (\ref{EUG2}) gives
\be
\frac {f(z)}{f(w)}\left[- \frac 43 i \left(\frac 1{1+zw}-\frac w{w-z}\right)
(1-p_i(z,w))+ \frac {z^2}{w}\, \frac{(1+w^2)^3}{(w-z)^2(1+wz)^2}\right]\label{GEU3}
\ee
The derivative in (\ref{GEU2}) gives
\be
\frac {f(z)}{f(w)}\left[ \frac 43 i \left(\frac 1{1+zw}-\frac w{w-z}\right)
(1-p_i(z,w))- \frac {z^2}{w}\, \frac{(1+w^2)^3}{(w-z)^2(1+wz)^2}\right]\label{EUG3}
\ee
The sum of these two terms vanishes.

The derivative in (\ref{GZ}) gives
\be
\frac{z^2(-1-3w^2+2wz)}{w(w-z)^2}\label{GZ1}
\ee
The derivative in (\ref{ZG}) gives
\be
\frac{z^2(1-3z^2+4wz)}{w(w-z)^2}\label{ZG1}
\ee
The sum of these two terms is
\be
-3\frac{z^2}w\label{badterm}
\ee

Therefore, apart from this term we get $ U_{(i)}G+G\,U_{(i)}=0$, or
\be
[\Ct U_{(i)}, G]_{nm}=-3 \delta_{n,2}\delta_{m,-1}\label{CUGcomm}
\ee
This is of course true also for $\Ct \bar U_{(-i)}$.

It is even simpler to prove that
\be
[\Ct E_{(\pm i)}, G]_{nm}=-3 \delta_{n,2}\delta_{m,-1}\label{CEGcomm}
\ee

The anomaly in the RHS of these commutators is a well--known effect of not
having included the last term in the RHS of (\ref{K1gh}) in the definition of
$G$ and $H$. We can easily cancel this anomaly by adding the $3\times 3$ matrix $z$ to
$E,U_{(i)}$ and $\bar U_{(-i)}$. For let us compute $[\Ct U'_{(i)}, G]$.
The only change with the commutator
$[\Ct U_{(i)}, G]$ is the addition of $- \hat g \,z$. This vanishes everywhere
except for the (2,-1) entry, which equals 3. Therefore
\be
[\Ct U'_{(i)},G]=0\label{CU'G}
\ee
Likewise one can prove that
\be
[\Ct \bar U'_{(-i)},G]=0, \quad\quad [\Ct E'_{(\pm i)}, G]=0\label{CE'G}
\ee
Therefore we conclude that $X_{(i)}^{rs}+ \delta^{rs}\hat c z$ commute with $G$.

Summarizing the results of this subsection: the matrices $X^{'rs}, X_{(i)}^{rs}$
as well as the matrix ${\mathfrak E}$ commute with $G$, therefore they will
be diagonal in the bases of eigenvectors of the latter matrix.
To conclude
let us recall once again that we can always add a matrix $z_{ij}$ to $V^{rr}_{ij}$,
since, as pointed out in sec. 2.4 this ambiguity is allowed by the formalism.

\subsection{$G$ and the $V_{(i)}^{rs}$ vertex}

We need to prove that $G$ commutes with $\Ct U_{(\pm i)}$.
The case $U_{(i)}$ has just been analyzed. The case  $U_{(-i)}$ requires
minor changes. $(\EU_{(-i)}G)_{nm}$ is the same as $(\EU_{(i)}G)_{nm}$,
eqs.(\ref{EUG1},\ref{EUG2})
with the simple replacement $p_i\rightarrow p_{-i}$, and
 $G\,(\EU_{(-i)})_{nm}$ is the same as $G\,(\EU_{(i)})_{nm}$ with the same replacement.

After this replacement the derivative in (\ref{EUG2}) gives
\be
\frac {f(z)}{f(w)}\left[- \frac 43 i \left(\frac 1{1+zw}-\frac w{w-z}\right)
(1-p_i(z,w))+ \frac {z^2}{w}\, \frac{(1+w^2)^3}{(w-z)^2(1+wz)^2}\right]\0
\ee
The derivative in (\ref{GEU2}) gives
\be
\frac {f(z)}{f(w)}\left[ \frac 43 i \left(\frac 1{1+zw}-\frac w{w-z}\right)
(1-p_i(z,w))- \frac {z^2}{w}\, \frac{(1+w^2)^3}{(w-z)^2(1+wz)^2}\right]\0
\ee
that is, they are the same as before replacement, and the sum of these two terms vanishes.

The derivatives in (\ref{GZ}) and (\ref{ZG}) of course remain the same, thus
their sum is again $-3\frac{z^2}w$.

Therefore we get
\be
[\Ct U_{(-i)}, G]_{nm}=-3 \delta_{n,2}\delta_{m,-1}\label{CUmiGcomm}
\ee
This is of course true also for $\Ct \bar U_{(i)}$. Moreover
\be
[\Ct E_{(-i)}, G]_{nm}=-3 \delta_{n,2}\delta_{m,-1}\label{CEmiGcomm}
\ee

Again we can eliminate the anomaly in the RHS of these commutators by adding
$z$. Indeed
\be
[\Ct U'_{(-i)},G]=0\label{CUmi'G}
\ee
Likewise
\be
[\Ct \bar U'_{(i)},G]=0, \quad\quad [\Ct \,E_{(\pm i)}', G]=0\label{CEmi'G}
\ee
Therefore we conclude that $X_{(\pm i)}^{rs}+ \delta^{rs} \hat c z$ commute with $G$.

Summarizing the results of this subsection: the matrices
 $X_{(\pm i)}^{'rs}, X_{(\pm i)E}^{rs}$
as well as the matrix ${\mathfrak E}$ commute with $G$, therefore they will
be diagonal in the bases of eigenvectors of the latter matrix.
Our next purpose is to introduce such bases.

\section{The weight 2 and -1 bases}

This section is devoted to the bases of eigenfunctions of
$G$. As it turns out the $b,c$ bases introduced in I were
incomplete, because only the continuous eigenvalues of $G$ were taken
into account, while the discrete ones were disregarded.
Once complete bases are introduced, we will be able to
write down spectral formulas for the $G$ matrix and for several different
Neumann coefficient matrices.

We will also write reconstruction formulas for the $A,B,C,D$ matrices (for
their definition see eq.(\ref{ABCD}) below)
introduced in I. This analysis was started in \cite{BST} (see also
\cite{Belov1,Belov2,BeLove}), where spectral
formulas were derived on a heuristic basis. We are now in the condition
to clarify to what extent those formulas are valid.

To start with let us recall the definitions of the weight 2 and -1
continuous bases of eigenvectors of $G$. The unnormalized weight
2 basis is given by
\be
f^{(2)}_\k(z) = \sum_{n=2} V_n^{(2)}(\k) \, z^{n-2}\label{bbasis}
\ee
in terms of the generating function
\be
f^{(2)}_\k(z) = \left(\frac 1{1+z^2}\right)^2 \, e^{\k \arctan (z)}=
1+\k z +\left(\frac {\k^2}2 -2\right) z^2+\ldots\label{genf}
\ee
Following \cite{Belov1,Belov2}, (see also Appendix B of \cite{BMST}),
we normalize the eigenfunctions as follows
\be
\tilde V_n^{(2)}(\k)= {\sqrt{A_2(\k)}} V_n^{(2)}(\k)
\label{nbbasis}
\ee
where
\be
 A_2(\k) = \frac {\k (\k^2+4)}
{2\sinh \left(\frac {\pi \k}2\right)}\0
\ee

The unnormalized weight -1 basis is given by
\be
f_\k^{(-1)}(z) = \sum_{n=-1} V_n^{(-1)}(\k) \, z^{n+1}\label{cbasis}
\ee
in terms of the generating function
\be
f^{(-1)}_\k(z) = (1+z^2) \, e^{\k \arctan (z)}=
1+\k z +\left(\frac {\k^2}2 +1\right) z^2+\ldots\label{f-1}
\ee
The normalized one is
\be
\tilde V_n^{(-1)}(\k)=\sqrt{A_{-1}(\k)} V_n^{(-1)}(\k),
\quad\quad \sqrt{A_{-1}(\k)} =  {\cal P} \frac 1{\k}
\frac {\sqrt{A_2(\k)}}{\k^2+4}\label{normV-1n}
\ee
where ${\cal P}$ denotes principal value. We reported in \cite{BMST}
the `bi--completeness'
\be
\int_{-\infty}^{\infty} d\k\, \tilde V^{(-1)}_{n}(\k)\,\tilde
V^{(2)}_{m}(\k)= \delta_{n,m}, \quad\quad n\geq 2\label{bicompl}
\ee
and bi--orthogonality relation
\be
\sum_{n=2}^\infty \, \tilde V^{(-1)}_{n}(\k)\, \tilde V^{(2)}_{n}(\k')=
\delta(\k,\k')\label{biortho}
\ee
taking them from \cite{BeLove}.

As for the first three elements of the -1 basis, $\tilde V_i^{(-1)}(\k)$, $i=-1,0,1$,
they can be expressed in terms of the others
(see \cite{BeLove} and Appendix B of \cite{BMST})
\be
\tilde V^{(-1)}_i (\k) = \sum_{n=2}^\infty b_{i,n}\, \tilde V_{n}^{(-1)}(\k)\label{VibVn}
\ee
One can easily show that
\be
b_{-1,2n+3}= (-1)^n (n+1),\quad\quad   b_{0,2n+2}= (-1)^n, \quad\quad
b_{1,2n+3}=(-1)^n (n+2)\label{b-1n}
\ee

However the `bi--completeness' relation (\ref{bicompl}) is not complete.
The reason can be understood by studying the spectrum of $G$.
The matrix $G$ looks as follows
\be
G=\left(\begin{matrix}  0&-2&0&0&0&...\\
                        1&0&-1&0&0&...\\
0&2&0&0&0&...\\
0&0&3&0&1&...\\
0&0&0&4&0&...\\
.&.&.&.&.&...
\end{matrix}\right)\label{Gmatrix}
\ee
It is easy to see that the $g_0$ matrix (the upper left $3\times 3$ block of $G$)
has left eigenvectors $(1,0,1), (1,\pm 2i,-1)$ with eigenvalues
$0, \pm 2i$, respectively. By adding to this eigenvectors a sequence of
zeroes in position $2,3,...$ they become left eigenvectors of the full
$G$ matrix, i.e.
\be
V^{(-1)}(0)=(1,0,1,0,0,0,\ldots),\quad\quad
V^{(-1)}(\pm 2i)=(1,\pm 2i,-1,0,0,0,\ldots)\label{V-1}
\ee
are left eigenvectors of $G$ with eigenvalues 0 and $\pm 2i$, respectively.
It is easy to see that they correspond to the
vectors $V^{(-1)}_n(\k)$ for $\k=0,\pm 2i$, respectively. In other words
the discrete eigenvectors are the same as the continuous eigenvectors
evaluated at the corresponding eigenvalue in the $\k$ plane.

$g_0$ has also right eigenvectors, with the same eigenvalues. One can
easily check that $\left(\begin{matrix} 1\\0\\1\end{matrix}\right),
\left(\begin{matrix} 1\\\mp i\\-1\end{matrix}\right) $ are right
eigenvectors with eigenvalues $0,\pm 2i$ respectively. However,
in order to get the right eigenvectors of $G$ it is not enough to
add an infinite sequence of zeroes to the eigenvectors of $g_0$, because
of the presence of a nonzero entry in position (2,1) of $G$.

The problem of finding the right eigenvectors of $G$ can however
be solved in an algebraic way. One adds unknowns in position $2,3,...$
of the vectors and imposes that the resulting vectors be
eigenvectors of $G$ with the above discrete eigenvalues. One easily gets
\be
v^{(2)}(0)= \left(\begin{matrix} 1\\0\\1\\0\\-3\\0\\ \dot :\end{matrix}\right),
\quad\quad
v^{(2)}(\pm 2i)= \left(\begin{matrix} 1\\\mp i\\-1\\ \pm i\\ 1\\ \mp i\\
\dot :\end{matrix}\right)\label{V(2)}
\ee
More precisely the entries $v_n^0$ of $v^{(2)}(0)$ are zero for $n$ even  and
equal $v^0_{2n+1} = (-1)^n (2n+1)=-(b_{-1,2n+1}+b_{1,2n+1})$ for $n$ odd.
The entries $v_n^{\pm}$ of $V^{(2)}(\pm 2i)$ are $v_{2n}^{\pm} = \mp i(-1)^n=
\pm i b_{0,2n}$ for $n$ even and $v^\pm_{2n+1} = -(-1)^n = b_{1,2n+1}-b_{-1,2n+1}$
for $n$ odd. The $b$ coefficients are the familiar ones
\be
b_{0,2n}= -(-1)^n, \quad\quad b_{1,2n+1}=-(-1)^n (n+1),
\quad\quad b_{-1,2n+1}=-(-1)^n n\label{bn}
\ee

Let us stress that $v^{(2)}(0), v^{(2)}(\pm 2i)$ are
{\it different} from the values taken by the continuous $V^{(2)}(\k)$ basis
evaluated at $\k=0,\pm 2i$. This is the reason why we use  for
these discrete eigenvectors different symbols form the continuous ones,
while for $V^{(-1)}$ we use the same notation for both.

Next we normalize the discrete eigenvectors as follows
\be
&&\tilde v^{(2)}(0) = \frac 1{\sqrt 2}  v^{(2)}(0),
\quad \quad \tilde v^{(2)}(\pm 2i) = \frac 12 v^{(2)}(\pm 2i)\0\\
&& \tilde V^{(-1)}(0) = \frac 1{\sqrt 2}  V^{(-1)}(0),
\quad \quad \tilde V^{(-1)}(\pm 2i) = \frac 12 V^{(-1)}(\pm 2i)\0
\ee
Using (\ref{VibVn}) it is easy to prove the following orthogonality
conditions\footnote{These
orthogonality conditions certainly hold for $\k$ away from the singularities
of the bases normalization factors (see the beginning of this section),
but must be otherwise used with extreme care.}
(we denote by $\xi$ the discrete eigenvalues $0,\pm 2i$)
\be
&& \sum_{n=-1}^\infty \tilde V_n^{(-1)}(\xi) \tilde v_n^{(2)}(\xi')=
\delta_{\xi,\xi'}\label{orthog}\\
&& \sum_{n=-1}^\infty \tilde V_n^{(-1)}(\k) \tilde v_n^{(2)}(\xi')=0\0\\
&&\sum_{n=-1}^\infty \tilde V_n^{(-1)}(\xi) \tilde V_n^{(2)}(\k)=0\0\\
&&\sum_{n=-1}^\infty \tilde V_n^{(-1)}(\k) \tilde V_n^{(2)}(\k')=
\delta(\k,\k')\0
\ee
To adopt this notation we have
added three zeroes to $V_n^{(2)}(\k)$ in position $n=-1,0,1$. In I,
see eq.(5.17), we showed that
\be
\sum_{m=-1}^\infty V_{m}^{(-1)}\, G_{mn} = \k\, V_n^{(-1)}\label{VGkV}
\ee
From the explicit proof it is evident that $G$ is diagonal on
$V^{(-1)}(\k)$ for {\it any complex value}
of $\k$. Therefore the second equation in (\ref{orthog}) holds as long as
$\k\neq \xi$. More about this issue later on.

Now let us consider the matrix
$I_{nm}= \sum_\xi \tilde v_n^{(2)}(\xi) \tilde V_m^{(-1)}(\xi) + \int d\k \,
\tilde V_n^{(2)}(\k) \tilde V_m^{(-1)}(\k)$. Using (\ref{orthog}) it is
easy to prove that, for instance,
$\sum_{m=-1} I_{nm} \tilde v_m^{(2)}(\xi) = \tilde v_n^{(2)}(\xi)$, etc.,
both from the right and from the left. Therefore we conclude that
\be
\sum_\xi \tilde v_n^{(2)}(\xi) \tilde V_m^{(-1)}(\xi) + \int_{-\infty}^\infty d\k \,
\tilde V_n^{(2)}(\k) \tilde V_m^{(-1)}(\k)=\delta_{nm},
\quad n,m\geq -1\label{bicomplete}
\ee
This is the correct bi--completeness relation. In this formula it is understood the the integration on $\k$
is along the real axis.

For future use we record
\be
v_n^{(2)}(\xi)+ \sum_{i=-1}^1\, b_{i,n}\, v_i^{(2)}(\xi)=0\label{VnxibV}
\ee

\subsubsection{Diagonalization of ${\mathfrak E}$}

We have already noticed that
the ${\mathfrak E}$ matrix of the previous subsection is diagonal
in the basis of $G$ eigenvectors. Indeed one easily realizes that
${\mathfrak E} V^{(2)}(\k)=0=V^{(-1)}(\k){\mathfrak E} $ for the
continuous eigenvectors, while
\be
&&{\mathfrak E} v^{(2)}(0)=-2 v^{(2)}(0), \quad\quad
{\mathfrak E} v^{(2)}(\pm 2i)=0\label{Eeigen1}\\
&& V^{(-1)}(0){\mathfrak E}= -2 V^{(-1)}(0),\quad
\quad V^{(-1)}(\pm 2i){\mathfrak E}=0 \0
\ee
for the discrete ones. Therefore the presence of ${\mathfrak E}$
in (\ref{X'fundid}) only affects the 0 discrete eigenvalue of $G$.

\subsection{Spectral formulas}

Using the spectral representation one can reconstruct $G$ from
its eigenvalues and eigenvectors:
\be
G_{nm} &=& \int_{-\infty}^{\infty} d\k \,\tilde V_n^{(2)}(\k)\,\k \,\tilde V_m^{(-1)}(\k)+\
\sum_\xi \tilde V_n^{(2)}(\xi)\, \xi\,\tilde V_m^{(-1)}(\xi)\label{Gspectral}\\
&=& \int d\k_{-\infty}^{\infty}  \,\frac {\k}{2 \sinh {\frac {\pi \k}2}}
V_n^{(2)}(\k) V_m^{(-1)}(\k)+ \frac i 2 \left(v_n^{(2)}(2i)\, V_m^{(-1)}(2i)
- v_n^{(2)}(-2i)\, V_m^{(-1)}(-2i)\right)\0
\ee
For instance, for $-1\leq i,j\leq 1$ we have
\be
G_{ij}= \frac i2\left( \left(\begin{matrix} 1\\-i\\-1\end{matrix}\right) \otimes
 (1,2i,-1)
- \left(\begin{matrix} 1\\i\\-1\end{matrix}\right) \otimes(1,-2i,-1)  \right)
= \left(\begin{matrix} 0&-2&0\\1&0&-1\\0&2&0\end{matrix}\right) \0
\ee
Next, using (\ref{V(2)}),
\be
G_{21} = \int d\k \,\frac {\k\left(\frac {\k^2}2+1\right)}
{2 \sinh {\frac {\pi \k}2}} +\frac i2 (i (-1) -(-i) (-1))= 2+1 =3\0
\ee
Similarly $G_{2,-1}=0, G_{2,0}=0$ and so on, as expected.

\subsubsection{Properties of the $G$ spectrum}

According to formula (5.17) of I (or (\ref{VGkV}) above), formally any value of $\k$ is a continuous eigenvalue
of $G$. Nothing prevents us from extending eqs.(\ref{orthog}) to complex $\k$,
provided we remain in a strip around the real axis.
Indeed for $\k=\pm 2 in$, with natural $n$, something happens: the $V^{(-1)}(\k)$ basis
has only a finite number of nonvanishing terms and the measure in the
spectral formula (\ref{Gspectral}) above has a simple pole. In fact,
if we compute
for instance the element $G_{21}$, we get 3 as above as long as
the integration contour
stretches from $-\infty$ to $+\infty$ in the strip $|\Im (\k)|<2$, it becomes
7 in the band $2<|\Im (\k)|<4$, -49 in the band $4<|\Im(\k)|<6$, etc.
These jumps are
due exactly to the contributions of the poles: as one moves the contour up
or down some poles may remain trapped inside the contour, giving rise to a
contribution which equals exactly the corresponding residue.

From this we learn that, unless we do not want to correct the results by hand
each time, the good region for the spectral formula of $G$ is $|\Im (\k)|<2$.

\subsubsection{The reconstruction of $A,B,C,D$}

The $A,B,C,D$ matrices are defined by the relation
\be
{\cal L}^{(g)}_0 +
{\cal L}_0^{(g)\dagger}\equiv c_M^\dagger A_{Mn}b_n^\dagger
+ c^\dagger_M  C_{MN}b_N+b_m^\dagger D_{mn}c_n-
c_mB_{mN}b_N \label{ABCD}
\ee
They were explicity calculated in I. Here we want to discuss their reconstruction
formulas. In \cite{BST} we numerically proved the reconstruction formulas
(4.28) and (4.31) for the bulk $\tilde A$ and $D^T$ matrices, using boundary data.
We show below that the boundary data information is contained in the
discrete basis.

To start with let us propose the spectral formulas:
\be
\tilde A_{nm}&=& \int_{-\infty}^{\infty}d\k  \tilde V_n^{(2)} (\k) {\mathfrak a}(\k)\tilde
V_m^{(-1)}(\k)
+ \sum_\xi  \tilde v_n^{(2)} (\xi) {\mathfrak  a}(\xi)
\tilde V_m^{(-1)}(\xi)\label{Arec}\\
 C_{nm}&=& \int_{-\infty}^{\infty}d\k  \tilde V_n^{(2)} (\k) {\mathfrak c}(\k)
\tilde V_m^{(-1)}(\k)
+ \sum_\xi  \tilde v_n^{(2)} (\xi) {\mathfrak  c}(\xi)
\tilde V_m^{(-1)}(\xi)\label{Crec}
\ee
Let us recall from I the continuous eigenvalues of $\tilde A$ and $C$ (see also the
re--derivation of these formulas in Appendix C)
\be
{\mathfrak a}(\k) = \frac {\pi \k}2 \frac 1{\sinh\left(\frac {\pi \k}2\right)},
\quad\quad {\mathfrak c}(\k) = \frac {\pi \k}2 \frac
{\cosh\left(\frac {\pi \k}2\right)}{\sinh\left(\frac {\pi \k}2\right)}\label{akck}
\ee
and notice that ${\mathfrak  a}(\k)$ becomes singular
on the discrete points of the spectrum $\k=\pm 2i$. Let us assume
for the time being that the discrete eigenvalues of $A$ coincide with
the continuous ones evaluated at $\xi$ (this is not obvious and will be
justified later on).

Taking an expansion about $\xi$ we have
\be
&&{\mathfrak a}(x)=1+O(x^2),
\quad\quad {\mathfrak a}(x\pm 2i)= \mp \frac {2i}x-1+O(x)\label{anearxi}\\
&& {\mathfrak  c}(x)=1+O(x^2),\quad\quad
{\mathfrak c}(x\pm 2i)= \pm \frac {2i}x+1+O(x)\label{cnearxi}
\ee
for small $x$. $V_n^{(2)} (\xi)$ and $V_n^{(-1)}(\xi)$ have been
defined above. In particular all entries of $V_n^{(-1)}(\xi)$
vanish in positions $n\geq 2$.
We have already remarked that their values coincide with the limit of $V_n^{(-1)}(\k)$
when $\k\to\xi$. If we use this definition the vanishing of all entries
in positions $n\geq 2$ is true only in the limit $x\to 0$, which is enough
in general, but not in (\ref{Arec}) and (\ref{Crec}), where these zeroes
are needed to cancel the poles in (\ref{anearxi},\ref{cnearxi}).
More precisely we have
\be
V_n^{(-1)}(0)= \oint \frac{dz}{2\pi i}  \,(1+z^2) \frac 1{z^{n+2}}=
\delta_{n,1}+\delta_{n,-1}\label{Vn-10}
\ee
and
\be
V_n^{(-1)}(2i+x) &=& \oint \frac{dz}{2\pi i}  \,(1+z^2)
\left(\frac {1+iz}{1-iz}\right)^{1+ \frac x{2i}} \frac 1{z^{n+2}}\0\\
&\approx& \oint \frac{dz}{2\pi i}  \,(1+iz)^2    \frac 1{z^{n+2}}
+\frac x{2i} \oint \frac{dz}{2\pi i}  \,(1+iz)^2 \ln \left(\frac {1+iz}{1-iz}\right)
\frac 1{z^{n+2}}  \label{Vn-12i}\\
& =& \delta_{n,-1}+ 2i \, \delta_{n,0}-  \delta_{n,1}+ \frac x{2i}
\left(-2i \,A_{0,n} +\,A_{-1,n}-\,A_{1,n}+2i \delta_{n,0} -
4  \delta_{n,1} \right)\0
\ee
Therefore, for $n\geq 2$,
\be
V_n^{(-1)}(2i+x)\approx
\frac x{2i}\left(-2i \,A_{0,n} +\,A_{-1,n}-\,A_{1,n}\right)\label{Vn-12ix}
\ee
In a similar way one can prove that
\be
V_n^{(-1)}(-2i+x)\approx \delta_{n,-1}- 2i  \delta_{n,0}-  \delta_{n,1}+ \frac x{2i}
\left(-2i \,A_{0,n} -\,A_{-1,n}+\,A_{1,n}+2i \delta_{n,0} +
4  \delta_{n,1} \right)\0
\ee
i.e., for  $n\geq 2$
\be
V_n^{(-1)}(-2i+x)\approx
\frac x{2i}\left(-2i \,A_{0,n} -\,A_{-1,n}+\,A_{1,n}\right)\label{Vn-1-2ix}
\ee
We recall that $A_{0,n}$ vanishes for odd $n$ while $A_{1,n}=-A_{-1,n}$
vanishes for even $n$. Now
\be
&& \sum_\xi  \tilde v_n^{(2)} (\xi) {\mathfrak  a}(\xi)\tilde V_m^{(-1)}(\xi)\0\\
&&= \lim_{x\to 0} \left( {\mathfrak a}(x+ 2i) \tilde v_n^{(2)}(2i) \tilde
V_m^{(-1)}(x+2i) + {\mathfrak a}(x- 2i) \tilde v_n^{(2)}(-2i) \tilde
V_m^{(-1)}(x-2i)\right)\0\\
&&= -\frac 14 (-2i \,A_{0,2m} +\,A_{-1,2m+1}-\,A_{1,2m+1})(ib_{0,2n}+
b_{1,2n+1}-b_{-1,2n+1})\0\\
&& +\frac 14 (-2i \,A_{0,2m} -\,A_{-1,2m+1}+\,A_{1,2m+1})(ib_{0,2n}+
b_{1,2n+1}-b_{-1,2n+1})\0
\ee
Therefore
\be
\sum_\xi  \tilde V_n^{(2)} (\xi) {\mathfrak  a}(\xi) \tilde V_m^{(-1)}(\xi)&=&
- b_{0,2n} A_{0,2m} +A_{1,2n+1} (b_{1,2n+1}-b_{-1,2n+1})\0\\
&=& - \sum_{a=-1,0,1} b_{a,n} \tilde A_{a,m}\0
\ee
Finally
\be
\tilde A_{nm}&=& \int d\k \tilde V_n^{(2)} (\k) {\mathfrak a}(\k)
\tilde V_m^{(-1)}(\k) - \sum_{a=-1,0,1} b_{a,n} \tilde A_{a,m}\label{verifA1}
\ee
This is precisely formula (4.28) of \cite{BST}.

Similarly, for $a=-1,0,1$ and $m\geq 2$, using the same equations and the fact
that $ V_a^{(2)} (\k)=0$, we find
\be
\tilde A_{a,m}&=& \sum_\xi  \tilde v_a^{(2)} (\xi)
{\mathfrak  a}(\xi) V_m^{(-1)}(\xi)\0\\
&=&\lim_{x\to 0} \left( {\mathfrak a}(x+ 2i) \tilde v_a^{(2)}(2i) \tilde
V_m^{(-1)}(x+2i) + {\mathfrak a}(x- 2i) \tilde v_a^{(2)}(-2i) \tilde
V_m^{(-1)}(x-2i)\right)\0\\
&=&\left( -\frac 14 \left( \begin{matrix} 1\\-i\\-1\end{matrix}\right)
(-2i \,A_{0,2m} +\,A_{-1,2m+1}-\,A_{1,2m+1})\right.\0\\
&&~~~~~~~\left.+\left(
\begin{matrix} 1\\i\\-1\end{matrix}\right)
(-2i \,A_{0,2m} -\,A_{-1,2m+1}+\,A_{1,2m+1})\right)\label{Aam}
\ee
This means
\be
\tilde A_{-1,m}= - A_{-1,m},\quad\quad \tilde A_{0,m}= A_{0,m}, \quad\quad
\tilde A_{1,m}= - A_{1,m}\label{Aamrec}
\ee
This completes the recostruction of the matrix A, including the first
three rows, which in \cite{BST} were called {\it boundary data}. These boundary data
turn out in fact to be stored in the discrete basis.

The previous result confirms that the guess for ${\mathfrak a}(x\pm 2i)$
was correct, but it does not say anything about the $\xi=0$ discrete
eigenvalue of $A$ (the latter has not been used in the previuous derivation).

Using the same method it is easy to prove from (\ref{Crec}) that
\be
C_{nm} = \int d\k \tilde V_n^{(2)} (\k) {\mathfrak c}(\k)\tilde V_m^{(-1)}(\k)
- \sum_{a=-1,0,1} b_{a,n} \,C_{a,m}\label{Crecanm}
\ee
for $n,m\geq 2$. And, using (\ref{Crec}) for $a=-1,0,1$ and $m\geq 2$,
one can show that
\be
C_{-1,m}= - A_{1,m},\quad\quad C_{0,m}= - A_{0,m},
\quad\quad C_{1,m}= - A_{-1,m}\0
\ee
i.e
\be
C_{a,m}= - A_{-a,m}\label{Camrec}
\ee
This is another set of boundary data of \cite{BST} which is contained in the
discrete basis. Again this confirm the validity of (\ref{cnearxi})
(except for $\xi=0$).

It remains for us to reconstruct the values of $B_{n,i}$ and $C_{n,i}$.
Applying flatly the same formulas above we obtain divergent results,
because the divergences of (\ref{anearxi}) and (\ref{cnearxi}) are
not compensated by vanishing basis vectors. As this does not interfere with
the subsequent developments we leave this problem open. Let us simply summarize
the following facts: {\it (Eq.\ref{Arec}) is a good representation of $A$, provided
we supplement it with three columns of zeroes; it of course provides a good
representation for the bulk of $B$; Eq.(\ref{Crec}) is a good representation for
$C$ excluding the first three columns and is a good representation for $D^T$
if we limit ourselves to the bulk of Eq.(\ref{Crec})}.

\section{Reconstruction of $X$, $X^{\pm}$, $X_{(i)}$, $X_{(i)}^{rs}$}

The first important test of the formalism is the reconstruction of the
three string vertex Neumann coefficient matrices. It is convenient to start
from the real (average) vertex.

The definition of $\Ct X=V^{rr}$, eq.(\ref{Vnmrs}), is
\be
V^{rr}_{nm}
=\frac 12 \oint\frac{dz}{2\pi i}\oint\frac{dw}{2\pi i}\frac{1}{z^{n-1}}
\frac{1}{w^{m+2}} \left( \frac 43 i\frac {1+w^2}{(1+z^2)^2} \frac {f(z)+f(w)}{f(z)-f(w)}-
\frac {1}{z-w}\right) \label{CX}
\ee
This must be compared with the reconstruction formula
\be
X'_{nm}&=& \int d\k \tilde V_n^{(2)} (\k) {\mathfrak x}(\k)\tilde V_m^{(-1)}(\k)
+ \sum_\xi  \tilde v_n^{(2)} (\xi) {\mathfrak  x}(\xi) \tilde V_m^{(-1)}(\xi)\label{Xrec}
\ee

In order to make the comparison we have to know the eigenvalues.
The continuous eigenvalue of the wedge states from the KP equation, \cite{KPot}
derived in I
(see also Appendix C), are
\be
{\mathfrak t}_t(\k)= \frac {\sinh \left(\frac {\pi \k}4 (2-t)\right)}
{\sinh \left(\frac {\pi \k}4 t\right)} \label{tt}
\ee
In particular for the case $n=3$, which must coincide with $X'$, this gives
\be
{\mathfrak x}(\k)=- \frac {\sinh \left(\frac {\pi \k}4\right)}
{\sinh \left(\frac {3\pi \k}4 \right)} \label{x(k)}
\ee
We will take this as the appropriate continuous eigenvalue for $X'$.
Next we face the problem of the discrete eigenvalues
${\mathfrak  x}(\xi)$. Evaluating the continuous eigenvalue at the discrete
points of the spectrum we would get ${\mathfrak  x}(0)=-\frac 13,
 {\mathfrak  x}(\pm 2i)=1$, but it turns out that this choice is wrong.
The appropriate discrete eigenvalues turn out to be (see Appendix C for a justification)
\be
{\mathfrak  x}(0)=-1,\quad\quad {\mathfrak  x}(\pm 2i)=1 \label{xdiscrete}
\ee

Using (\ref{Xrec},\ref{xdiscrete}) we get immediately $X'_{i,m}=0,\quad m\geq 2$,
because $V_i^{(2)}(\k)=0$ and $V_m^{(-1)}(\xi)=0$, for $x\to 0$, and no
singularity to be compensated. Instead
\be
X'_{ij}&=& 0 -\frac 12 \left(\begin{matrix} 1\\0\\1\end{matrix}\right)\otimes (1,0,1)
 +\frac 14 \left(\left(\begin{matrix} 1\\-i\\-1\end{matrix}\right)\otimes
(1,2i,-1)  + \left(\begin{matrix} 1\\i\\-1\end{matrix}\right)\otimes
(1,-2i,-1)   \right)\0\\
&=& \left(\begin{matrix}0&0&- 1\\0&1&0\\-1&0&0\end{matrix}\right)\label{Xij}
\ee
$X$, given by (\ref{CX}), has vanishing first three rows. Therefore the matrix
in (\ref{Xij}) corresponds rather to the matrix $z$ discussed in sec. 2.4.
But we have already seen that to the Neumann matrices of the left (ghost number 3)
vertex we can always add a matrix like $z$.

Let us see now a sample of the other entries
\be
X_{2,1}&=& 0 + \frac 14 (i\cdot 1 -i\cdot 1)=0\0\\
X_{2,0}&=& -\int d\k \frac {\k} {2 \sinh \left(\frac {\pi \k}2\right)}
\frac {\sinh \left(\frac {\pi \k}4\right)}{\sinh \left(\frac {3\pi \k}4\right)}
 + \frac 14 (i\cdot 2i - i \cdot(-2i)) = -\frac {32}{27}\0\\
X_{3,-1}&=& -\int d\k \frac {\k} {2 \sinh \left(\frac {\pi \k}2\right)}
\frac {\sinh \left(\frac {\pi \k}4\right)}{\sinh \left(\frac {3\pi \k}4\right)}
 -\frac 12 (-3) 1+ \frac 14 (1\cdot 1+ 1 \cdot 1) =
\frac {49}{27}\0\\
X_{4,0}&=& -\int d\k \frac {\k} {2 \sinh \left(\frac {\pi \k}2\right)}\,
\frac{\k^2-1}2 \,\frac {\sinh \left(\frac {\pi \k}4\right)}
{\sinh \left(\frac {3\pi \k}4\right)}
 + \frac 14 (-i\cdot 2i+ i \cdot (-2i)) = \frac {320}{243}\0
\ee
and in the bulk, where only the continuous spectrum contributes,
\be
X_{3,3}&=& -\int d\k \frac {\k} {2 \sinh \left(\frac {\pi \k}2\right)}
\frac {\k^2(\k^2+4)}{24}
\frac {\sinh \left(\frac {\pi \k}4\right)}{\sinh \left(\frac {3\pi \k}4\right)}
=-\frac {541}{19683}\0
\ee
and so on. These are precisely the values expected from (\ref{CX}).

Analogously, one can reconstruct the $X^{\pm}$ matrices.
In that case, the discrete spectrum does not contribute, that is,
${\mathfrak x}^{\pm}(\xi)=0$ for $\xi = 0, \pm 2 i$. This implies,
correctly, that $X^{\pm}_{ij}=0$. For the continuous spectrum, we use
the same expressions as for the matter part, see \cite{RSZ1} :
\begin{equation}
{\mathfrak x}^{+}(\k) = -(1 +   e^{\frac{\pi \k}{2}}){\mathfrak x}(\k)
~~~ ; ~~~ {\mathfrak x}^{-}(\k) =
-(1 + e^{-\frac{\pi \k}{2}}){\mathfrak x}(\k)
\end{equation}
Here is a sample of components of $X^{+}$\footnote{In computing some
of the components, for example, $X^{+}_{2,1}$, $X^{+}_{2,-1}$ and
$X^{+}_{4,-1}$, the integrals must be regularized. This has been done
using the principal value prescription.}:

\begin{eqnarray}
X^{+}_{2,0} &=& \int d\k \frac{{\mathfrak x}^{+}(\k) \k}{2 {\rm sinh}
(\frac{\pi \k}{2})} = \frac{16}{27} \0 \\
X^{+}_{3,-1} &=& \int d\k \frac{{\mathfrak x}^{+}(\k) \k}{2 {\rm sinh}
(\frac{\pi \k}{2})} = \frac{16}{27} \0 \\
X^{+}_{3,0} &=& \int d\k \frac{{\mathfrak x}^{+}(\k) \k^2}{2 {\rm sinh}
(\frac{\pi \k}{2})} = \frac{64}{81 \sqrt{3}} \0 \\
X^{+}_{4,0} &=& \int d\k \frac{{\mathfrak x}^{+}(\k) \k}{2 {\rm sinh}
(\frac{\pi \k}{2})} \left( \frac{\k^2}{2}-2 \right) = -\frac{160}{243} \0 \\
X^{+}_{2,2} &=& \int d\k \frac{{\mathfrak x}^{+}(\k)}{2 {\rm sinh}
(\frac{\pi \k}{2})} \left( \frac{\k^3}{6} + \frac{2 \k}{3} \right)
= \frac{416}{729} \0 \\
X^{+}_{3,2} &=& \int d\k \frac{{\mathfrak x}^{+}(\k) \k}{2 {\rm sinh}
(\frac{\pi \k}{2})} \left( \frac{\k^3}{6} + \frac{2 \k}{3} \right)  =
\frac{896}{729 \sqrt{3}} \0
\end{eqnarray}
For $X^{-}$ one can do the same using ${\mathfrak x}^{-}(\k)$ and see
that it also works perfectly.

Let us conclude with some remarks. In the previous section
we have seen that the twisted Neumann matrices $X^{'rs}$ commute with $G$,
and thus are diagonal in the bases of its eigenvectors. In this section
we have written down such bases and we have shown that by their means
we can construct spectral representations of $X^{'rs}$ which faithfully
reconstruct the latter. We remark that it is easier to identify the
discrete eigenvalues of $X^{'rs}$  by this indirect method rather
than by a direct approach.

Let us consider now $X_{(i)}$. This matrix is obtained by twisting
\be
V^{rr}_{(i)nm}
=\frac 12 \oint\frac{dz}{2\pi i}\oint\frac{dw}{2\pi i}\frac{1}{z^{n-1}}
\frac{1}{w^{m+2}} \left( \frac 43 i\frac {1+w^2}{(1+z^2)^2} \frac 1{f(z)-f(w)}\left(
\frac {f(w)}{f(z)}\right)^3-\frac {1}{z-w}\right) \label{Vinm}
\ee
The corresponding $X_{(i)}= \Ct V^{rr}_{(i)}$ matrix can be reconstructed from
the matrix $X$ above by adding additional contributions (the primed matrices
are obtained by adding $z$)
\be
X_{(i)nm}=X_{nm}- \frac 43 i \, V^{(2)}_n \left(\frac 43 i\right)\,
V^{(-1)}_m\left(\frac 43 i\right)-  \frac 23 i \, V^{(2)}_n (0)\,
V^{(-1)}_m(0)\label{reconXi}
\ee
The last addend affects only the first three columns. In this equation
$V^{(2)}$, as well as $V^{(-1)}$, stands for the continuous basis evaluated
at the corresponding points. Let us see some examples of the validity of
(\ref{reconXi})
\be
(2,0)&:& \frac{16}{27}=-\frac{32}{27}+\frac {16}{9}+0\0\\
(2,-1)&:& -2i=0-\frac{4}{3}i-\frac {2}{3}i\0\\
(3,0)&:& \frac{64}{27}i=0+\frac {64}{27}i+0\0\\
(3,3)&:& -\frac{6301}{19683}=-\frac{541}{19683}-\frac {640}{2187}+0\0
\ee

In a similar way one can deal with $X^{\pm}_{(i)}$, see Appendix D.

In order to understand the origin of the correction in (\ref{reconXi}) with
respect to (\ref{Xrec}) let us return to the latter, which is the classical
spectral formula one would expect, i.e. the summation over the eigenvalues
(both continuous and discrete) multiplied by the appropriate eigenprojectors.
In that formula the continuous eigenvalues are real.  However we have noticed
above that the continuous eigenvalues may as well be complex. Therefore we could
consider the spectral formula with a contour away from the real axis. If there are
poles between the new and the old contour the final results will be different.
This is precisely what happens in the passage from (\ref{Xrec}) to (\ref{reconXi}).
The difference corresponds to a modification of the integration contour over the
continuous spectrum.

The continuous part of the spectrum is
\be
\Delta X_{nm}=\int d\k \,\mu(3,\k) V^{(2)}_n(\k) V^{(-1)}_m(\k)
\label{deltaXnm}
\ee
where the measure is
\be
\mu(3,\k)= \frac{{\mathfrak x}
(\k)}{2 {\rm sinh} (\frac{\pi \k}{2})}=-\frac 1{2 {\rm sinh} (\frac{\pi \k}{2})}
\frac {\sinh \left(\frac {\pi \k}4 \right)}
{\sinh \left(\frac {3\pi \k}4\right)}\label{measX}
\ee
This measure has poles at $\k=\pm \frac 43 i n$ for natural $n$.
If in (\ref{deltaXnm}) the integration contour is along the real axis we
get back $X_{nm}$. But let us suppose that
\be
\Delta X_{nm}=\int_{C_1} d\k\, \mu(3,\k) V^{(2)}_n(\k) V^{(-1)}_m(\k)
\label{deltaXnm1}
\ee
where $C_1$ is a straight contour from $-\infty$ to $+\infty$ with
$\frac 43 < \Im (\k) < 2$. If we move the upper contour toward the
real axis we are bound to meet two poles of the measure,
one at $\k=\k_1= \frac {4 i }3$
and another at $\k=\k_0=0$. So finally we obtain the usual integral along
the real axis (which corresponds to $X$) plus two contributions
from the two poles that remain trapped inside the contour. The latter
are clockwise oriented, so we have to change the sign when calculating the
residues.

It is easy to show that near the poles $\k_i$, see (\ref{muNx}),
\be
\mu(\k_1+x)\approx\frac 23 \frac 1 {\pi x},\quad\quad  \mu(\k_0+x)\approx
\left(\frac 23 -1\right)\frac 1 {\pi x}\0
\ee
where we have kept distinct the contribution represented by -1 for a reason
that will become clear later on. Therefore
\be
&&-\oint_{\rm poles} d\k\, \mu(3,\k) V^{(2)}_n(\k) V^{(-1)}_m(\k)\label{poles}\\
&=&-\oint dx\,\frac {4i}{6\pi i}\frac 1x \,V^{(2)}_n(\k_1+x) V^{(-1)}_m(\k_1+x)
-\oint dx\,\frac 12\,\frac {4i}{6\pi i}\frac 1x \,V^{(2)}_n(x) V^{(-1)}_m(x)\0\\
&=& -\frac {4i}3\,V^{(2)}_n(\k_1) V^{(-1)}_m(\k_1)-
\frac {2i}3\,V^{(2)}_n(0) V^{(-1)}_m(0)\0
\ee
The contribution of the pole at $\k=0$ has been divided by two because
only `half' pole contributes (this is consistent with the remaining
calculations). In this way the contribution (\ref{poles}) accounts
precisely for the difference between $X$ and $X_{(i)}$, except for the contribution
$i\, V^{(2)}_n(0) V^{(-1)}_m(0)$. Taking it into account we can write
\be
X_{(i)nm}=\int_{\Im(\k)> \frac 43} d\k \,\mu(3,\k) V^{(2)}_n{\k} V^{(-1)}_m(\k) -
i\, V^{(2)}_n(0) V^{(-1)}_m(0)\,+(...)\label{Xnmi}
\ee
where the integration contour runs parallel to the real axis just above the pole
at $\k=\frac {4i}3$. The second piece in the
RHS of (\ref{Xnmi}) is a `necessary scar' of that formula we will comment
about later on.
The omitted terms $(...)$ are the contribution from the discrete spectrum
(which is not touched by the shift in the $\k$ integration).

\section{The spectral argument}

It is time to see our three strings vertex at work.
Let us consider the star product of two wedge states as in the RHS of
(\ref{ghostwedge})
\be
|S\rangle={\cal N}
\exp \left(c^\dagger S b^\dagger\right)|0\rangle\label{squeezed}
\ee
i.e.
\be
\langle\hat V_3|S_1\rangle |S_2 \rangle = \langle \hat S_{12}|\label{S12}
\ee
We remark that the states like (\ref{squeezed}) are defined
on the ghost number 0 vacuum $|0\rangle$, while the resulting state in
the RHS of (\ref{S12}) is defined in the ghost number 3 vacuum
$\langle \hat 0|$. Therefore $\langle \hat S_{12}|$ is not yet
the star product. We will discuss in III on how to recover
$ |S_{12}\rangle$.

The matrix $S_{12}=\Ct T_{12}$ is given by the familiar formula
\be
T_{12}= X + (X^+,X^-) \,\frac 1{1- \Sigma_{12} {\cal V}} \,\Sigma_{12}
\,\left(\begin{matrix} X^-\\ X^+ \end{matrix}\right)\label{T12}
\ee
where
\be
\Sigma_{12} =\left(\begin{matrix}\Ct S_1&0\\0&\Ct S_2
\end{matrix}\right),
\quad\quad
{\cal V}= \left(\begin{matrix}X & X^+\\X^-& X\end{matrix}\right)
\label{SigmaV}
\ee
In these formulas the matrices $X,X^\pm$ represent
$X',X^{'\pm}$ or $X_{(\pm i)}',X_{(\pm i)}^{'\pm}$.
As for the matrices $T_1=\Ct S_1, T_2, T_{12}$ they are supposed to
represent wedge states. The latter, denoted simply by
$|n\rangle\equiv |S_n\rangle$, must satisfy the recursive star product formula
\be
|n\rangle \star |m\rangle = |n+m-1\rangle\label{wedgenm}
\ee

Our purpose in the sequel is to prove that the squeezed states at the RHS of
(\ref{ghostwedge}), when star--multiplied with our three strings ghost vertex,
do obey the recursive formula (\ref{wedgenm}). To this end we will proceed as follows.
After determining (which we have done in the previous section) the eigenvalues
of the twisted
Neumann matrices of the vertex we will determine those of the squeezed states
at the RHS of (\ref{ghostwedge}), by inferring them from
the properties of the $gh=0$ wedge states via the KP equation. We will show that
the recursion relations of the wedge states ensuing from (\ref{wedgenm})
are satisfied. Finally we will show how to reconstruct
the ghost $gh=3$ results of the star product with
the appropriate spectral formulas. The $gh=0$ states corresponding
to them will be reconstructed in III. This argument is based on the
prejudice that $gh=3$ and $gh=0$ wedge Neumann functions have the same
continuous and discrete eigenvalues. This fact is not at all obvious a
priori. But it will be justified beyond any doubt with the reconstruction
formulas of the $gh$=0 states in III.\\

\noindent The argument is anything but simple. To facilitate the
comprehension let us
for the time being assume that
$T_1,T_2$ commute with $X,X^\pm$ (which is not true!). In such a case, setting
$T_1=T_n$ and $T_2=T_m$ we would get that, if (\ref{wedgenm}) is true, it follows
from (\ref{T12}) that
\be
T_{n+m-1}= \frac {X-(T_n+T_m)X+T_nT_m (1+{\mathfrak E})+ (T_n+T_m)
{\mathfrak E}}{1- (T_n+T_m)X+T_nT_m(X-{\mathfrak E})}\label{TnTm}
\ee
Setting $T_2=0$ we can write the recursion formula
\be
T_{n+1}= \frac {X(1-T_n)+T_n{\mathfrak E}}{1 - T_n X}\label{Tn+1}
\ee
This result is not true for the matrices but we will show it to be true
for their eigenvalues. This is due to the fact that the
Neumann matrices of the ghost number 0 wedge states have a subset
of eigenvectors in common with $G$, while the remaining ones are different
\footnote{Two matrices can
of course have the same eigenvalues without commuting. An elementary example
is given by the two matrices ${M_0} =  \left(\begin{matrix} 1&0\\
                                                  0& 2\\
\end{matrix}\right)$ and $ {M_b} =  \left(\begin{matrix} 1& b\\
                                                  0& 2\\
\end{matrix}\right)$ with $b\neq 0$. They have the same eigenvalues but do
not commute. Moreover (counting left and right eigenvectors) they have
two eigenvectors in common, while the
other two are different.}. Once again this fact
will be entirely clear only at the end of III, where it will appear that
$gh= 0$ and $gh=3$ wedge states Neumann matrices have the same spectrum.
The next thing to be done therefore is to evaluate the eigenvalues of $T_n$.

\subsection{The recursion relations for eigenvalues}

The recursion relations for matrices (\ref{Tn+1}) are not expected to hold, but we wish
to show them to be true
for their eigenvalues. Applying (\ref{T12}) to the bases
$V^{(2)}(\k)$ and $V^{(-1)}(\xi)$ one can see that the continuous eigenvalues
must satisfy
\be
{\mathfrak t}_{n+1}(\k)=
{\mathfrak x}(\k)\frac {1- {\mathfrak t}_n(\k)}
{1- {\mathfrak t}_n(\k){\mathfrak x}(\k)},\quad\quad
{\mathfrak t}_3(\k)={\mathfrak x}(\k)\label{recur}
\ee
while for the discrete eigenvalues one should get
\be
{\mathfrak t}_{n+1}(\xi)=
\frac{{\mathfrak x}(\xi) (1- {\mathfrak t}_n(\xi))-2 {\mathfrak t}_n (\xi)
\, \delta_{\xi,0}}
{1- {\mathfrak t}_n(\xi){\mathfrak x}(\xi)},\quad\quad
{\mathfrak t}_3(\xi)={\mathfrak x}(\xi)\label{recurdiscrete}
\ee
where the -2 addend in the numerator comes from the eigenvalue
of ${\mathfrak E}$. The values of ${\mathfrak t}_n(\k), {\mathfrak t}_n(\xi)$
are determined in Appendix C (using the results of I). ${\mathfrak t}_n(\k)$ has already
been reported in eq.(\ref{tt}), while the discrete eigenvalues are given by
\be
{\mathfrak t}_n(\xi=0)=-1, \quad\quad {\mathfrak t}_n(\pm 2i)=1\label{discretn}
\ee
It has been shown in I that (\ref{tt}) does indeed satisfy the recursion relation
(\ref{recur}).
It is easy to see that  (\ref{recurdiscrete}) is also satisfied by the
above found values (\ref{discretn}), provided one observes
the following procedure: one first replaces the values $
{\mathfrak x}(\pm 2i)=1, \quad\quad {\mathfrak x}(0)=-1
$
while keeping ${\mathfrak t}_n(\xi)$ generic. After simplifying the expression
one inserts the values (\ref{discretn}). We remark that the presence of
the ${\mathfrak E}$ matrix in (\ref{Tn+1}) is essential in this respect.

It is well--known that (\ref{recur}) can be solved in terms of the sliver
eigenvalue, \cite{Furu,Kishimoto}.
We repeat here this derivation to stress its uniqueness.
We require that $|2\rangle$ coincide with the vacuum $|0\rangle$,
both for the matter and the ghost sector.
This implies in particular that ${\mathfrak t}_2=0$ which
entails from (\ref{recur}) that ${\mathfrak t}_3={\mathfrak x}$,
${\mathfrak t}_4= \frac {\mathfrak x}{1+{\mathfrak x}}$, etc.
That is, ${\mathfrak t}_n$
is a uniquely defined function of ${\mathfrak x}$.
But ${\mathfrak x}$ can be uniquely expressed in terms of
the sliver matrix ${\mathfrak t}$
\be
{\mathfrak x}=\frac {\mathfrak t}{{\mathfrak t}^2-{\mathfrak t}+1}\label{XT}
\ee
a formula whose inverse is well--known, \cite{KPot,RSZ2}
\be
{\mathfrak t}= \frac 1{2{\mathfrak x}} \left(1+{\mathfrak x}-
 \sqrt{(1-{\mathfrak x})(1+3{\mathfrak x})}\right)\label{sliver}
\ee
Therefore ${\mathfrak t}_n$ can be expressed as a uniquely defined function
of ${\mathfrak t}$.
Now consider the formula
\be
{\mathfrak t}_n= \frac {{\mathfrak t}+(-{\mathfrak t})^{n-1}}{1-(-{\mathfrak t})^n}\0
\ee
It satisfies (\ref{recur}) as well as the condition ${\mathfrak t}_2=0$, therefore
it is the unique solution to (\ref{recur}) we were looking for.

For completeness we recall also the recursion relation
for the normalization constants
\be
{\cal N}_{n+1}={\cal N}_n\, {\cal K} \, \det \left(1-T_n X\right)
\label{normrecur}
\ee
It is easy to see that the discrete eigenvalues give a vanishing contribution to
the determinant, therefore the discussion reduces to the continuous eigenvalues, and
this was done in I.

In III we will also give evidence that (when the matter sector is coupled) this overall normalization will
be 1.

\section{Reconstruction of the dual wedge states}

In the previous sections we have defined three strings vertices for the ghost part.
We have consequently defined a midpoint--star product. It is not possible to do the star
product in a single step, i.e. it is not possible to start from two  $gh=0$
states and end up with the resulting star product as a $gh=0$ state (in the matter
case this is possible up to a $bpz$ conjugation). In this case we must go through
a two step process. We first compute the $gh=3$ state which is the result of the
operation in eq.(\ref{S12}). The second step consists in computing the $gh=0$ ket
corresponding to this result (this operation turns out to be far more complicated
than the simple $bpz$ conjugation of the matter sector).

In the rest of this paper we will be concerned with the first step, while the second will be
the subject of III.
In the previous section we have calculated the eigenvalues of the resulting ($gh=3$)
object (which we will call the dual or bra star product wedge state) in the weight 2 and -1
discrete and continuous basis. Now we wish to express this state as a squeezed
state in the oscillator form. To this end we resort to the reconstruction formulas,
which are nothing but the ordinary spectral formulas in which, however,
the integration contour for the continuous spectrum must be specified. As we
will see, different contours give different results with different characteristics:
they may or may not be surface states and may or may not be BRST invariant.

In parallel with section 5 let us write down the spectral
representation for the left $gh=3$ wedge states:
\be
\hat T^{'(N)}_{nm}&=& \int d\k
\tilde V_n^{(2)} (\k){\mathfrak  t}_N(\k)\tilde V_m^{(-1)}(\k)
+ \sum_\xi  \tilde v_n^{(2)} (\xi) {\mathfrak  t}_N(\xi) \tilde V_m^{(-1)}(\xi)
\label{Tnrec}
\ee
where for practical reasons we have slightly changed the notation:
$\hat T^{(N)}={\Ct}{ {\hat S}}_N$ and the prime denotes the addition of the
$z$ matrix. Let us recall that the discrete eigenvalues are
\begin{equation}
{{\mathfrak t}}^{(N)}(0)={\mathfrak x}(0)=-1 ~~~;~~~
{{\mathfrak t}}^{(N)}(\pm 2i)={\mathfrak x}(\pm 2i)=1
\end{equation}
while the continuous eigenvalue is given by (\ref{tt}). We thus have
\begin{eqnarray}
{\hat T}^{(N)}_{nm} &=& \int_{-\infty}^{\infty} d\k \frac{{ {\mathfrak t}}^{(N)}
(\k)}{2 {\rm sinh} (\frac{\pi \k}{2})} V^{(2)}_{n}(\k) V^{(-1)}_{m}(\k)- \0 \\
&-& \frac{1}{2} v^{(2)}_{n}(0) V^{(-1)}_{m}(0) + \frac{1}{4}(v^{(2)}_{n}
(2i) V^{(-1)}_{m}(2i) + v^{(2)}_{n}(-2i) V^{(-1)}_{m}(-2i))
\end{eqnarray}
where the integral is, for the time being, along the real axis.

Let us compute  a sample of the entries of ${\hat T}_4$
\begin{eqnarray}
{\hat T}^{'(4)}_{2,0} &=&
\int_{-\infty}^{\infty}   d\k \frac{{\hat {\mathfrak t}}^{(4)}(\k)}{2 {\rm sinh}
(\frac{\pi \k}{2})} \k + \frac{1}{4} (i \cdot 2i - i \cdot (-2i))
= -\frac{1}{4} - 1 = -\frac{5}{4} \0 \\
{\hat T}^{'(4)}_{3,-1} &=&
\int_{-\infty}^{\infty} d\k \frac{{\hat {\mathfrak t}}^{(4)}(\k)}{2 {\rm sinh}
(\frac{\pi \k}{2})} \k -\frac{1}{2} \cdot (-3) + \frac{1}{4} (1 + 1)
= -\frac{1}{4} + 2 = \frac{7}{4} \0 \\
{\hat T}^{'(4)}_{4,0} &=& \int_{-\infty}^{\infty}   d\k \frac{{\hat {\mathfrak t}}^{(4)}
(\k)}{2 {\rm sinh} (\frac{\pi \k}{2})} \k \left( \frac{\k^2}{2} -
2 \right) + \frac{1}{4} ((-i) \cdot 2i + i \cdot (-2i)) = \frac{7}{16}
+ 1 = \frac{23}{16} \0 \\
{\hat T}^{'(4)}_{2,2} &=& \int_{-\infty}^{\infty}   d\k \frac{{\hat {\mathfrak t}}^{(4)}
(\k)}{2 {\rm sinh} (\frac{\pi \k}{2})} \left( \frac{\k^3}{6} +
\frac{2 \k }{3} \right) = -\frac{3}{16} \0 \\
{\hat T}^{'(4)}_{3,3} &=& \int_{-\infty}^{\infty}   d\k \frac{{\hat {\mathfrak t}}^{(4)}
(\k)}{2 {\rm sinh} (\frac{\pi \k}{2})} \k \left( \frac{\k^4}{24} +
\frac{\k^2 }{6} \right) = -\frac{1}{32} \0
\end{eqnarray}
These perfectly agree with the formula for the Neumann coefficients of the
left $gh=3$ states
\be
\bra {\hat n}= \bra{\hat0} e^{-c_p\,\hat S^{(n)}_{pM}\,b_M}\label{hatn}
\ee
where
\be
\hat S^{(n)}_{pM}&=&\oint_0\frac{dz}{2\pi i}\oint_0\frac{dw}{2\pi i}
\frac{1}{z^{p-1}}\frac{1}{w^{M+2}}\,\left(\frac{2i}{n}\,
\frac{1+w^2}{(1+z^2)^2}\,
\frac{f_n(z)+f_n(w)}{f_n(z)-f_n(w)}-\frac1{z-w}\right)\label{Sn}
\ee
and
\be
f_n(z)= \left(\frac {1+iz}{1-iz}\right)^{\frac 2n}\label{fnz}
\ee
The only exception is the insertion of the $z$ matrix in the upper left corner
(which, however, can always be done due to an intrinsic ambiguity of the oscillator
formalism, as explained in sec. 2.4).

\noindent A similar numerical agreement has been checked also
for ${\hat T}^{(5)}$ and higher states. The reconstruction formula has given us back
squeezed states that belong to the same family as the average three vertex (see section
5). These states however are not surface states with insertions, which creates problems with BRST invariance
(see Appendix A and especially III for a discussion of these issues).

\subsection{The dual wedge states with $Y(i)$ insertion}

In order to get BRST invariant surface states as $gh=3$ wedge states we have
to change the integration contour. To this end we follow the recipe of
the second part of section 5. That is we use again (\ref{Tnrec}), but redefine
the integration contour over the continuous spectrum
\be
( T_c^{(N)})_{nm}= \int_{C_N} d\k\,
\tilde V_n^{(2)} (\k){\mathfrak  t}_N(\k)\tilde V_m^{(-1)}(\k)
=\int_{C_N} d\k\, \mu(N,\k) \,V_n^{(2)} (\k)V_m^{(-1)}(\k)
\label{DeltaTN}
\ee
where the subscript $c$ stands for the continuous part of the spectral formula,
$C_N$ is the contour to be specified and the measure is
\be
\mu(N,\k)= \frac{{\mathfrak t}_N
(\k)} {2 {\rm sinh} (\frac{\pi \k}{2})} =\frac 1{2 {\rm sinh} (\frac{\pi \k}{2})}
\frac {\sinh \left(\frac {\pi \k}4 (2-N)\right)}
{\sinh \left(\frac {\pi \k}4 N\right)}\label{meas}
\ee
This measure has poles at $\k=\pm \frac {4 i n}N$ for natural $n$.
If in (\ref{Tnrec}) the integration contour is along the real axis
and we move it up, we are bound to meet the first pole at
$\k=\k_1= \frac {4 i }N$.
In (\ref{DeltaTN}) the contour $C_N$ stretches from $-\infty$ to $\infty$ in
the upper $\k$ plane
with $\Im(\k)$ just above $\frac 4 N$. This traps two poles of ${\mu}(N,\k)$,
i.e. the poles at $\k\equiv \k_1= \frac {4i}N$ and $\k=\k_0=0$, lying between this
contour and the real axis. Therefore the integral over $C_N$ reduces
to the usual integral along the real axis plus the contributions of clockwise
oriented contours around $\k=\k_1$ and $\k_0$, i.e.
\be
(T_c^{(N)})_{nm}&=&\int_{-\infty}^{\infty} d\k \mu(N,\k) V^{(2)}_n(\k) V^{(-1)}_m(\k)
\label{deltaTnm1}\\
&&-\oint_{C_{\k_1}} d\k \mu(N,\k) V^{(2)}_n(\k) V^{(-1)}_m(\k)-\frac 12
\oint_{C_{\k_0}} d\k \mu(N,\k) V^{(2)}_n(\k) V^{(-1)}_m(\k)\0
\ee
where $C_{\k_1}$ and $C_{\k_0}$ are small anticlockwise contours around
$\k_1$ and $\k_0$, respectively (taking half of the latter contribution
for the reason explained in section 5).
Let us write $\k =\frac {4i}N n+x$ for small $x$. It is easy to show that
near the pole and for $N\neq 2$
\be
\mu(N,\k)\approx \left\{\begin{matrix}\frac 2{\pi N} \frac 1x, \quad\quad\quad
\sin \left(\frac {2\pi n}N\right) \neq 0\\
\frac 1{\pi N}\frac 1 x (2-N),\quad\quad\quad \sin \left(\frac {2\pi n}N\right)=0
\end{matrix}\right.\label{muNx}
\ee
When $N=2$ the measure $\mu(2,\k)=0$, as it should be. However, in that case
the relevant measure becomes $\mu(2,\k) = \frac 1 {2 {\rm sinh} (\frac{\pi \k}{2})}$ and
the poles are at $\k= 2 i n$, and near them we have
\be
\mu(2,\k)\approx \frac  {(-1)^n}{\pi x}\label{mu2k}
\ee
Returning to $N\neq 2$ we have, for instance,
\be
\oint_{C_{\k_1}} d\k \mu(N,\k) V^{(2)}_n(\k) V^{(-1)}_m(\k)&=&
\oint dx\,\frac {4i}{2\pi i N}\frac 1x \,V^{(2)}_n(\k_1) V^{(-1)}_m(\k_1)\0\\
&=& \frac {4i}N\,V^{(2)}_n(\k_1) V^{(-1)}_m(\k_1)\label{correct1}
\ee
and
\be
\oint_{C_{\k_0}} d\k \mu(N,\k) V^{(2)}_n(\k) V^{(-1)}_m(\k)=
{2i}\left(\frac 2N-1\right)\,V^{(2)}_n(\k_0) V^{(-1)}_m(\k_0)\label{correct0}
\ee
The factor $-2i$ in the RHS of this equation, which is absent in the case $N=2$,
is the contribution
of the pole coming from the first factor in the RHS of (\ref{muNx}), which is the
measure appearing in the orthogonality relations. We will forget about this
additional factor for the time being and comment about it later on.

\noindent Finally
\be
T^{(N)}_{(i)nm}=\hat T^{(N)}_{nm}- \frac 4N i \, V^{(2)}_n \left(\frac 4N i\right)\,
V^{(-1)}_m\left(\frac 4N i\right)-  \frac {2i}N  \, V^{(2)}_n (0)\,
V^{(-1)}_m(0)\label{reconTN}
\ee
where $V^{(2)}$ and $V^{(-1)}$ stand for the continuous bases evaluated at the
corresponding points. Here are some examples of this formula for $N=4$ (the $N=3$
coincides with the ghost vertex Neumann coefficients of section 5)
\be
(2,-1)&:& -\frac 32 i = 0- i- \frac i2\0\\
(3,1)&:& \frac {19}{16} = \frac {11}{16} +\frac 12 +0\0\\
(4,2)&:& -\frac {15}{16}= \frac {5}{16} -\frac 54 +0\0\\
(5,4)&:& \frac {5i}{16} = 0+ \frac {5i}{16}+0\0
\ee
This is precisely what is expected for the BRST invariant dual wedge state
specified by the following Neumann coefficients
\be
{S_{(i)}}^{(N)}_{pM}&=&\oint_0\frac{dz}{2\pi i}\oint_0\frac{dw}{2\pi i}
\frac{1}{z^{p-1}}\frac{1}{w^{M+2}}\label{SiN}\\
&&\cdot\left[\frac {f'_N(z)^2}{f'_N(w)}
\frac 1{f_N(z)-f_N(w)}
\left(\frac {f_N(w)}{f_N(z)}\right)^3-\frac 1{z-w}\right]\0
\ee
These are surface states with $Y(i)$ insertion. They are obtained by setting
$t=i$ in the appropriate formulas for the Neumann matrix in Appendix A.

\noindent In Appendix D we show how to reconstruct also $T^{(N)}_{(-i)}$.\\

Now let us make a comment about the factor of $+i$ we disregarded
above in the RHS of (\ref{correct0}). This factor gives rise in the spectral
formulas (\ref{reconTN}) to a term proportional to
$P_{ni}=V^{(2)}_n (0)\,V^{(-1)}_i(0)$ (remember that $V^{(-1)}_n(0)=0$ for $n\geq 2$).
Putting everything together, the reconstruction formula for $T^{(N)}_{(i)}$ in terms of
contour integration is (see also (\ref{Xnmi})
\be
( T_{(i)}^{(N)})_{nm}= \int_{\Im (\k)> \frac 4N} d\k\,
\tilde V_n^{(2)} (\k){\mathfrak  t}_N(\k)\tilde V_m^{(-1)}(\k)
+ \sum_\xi  \tilde v_n^{(2)} (\xi) {\mathfrak  t}_N(\xi) \tilde V_m^{(-1)}(\xi)-i P_{nm}\label{TiNnm}
\ee
where the contour is a straight line from $-\infty$ to $\infty$ just above the
pole at $\k = \frac {4i}N$. The rank 1 matrix $P$ commutes with
everything else in the spectral formulas and one would be tempted to drop it;
however this piece will turn out to be
godgiven in paper III.

Let us end with a few important remarks.

{\bf Remark 1.} In functional analysis the spectral formulas for operators
are the sums (integral) of their eigenvalues multiplied by the
corresponding eigenprojectors. In (\ref{reconTN}) this corresponds to the
first term, $\hat T^{(N)}$, in the RHS. The other terms in the RHS are still diagonal
and made of eigenprojectors, but the corresponding eigenvalues are infinite
and are replaced by the residues of the relevant poles. This is
the real novelty of such formulas.
We call the former part, the genuinely spectral representation, the {\it principal part}
and the latter {\it the residual part}. With some abuse of language we
will keep referring to formulas like (\ref{TiNnm}) as spectral representations,
since they are diagonal and contain only information about the spectrum.
It is important to notice that all the spectral representations
considered in section 5,6 and 7 represent matrices which are completely diagonal
in both the continuous and discrete bases of eigenvectors of the matrix $G$.
This characterizes all the ghost number 3 wedge states and marks a
sharp difference with the ghost number 0 wedge states, characterized by
Neumann matrices which are not completely diagonalizable in the same bases.

{\bf Remark 2.} The wedge states we have considered in this section are
characterized by the fact that they can be represented as squeezed states,
but only those with $Y(\pm i)$ insertions are BRST invariant surface states;
for the remaining ones
the latter properties are open questions and will be rediscussed in III.
However we would like to notice that the reconstruction formulas and commutativity
of their Neumann matrices hold for all of them.

{\bf Remark 3.} The spectral formulas are much more effective than the
analytic methods from the calculational point of view. In Appendix E,
where the equation $U^2=1$ is proved using the reconstruction formulas, one
can find an example of their potential by comparison with the long derivations
of section 2.

\section{Conclusion}

Let us conclude this paper by recalling the main results we have obtained.
The first is the construction of the ghost number 9 vertices,
eqs.(\ref{Vnmrsi},\ref{Vnmrs}).
The second important result is the construction of the discrete bases
of eigenvectors of $G$ as well as the bi--orthogonality and bi--completeness relations
(\ref{orthog}) and (\ref{bicomplete}), and the analysis of the
highly nontrivial properties of this spectrum.
Then we have completed the argument of I, showing that the squeezed states
appearing in the midterm of (\ref{ghostwedge}) do satisfy the recursion
relations of the wedge states. The way we have done it is somewhat different
from the one envisaged in I. The idea behind I was that all the involved
Neumann matrices could be simultaneously and completely diagonalized. In this paper
we have realized that
this is not possible. Not all the Neumann matrices entering the problem
can be completely diagonalized (this will be evident in III). Nevertheless
it is still possible to carry out the program started in I. We have shown that
the wedge states recursion relations can be proved for the eigenvalues,
and that
on the basis of this knowledge it is possible to reconstruct ghost number 3
Neumann matrices which can be identified with surface states
representing the wedge states expected as a result of the star product.
This is enough to guarantee that the three
strings vertex we have introduced in section 2 does the job, that is
by $*$--multiplying two squeezed states like the ones in the RHS of
(\ref{ghostwedge}) we obtain in the LHS the wedge state required by the
recursion relation (\ref{wedgedef}). What is still missing
is how to recover the the ghost number 0 wedge states from the
so obtained ghost number 3 states. This task, which is simply the
$bpz$ conjugation in the matter sector, requires a very involved and
roundabout treatment in the ghost sector and will be dealt with in the next paper.

\acknowledgments

L.B. would like to thank the GGI in Florence for hospitality and
financial support during this research.
C.M. and D.D.T. would like to thank SISSA for the kind hospitality during part of this research.
The work of D.D.T. was supported by the Korean Research Foundation Grant funded by the Korean Government with
grant number KRF 2009-0077423.
R.J.S.S is supported by CNPq-MCT-Brasil.

\section*{Appendix}
\appendix

\section{Ghost insertions and correlators}

The two--point function for a $b-c$ system can take different forms depending
on the way we insert the zero modes to soak the background ghost charge
at $\infty$, which is necessary in order to get a nonvanishing result.
A generic way is to define, as in \cite{leclair, LeClair:1988sj},
\be
\ll c(z) b(w)\gg_{(t_1,t_2,t_3)} &=& \langle 0|c(z)
b(w) c(t_1) c(t_2)c(t_3) |0\rangle \0\\
&=& \frac 1{z-w} \prod_{i=1}^3 \frac {t_i-z}{t_i-w}\,
(t_1-t_2)(t_1-t_3)(t_2-t_3)\label{leclair}
\ee
Another way of inserting the zero modes is by means of the weight 0 operator
$Y(t)= \frac 12 \partial^2 c(t) \partial c(t) c(t)$.
We have
\be
\ll c(z)\, b(w)\gg_t =\langle c(z)\, b(w)\, Y(t)\rangle = \frac 1{z-w}
\frac {(t-z)^3}{(t-w)^3} \label{Yins}
\ee
In this appendix we would like to study the relation,  between the
normal ordering in the $b-c$ correlator and the ordering term in the
matrices of Neumann coefficients of the three strings vertex.
The radial ordering of the $b,c$ fields can be expressed
as follows in terms of the natural normal ordering ($:\,:$):
\be
R(c(z) b(w)) &=& c(z) b(w),\quad\quad |z|>|w|\label{cbnn}\\
&=& \sum_n c_n z^{-n+1} \, \left(\sum_{k\leq -2} b_k w^{-k-2}+
\sum_{k\geq -1} b_k w^{-k-2}\right)\0\\
&=& :c(z)\, b(w) : + \sum_{n\geq 2} z^{-n+1} \,w^{-k-2}\0\\
&=& :c(z)\, b(w) : + \frac 1 {z-w}\0
\ee
The same expression is obtained for $|z|<|w|$.

\noindent We can use the above radial ordering in order to get
the ordering terms for the Neumann coefficients.
\be
R(c(z) b(w))&=& :c(z)\, b(w) : + \frac 1 {z-w}\longrightarrow
\hat V_{nm}^{rs} = \ldots -\frac {\delta^{rs}}{z-w} \0\\
 &&\longrightarrow U_{nm} =\ldots - \frac z{z-w} \label{cn1}
\ee
where only the relevant parts are written down.

\noindent Now let us consider a ghost surface state determined by a map $g(z)$,
\be
\langle g| =  \langle 0| e^{-\sum c_n S^{(g)}_{nm} b_m}
\label{surface}
\ee
In order to find the matrix $S^{(g)}$ we proceed as follows:
using (\ref{Yins}) we identify (see \cite{GRSZ2,Schnabl05}) up to constant
factors
\be
\langle g| c(z)\, b(w)\, Y(t)\rangle=
\frac {(g'(w))^2}{g'(z)} \, \frac 1{g(z)-g(w)} \,
\left(\frac {g(t)-g(z)}{g(t)-g(w)}\right)^3\label{surface2}
\ee
with $Y$ insertion at the generic point $t$.
The wedge states are generated by the well--known functions
\be
g(z)\equiv g_N(z) = \left(\frac{1+iz}{1-iz}\right)^{\frac{2}{N}}\0
\ee

\noindent If we set the insert $Y$ at $t=0$ we get
\be
S_{nm}^{(g_N)} &
=&\oint\frac{dz}{2\pi i}\oint\frac{dw}{2\pi i}\frac{1}{z^{n-1}}
\frac{1}{w^{m+2}}\0\\
&&\cdot \left(\frac{(g_N'(z))^2}{(g_N'(w))}\,
\frac{1}{g_N(z)-g_N(w)}\left(\frac {1-g_N(w)}{1-g_N(z)}\right)^3 -
\frac {w^3}{z^3(z-w)}\right)\label{Snmf2}
\ee
This is the Neumann matrix for the ghost number 0 wedge states. The others,
which represent $gh=3$ states with a $Y(\pm i)$--insertion, are
just obtained by setting $t=\pm i$.

\section{Quadratic expressions involving $\EE$, $\EU_{(\pm i)}$ and $Z$}

This appendix is devoted to complete the derivation of the fundamental
properties of Neumann coefficients started in section 6.

\subsection{Quadratic expressions involving $\EE_{(i)}$ and $\EU_{(i)}$}

The product $\EE_{(i)}\EU_{(i)} $ can be carried out as in subsection 2.4.1,
forgetting the $f$ factors on the first matrix. It is easy to see that it
leads to
\be
(\EE_{(i)}\EU_{(i)})_{nm} &=&  \oint\frac{dz}{2\pi i}\frac{1}{z^{n+1}}
\oint\frac{dw}{2\pi i} \frac{1}{w^{m+1}}\,\left[ -
\left(\frac 1{1-zw} -\frac w{z+w}\right) \,
 \Big{(}1- p_i(z,-w)\Big{)}\right. \0\\
&& + \frac {z^2 w^2}{1-zw}\label{EEiEUi}\\
 &+&  \left(\frac 1{1-zw} -\frac w{z+w}\right)
 \Big{(}1- p_i(-z,w) \Big{)}\ \,
\left.\frac{f(-z)}{f(w)}\right]\0\\
 &=& (-\EE_{(i)} \Ct\,+ {\bf 1}_{ss}+ \Ct \, \EU_{(i)})_{nm}\0
\ee
which differs from $\Ct\EU_{(i)}$ only in the zero mode sector. Similarly
we can prove that
\be
\EE_{(-i)}\EU_{(i)}  = -\EE_{(-i)} \Ct\,+ {\bf 1}_{ss}+ \Ct \, \EU_{(i)}
\label{EEmiEUi}
\ee
Taking the average of these two we obtain
\be
\EE\,\EU_{(i)} = -\EE \Ct\,+ {\bf 1}_{ss}+ \Ct \, \EU_{(i)}\label{EEEUi}
\ee
In a similar way
\be
\EE_{(\pm i)}\bar\EU_{(i)} = -\EE_{(\pm i)} \Ct\,+ {\bf 1}_{ss}+ \Ct \,\bar \EU_{(i)}
\label{EEpmiEUi}
\ee
and
\be
\EE_{(\pm i)}\EE_{(i)}  = -\EE_{(\pm i)} \Ct\,+ {\bf 1}_{ss}+ \Ct \, \EE_{(i)}
\label{EEpmiEEi}
\ee
Using twist conjugation we finally get
\be
\EE\,\EU = {\bf 1}_{ss}\label{EEEE}
\ee
Let us now consider $\EU_{(i)}\,\EE_{(i)}$.
\be
\sum_{k=-1}^\infty \EU_{(i)nk}\,\EE_{(i)km}&=&\oint\frac{dz}{2\pi i}\frac{1}{z^{n+1}}
\oint\frac{d\z}{2\pi i}\oint\frac{d\zh}{2\pi i} \oint\frac{dw}{2\pi i}
\frac{1}{w^{m+1}}\frac {\z\zh}{\z\zh-1} \label{EUiEEi1}\\
&&\cdot \frac{f(z)}{f(\z)} \Big{(}\frac{1}{1+z\z} -\frac{\z}{\z-z}\Big{)}
  \Big{(}1- p_i(z,\zeta) \Big{)} \cdot\0\\
&&\cdot \left(\frac 1{1+\zh w} -\frac w{w-\zh}\right) \,
 \Big{(}1- p_i(\zh,w)\Big{)}=* \0\\
\ee
Here it is more convenient to integrate first with respect to $\zh$.
There are two poles
at $\zh=\frac 1\z, w$ (the poles at $\zh=\pm i$ can be avoided with a
regulator).
We get
\be
\{\zh= \frac 1\z\}\quad\quad\quad * &=& \oint\frac{dz}{2\pi i}\frac{1}{z^{n+1}}
\oint\frac{d\z}{2\pi i} \oint\frac{dw}{2\pi i}
\frac{1}{w^{m+1}}\,\left[ \frac{f(z)}{f(\z)}\Big{(}\frac{1}{1+z\z} -
\frac{\z}{\z-z}\Big{)}\right.\0\\
&&\cdot \Big{(}1- p_i(z,\zeta) \Big{)}\left(\frac \z{\z+ w} +
\frac {\z w}{1-w\z}\right) \,
 \Big{(}1- p_i(\frac 1\z,w)\Big{)}\
 \Big{)}\,\label{EUiEEi2}\\
\{\zh= w \}\quad\quad\quad &+& \frac{\z w^2} {\z w-1} \frac {f(z)}{f(\z)}
\left.\Big{(}\frac{1}{1+z\z} - \frac{\z}{\z-z}\Big{)}\Big{(}1- p_i(z,\zeta)
\Big{)}\right]\0\\
&=& \oint\frac{dz}{2\pi i}\frac{1}{z^{n+1}}
 \oint\frac{dw}{2\pi i} \frac{1}{w^{m+1}}\,\left[ \frac{f(z)}{f(-w)}
\Big{(}\frac{1}{1-zw} -\frac{w}{w+z}\Big{)}\right.\0\\
&&\cdot  \Big{(}1- p_i(z,-w) \Big{)}
+\frac {w^2 z^2}{1-zw}\0\\
&&-\left. \left( \frac w{z+w}-\frac 1{1-zw}\right)
\Big{(}1- p_i(-z,w) \Big{)}\right]\0\\
&=& (\EU_{(i)}\,\Ct -\Ct\,\EE_{(i)}+ {\bf 1}_{ss})_{nm}\0
\ee
Similarly
\be
\EU_{(-i)}\EE{(i)}&=& \EU_{(-i)}\,\Ct -\Ct\,\EE_{(i)}+ {\bf 1}_{ss}\0\\
\bar\EU_{(i)}\EE{(i)}&=& \bar\EU_{(i)}\,\Ct -\Ct\,\EE_{(i)}+ {\bf 1}_{ss}\label{barEUiEEi}\\
\bar\EU_{(-i)}\EE{(i)}&=&\bar \EU_{(-i)}\,\Ct -\Ct\,\EE_{(i)}+ {\bf 1}_{ss}\0
\ee
from which it is easy to deduce the quadratic relations involving $\EE$.

\subsection{Quadratic expressions involving $Z$}

Let us now compute
\be
&&\sum_{k=-1}^\infty Z_{nk}\,\EU_{(\pm i)km}= \oint\frac{dz}{2\pi i}\frac{1}{z^{n+1}}
\oint\frac{d\z}{2\pi i}\oint\frac{d\zh}{2\pi i} \oint\frac{dw}{2\pi i}
\frac{1}{w^{m+1}}\frac {\z\zh}{\z\zh-1}\\
&&\cdot  \Big{(}  -\frac {z^2}{\z}\frac 1{z-w}\Big{)}
\frac{f(\zh)}{f(w)} \Big{(}\frac{1}{1+\zh w}
-\frac{w}{w-\zh}\Big{)}\,
  \Big{(}1- p_{\pm i}(\zh,w) \Big{)} =*\label{ZEUi1}
\ee
Now we proceed as above and the result is
\be
\{\z= \frac 1\zh\}\quad \quad\quad *&=& \oint\frac{dz}{2\pi i}\frac{1}{z^{n+1}}
\oint\frac{d\z}{2\pi i} \oint\frac{dw}{2\pi i}
\frac{1}{w^{m+1}}\,\left[
\frac{\zh z^2}{1-z\zh} \right.
\frac{f(\zh)}{f(w)} \Big{(} \frac1{1+\zh w} -   \frac{w}{ w-\zh}
\Big{)}\cdot\0\\
&&\cdot   \Big{(}1- p_{\pm i}(\zh,w) \Big{)}\label{ZEUi2}\\
\{\z=z\}\quad\quad\quad &+& \frac {\zh z^2}{\zh z-1} \, \frac {f(\zh)}{f(w)} \,
\Big{(}\frac{1}{1+\zh w} -\frac{w}{w-\zh}\Big{)}\0\\
&&\left.\cdot   \Big{(}1- p_{\pm i}(\zh,w) \Big{)}\right]=0\0
\ee
By twist conjugation $Z\bar \EU_{\pm i}=0$.

\noindent The calculation of $Z\EE_{\pm i}$ is similar but simpler (no $f$ factor) and the
conclusion is the same: $Z\EE_{\pm i}=0$.

\noindent Consider now $\EU_{(i)} \,Z$:
\be
\sum_{k=-1}^\infty \EU_{(i)nk}\,Z_{km}&=&\oint\frac{dz}{2\pi i}\frac{1}{z^{n+1}}
\oint\frac{d\z}{2\pi i}\oint\frac{d\zh}{2\pi i} \oint\frac{dw}{2\pi i}
\frac{1}{w^{m+1}}\frac {\z\zh}{\z\zh-1} \label{EUiZ1}\\
&&\cdot \frac{f(z)}{f(\z)} \Big{(}\frac{1}{1+z\z} -\frac{\z}{\z-z}\Big{)}
  \Big{(}1- p_i(z,\zeta) \Big{)} \cdot\0\\
&&\cdot \Big{(} -\frac{\zh^2}{w}\frac 1{\zh-w}  \Big{)}=*\,\0
\ee
Here it is more convenient to integrate first with respect to $\zh$.
There are two poles
at $\zh=\frac 1\z, w$.
\be
\{\zh= \frac 1\z\}\quad\quad\quad *&=& \oint\frac{dz}{2\pi i}\frac{1}{z^{n+1}}
\oint\frac{d\z}{2\pi i} \oint\frac{dw}{2\pi i}
\frac{1}{w^{m+1}}\,\left[ \frac{f(z)}{f(\z)}\right. \0\\
&\cdot& \Big{(}\frac{1}{1+z\z} - \frac{\z}{\z-z}\Big{)}
  \Big{(}1- p_i(z,\zeta) \Big{)}\Big{(}-\frac 1{w\z^2}\frac 1{1-\z w} \Big{)}\,
\label{EUiZ2}\\
\{\zh= w \}\quad\quad\quad &-& \frac{\z w^2} {\z w-1} \frac {f(z)}{f(\z)}
\left.\Big{(}\frac{1}{1+z\z} - \frac{\z}{\z-z}\Big{)}
\Big{(}1- p_i(z,\zeta) \Big{)}\right]\0\\
&=& \oint\frac{dz}{2\pi i}\frac{1}{z^{n+1}}
\oint\frac{d\z}{2\pi i} \oint\frac{dw}{2\pi i}
\frac{1}{w^{m+1}}\, \frac{f(z)}{f(\z)}\0\\
&\cdot& \Big{(}\frac{1}{1+z\z} - \frac{\z}{\z-z}\Big{)}
  \Big{(}1- p_i(z,\zeta) \Big{)}\Big{(}-\frac 1{w\z^2}(1+w\z+w^2\z^2) \Big{)}=**\0
\ee
This gives
 \be
** &=& \oint\frac{dz}{2\pi i}\frac{1}{z^{n+1}}
\oint\frac{dw}{2\pi i}
\frac{1}{w^{m+1}}\,\left\{\frac 12 \frac {d^2}{d\z^2}
\left[  \z \frac{f(z)}{f(\z)}\right.\right.\0\\
&\cdot&\left. \Big{(}\frac{1}{1+z\z} - \frac{\z}{\z-z}\Big{)}
  \Big{(}1- p_i(z,\zeta) \Big{)}
\Big{(} -\frac 1{w}(1+w\z+w^2\z^2) \Big{)}\right]_{\z=0}\0\\
&+&\left.\frac 1{wz}(1+wz+w^2z^2)\right\}\0\\
&=&  \oint\frac{dz}{2\pi i}\frac{1}{z^{n+1}}
\oint\frac{dw}{2\pi i} \frac{1}{w^{m+1}}\,
\left(\frac {f(z)}{zw}\,\frac {P_i(z,w)}{9(z-i)^2}
+\frac 1{wz}+1+wz\right)\label{EUiZ3}
\ee
where
\be
P_i(z,w)= {9+7z^2+9z^4+ 6iz(1-z^2) +3wz(3-3z^2+2iz)+9w^2z^2}
\label{Pizw}
\ee
$P_i(z,w)$ is a polynomial of fourth order in $z$ and second order in $w$.
Therefore the $\EU_{(i)} Z$ matrix vanishes except for the first three columns.

\noindent The result for $\bar \EU_{(i)} Z$ is the same with the substitution
$P_i\to P_{-i}$,
\be
P_{-i}(z,w)= 9+30 i z -41 z^2 -30iz^2+9z^4 +w(9z +30i z^2-9 z^3)+ 9w^2z^2\label{Pmizw}
\ee
The calculation for $\EE_{(i)}\,Z$ is similar but simpler and the conclusion is
\be
(\EE_{(i)}Z)_{nm}= \oint\frac{dz}{2\pi i}\frac{1}{z^{n+1}}
\oint\frac{dw}{2\pi i} \frac{1}{w^{m+1}}\,
\left(\frac 1{zw}\,\frac {Q_i(z,w) }{(z-i)^2}
+\frac 1{wz}+1+wz\right)\label{EEiZ}
\ee
where
\be
Q_i(z,w)= 1+2iz-z^2-2iz^3+z^4 +w(z+ 2iz^2-z^3) +w^2z^2\label{Qizw}
\ee
$Q_i(z,w)$ is a polynomial of fourth order in $z$ and second order in $w$.
The $\EE_{(i)} \, Z$ matrix vanishes except for the first three columns.
By twist conjugation one gets $\EE_{(-i)} \, Z$.

\noindent It is easy to see that $Z^2=0$.

In the sequel we will need more explicit expressions for the RHS of
(\ref{EUiZ3}) and (\ref{EEiZ}). To this end let us compute $U_{(i)n,i}$
for $-1\leq i \leq 1$ in a more explicit form
\be
U_{(i)n,-1}&=&\oint\frac{dz}{2\pi i}\oint\frac{dw}{2\pi
i}\frac{1}{z^{n+1}} \left(\frac{f(z)}{f(w)}
\Big{(}\frac{1}{1+zw}-\frac{w}{w-z}\Big{)}
\, (1-p_i(z,w))  -\frac {z^2}w \frac 1{z-w}\right) \0\\
&=& \oint\frac{dz}{2\pi i}\frac 1{z^{n+1}}\,
\left(-\frac {zf(z)}{(z-i)^2}-z\right)\label{U-1}
\ee
In this calculation only the pole at $w=0$ matters, while the pole at $w=z$
has a vanishing residue.
Similarly
\be
U_{(i)n,0}&=&\oint\frac{dz}{2\pi i}\oint\frac{dw}{2\pi
i}\frac{1}{z^{n+1}}\,\frac 1w\, \left(\frac{f(z)}{ f(w)}
\Big{(}\frac{1}{1+zw}-\frac{w}{w-z}\Big{)}
\, (1-p_i(z,w))  -\frac {z^2}w \frac 1{z-w}\right)\0\\
&=& \oint\frac{dz}{2\pi i}\frac{f(z)}{z^{n+1}}\,\frac d{dw}
\left(w\,f(-w)  \Big{(}\frac{1}{1+zw}-\frac{w}{w-z}\Big{)}
\, (1-p_i(z,w)) \right)_{w=0}\0\\
&=& \oint\frac{dz}{2\pi i}\frac 1{z^{n+1}}\,
\left(f(z)\frac { -3-2iz +3z^2}{3(z-i)^2}-1\right)\label{U0}
\ee
Finally
\be
U_{(i)n,1}&=&\oint\frac{dz}{2\pi i}\oint\frac{dw}{2\pi
i}\frac{1}{z^{n+1}}\frac 1{w^2} \left(\frac{f(z)}{f(w)}
\Big{(}\frac{1}{1+zw}-\frac{w}{w-z}\Big{)}
\, (1-p_i(z,w))  -\frac {z^2}w \frac 1{z-w}\right) \0\\
&=& \oint\frac{dz}{2\pi i}\frac{f(z)}{z^{n+1}}\,\left[\frac 12 \frac {d^2}{dw^2}
\left(w\,f(-w)  \Big{(}\frac{1}{1+zw}-\frac{w}{w-z}\Big{)}
\, (1-p_i(z,w))\right)_{w=0}-\frac 1z\right] \0\\
&=& \oint\frac{dz}{2\pi i}\frac 1{z^{n+1}}\,\left(
-f(z)\frac {9+7z^2+9z^4+6iz(1-z^2)}{9z (z-i)^2} -\frac 1z\right)\label{U1}
\ee
Comparing these expressions with (\ref{EUiZ3}) and (\ref{Pizw}) it is evident
that
\be
(\EU_{(i)} Z)_{n,i}= - U_{(i)n,-i}, \quad {\rm i.e.} \quad (U_{(i)} Z)_{n,i}= -
U_{n,-i}\label{EUni}
\ee
As for $\bar\EU_{(i)}$, we have
\be
\bar U_{(i)n,-1}&=&\oint\frac{dz}{2\pi i}\frac 1{z^{n+1}}\,
\left(-\frac {zf(-z)}{(z-i)^2}-z\right)\label{barU-1}\\
\bar U_{(i)n,0}&=&\oint\frac{dz}{2\pi i}\frac 1{z^{n+1}}\,
\left(f(-z)\frac { -9-30iz +9z^2}{9(z-i)^2}-1\right)\label{barU0}\\
\bar U_{(i)n,1}&=& \oint\frac{dz}{2\pi i}\frac 1{z^{n+1}}\,\left(
f(-z)\frac {-9-30iz+41z^2+30iz^3-9z^4}{9z (z-i)^2} -\frac 1z\right)\label{barU1}
\ee
Therefore
\be
(\bar\EU_{(i)} Z)_{n,i}= - \bar U_{(i)n,-i}, \quad {\rm i.e.} \quad
(\bar U_{(i)} Z)_{n,i}= -
\bar U_{n,-i}\label{EbarUni}
\ee

\noindent On the other hand it is even easier to prove that
\be
E_{(i)n,-1}&=&\oint\frac{dz}{2\pi i}\oint\frac{dw}{2\pi i}\frac{1}{z^{n+1}}
 \left[\Big{(} \frac{1}{1+zw}-\frac{w}{w-z}\Big{)}
\,(1-p_i(z,w))  -\frac {z^2}w \frac 1{z-w}\right] \0\\
&=&\oint\frac{dz}{2\pi i}\oint\frac{dw}{2\pi i}\frac{1}{z^{n+1}}
\left(-\frac z{(z-i)^2}-z\right)= E_{(i)n,1}\label{E1}\\
E_{(i)n,0}&=&0\0
\ee
Comparing with the RHS of (\ref{EEiZ}) and the expression for $Q_i(z,w)$
we can conclude that
\be
(\EE_{(i)} \, Z)_{n,1}= (\EE_{(i)}  \, Z)_{n,-1}= - E_{(i)n,1}=-E_{(i)n,-1}, \quad\quad
(\EE_{(i)}  \, Z)_{n,0}=0\label{EEi01-1}
\ee
i.e.
\be
(E_{(i)}  \, Z)_{n,1}= (E_{(i)}  \, Z)_{n,-1}= - E_{(i)n,1}=-E_{(i)n,-1}, \quad\quad
(E_{(i)}  \, Z)_{n,0}=0\label{Ei01-1}
\ee
Analogously one can show that
\be
(E \, Z)_{n,1}= (E \, Z)_{n,-1}= - E_{n,1}=-E_{n,-1}, \quad\quad
(E  \, Z)_{n,0}=0\label{E01-1}
\ee
For future use we record also
\be
&&E_{n,0}=0\0\\
&& E_{2n,1}= E_{2n,-1}=0\label{Eni}\\
&& E_{2n+1,1}= E_{2n+1,-1}= (-1)^n (2n+1)\0
\ee

\section{The eigenvalues of $T_n$}

The explicit form of the squeezed states in the midterm of (\ref{ghostwedge})
was derived from the LHS of the same equation in I. It required solving the
equation we dubbed KP, \cite{KPot}. The latter arises from writing
\be
e^{t\left({\cal L}^{(g)}_0 +
{\cal L}_0^{(g)\dagger}\right)}\equiv e^{t\left(c_M^\dagger A_{Mn}b_n^\dagger
+ c^\dagger_M  C_{MN}b_N+b_m^\dagger D_{mn}c_n-
c_mB_{mN}b_N \right)}\label{L0L0}
\ee
where $t= \frac {2-n}2$, and equating this to
\be
e^{\eta(t)} e^{c^\dagger \alpha(t) b^\dagger}
e^{c^\dagger \gamma(t)b} e^{b^\dagger \delta(t)c} e^{c\beta(t) b}
\label{factorexp1}
\ee
The matrix $\alpha$ and the parameter $\eta$, in particular, must satisfy
\be
\dot \alpha=A+C\,\alpha +\alpha\,D^T +\alpha\,B\,\alpha\label{KP}
\ee
and
\be
\dot \eta = -{\rm Tr}\left(B,\alpha\right)\label{etanat}
\ee
In I we solved this equation. However, for reasons explained later on,
we will proceed here in a different way. Instead of solving (\ref{KP})
and then diagonalizing the solution, we will diagonalize (\ref{KP})
and then solve the equation for the eigenvalues. Therefore
our first task is to find the eigenvalues of the various matrices
appearing in (\ref{KP}). In I we showed that $\tilde A,D^T$
and $(D^T)^2-BA$ are diagonal in the continuous weight 2 basis and computed the
explicit eigenvalues. The analysis below will confirm this.

\subsection{Revisiting I}

This subsection is devoted to re--deriving the solution to the KP equation
with respect to I. The reason for it is the following.
In I we used the following `commutation rules' for lame matrices:
\be
A D^T = C A \label{ADT=CAnat}
\ee
and
\be
BC = D^T B\label{BC=DTBnat}
\ee
The latter has to be qualified. It is true except for the terms
\be
(BC - D^T B)_{2,0}= - \frac 32 \pi^2,
\quad\quad (BC - D^T B)_{3,-1}= -3\pi^2\label{toxic}
\ee
Therefore in the sequel, differently from I, we will not use eq.(\ref{BC=DTBnat}).
We will only use
(\ref{ADT=CAnat}). But, of course, we have to change the strategy with respect to I.
Instead of solving (\ref{KP}) and then diagonalizing it, we will first
diagonalize (\ref{KP}) and then solve for its eigenvalues.
Let us start from eq.(\ref{KP}) with the initial condition $\alpha(0)=0$.
The solution to (\ref{etanat}) is obvious once we know $\alpha(t)$.
Let us make the following ansatz for $\alpha(t)$
\be
\alpha_1(t) = A\,Q_1(t) \label{ansatz1}
\ee
Using (\ref{ADT=CAnat}) we get
\be
A\,\dot Q_1= A\,(1+D^T Q_1+Q_1D^T +Q_1BA\,Q_1)\label{eq1}
\ee
It is obvious that, if $Q_1$ satisfies
\be
\dot Q_1= 1+D^T Q_1+Q_1D^T +Q_1BA\,Q_1\label{intermediate1}
\ee
with $Q_1(0)=0$, $\alpha_1$ will satisfy (\ref{KP}).

Next we wish to solve (\ref{intermediate1}) for the continuous eigenvalues
of the matrix $Q_1(t)$. We have recalled above that the matrix $D^T$ is diagonal in the
$V^{(2)}(\k)$ basis, with eigenvalue ${\mathfrak c}(\k)$ and that $(D^T)^2-BA$ is also
diagonal. This means that $BA$ itself is diagonal in the same basis.
Looking at (\ref{intermediate1}) we see that since the solution $Q(t)$ will be a function
of $D^T$ and $BA$ it will also be diagonal in the same basis. So in
(\ref{intermediate1}) we can replace the matrices with their eigenvalues.
At this point solving the equation is elementary.
\be
Q_1(t)& =&\frac {{\rm sinh}\left(\sqrt{(D^T)^2-BA}\,t\right)}
{ \sqrt{(D^T)^2-BA}\, {\rm cosh} \left(\sqrt{(D^T)^2-BA}\,t\right)
-D^T\, {\rm sinh}\left(\sqrt{(D^T)^2-BA}\,t\right)}\label{Q1t}
\ee
where, for economy of notation, we assume that the matrices represent their
continuous eigenvalues. Since $A$ is also diagonal on the $V^{(2)}(\k)$ basis,
we can mutiply this solution by its eigenvalue and get
\be
\alpha_1(t) = A\, Q(t)\label{alpha1}
\ee
for the corresponding continuous eigenvalues. Now from the continuous eigenvalues of
$A, BA$ and $D^T$ we can construct the continuous eigenvalue of $\alpha_1(t)$.
This has already been done in I, and the result, ${\mathfrak t}_t(\k)$, is
that reported in eq.(\ref{tt}).

\subsection{The discrete eigenvalues}

Coming now to the discrete eigenvalues we would like to be able to write
\be
V^{(-1)} (\xi)\tilde\alpha(t) =
{\mathfrak t}_t(\xi) V^{(-1)}(\xi)\label{alphadiag-1}
\ee
or
\be
\tilde\alpha(t) \, V^{(2)}(\xi) =
{\mathfrak t}_t(\xi) \,V^{(2)}(\xi),\label{alphadiag2}
\ee
calculate the explicit expression of ${\mathfrak t}_n(\xi)$ and justify
our final statements ${\mathfrak t}_n(\pm 2i)=1$ and
${\mathfrak t}_n(\xi=0)=-1$, eqs.(\ref{discretetn0}) and (\ref{discretetn}).
But if we apply the previous matrices to
the discrete eigenvectors of $G$ we immediately run into difficulties.
If we use the same ansatz (\ref{ansatz1}) and eq.(\ref{intermediate1})
we see that, for instance, $D^T$ has identically vanishing eigenvalues
in the $V^{(-1)}(\xi)$ basis. This is a consequence of the
singular nature of the discrete eigenvalues of the $\tilde A,\tilde B,C,D$ matrices,
which creates serious problems when we try to compute the corresponding discrete
eigenvalues of the wedge states by integrating the KP equation. These problems
will be understood in the next paper, when we will be able to produce
the reconstruction formula for the ghost number 0 wedge states.
For the time being we proceed blindly in search of these eigenvalues.

As a preliminary step let us apply $C$ from the
right to the discrete basis
${ V}^{(-1)}(\xi)=(V_{-1}^{(-1)}(\xi),V_0^{(-1)}(\xi),\ldots)$.
We get, for $\xi=0$,
\be
({V}^{(-1)}(0)\,C)_n= C_{1,n}+C_{-1,n}=-A_{-1,n}-A_{1,n}=0\label{V-10C}
\ee
while for $\xi=\pm 2i$,
\be
(V^{(-1)}(\pm 2i)\,C)_n&=& C_{1,n}-C_{-1,n}\pm 2i C_{0,n}=
-A_{-1,n}+A_{1,n}\mp 2i A_{0,n}\0\\
&&= \lim_{x\to 0} \left( \pm \frac {2i}x
\,V^{(-1)}_n(\pm 2i +x)\right)\label{V-12iC}
\ee
for $n\geq 2$. Therefore we can write
\be
(V^{(-1)}(\xi)\,C)_n= \lim_{x\to 0} \left( \frac {\xi}x\,
V^{(-1)}_n(\xi +x)\right)\label{V-1xiC}
\ee
but
\be
(V^{(-1)}(\xi)\,C)_j= \xi^2 V^{(-1)}(\xi=0)\label{V-1Cj}
\ee
In summary we have the eigenvalue equation
\be
(V^{(-1)}(\xi)\,C)_N= \lim_{x\to 0} \left( \frac {\xi}x\,
V^{(-1)}_N(\xi +x)\right), \quad N\geq -1\label{V-1xi0C}
\ee
for $\xi=0$, but we have to get along with (\ref{V-1xiC},\ref{V-1Cj})
for $\xi\neq 0$. To give a meaning to (\ref{V-1xi0C}) we agree
that the eigenvalue $\xi =0$ will be represented by a small
number $\epsilon$, and that the latter will be sent to 0 at the
end of the calculations.

Likewise we get
\be
({V}^{(-1)}(0)\,A)_n= A_{1,n}+A_{-1,n}=0\label{V-10A}
\ee
and
\be
(V^{(-1)}(\pm 2i)\,A)_n= A_{1,n}-A_{-1,n}\pm 2i A_{0,n}=
 \lim_{x\to 0} \left( \mp \frac {2i}x\,
V^{(-1)}_n(\pm 2i +x)\right)\label{V-12iA}
\ee
Summarizing
\be
&&(V^{(-1)}(\pm 2i)\,A)_n= \lim_{x\to 0} \left(\mp\frac {2i}x\,
V^{(-1)}_n(\pm 2i +x)\right),\quad\quad n\geq 2\label{V-1xiA}\\
&&(V^{(-1)}(0)\,A)_n= \lim_{x\to 0,\xi\to 0} \left(\frac {\xi}x\,
V^{(-1)}_n(\xi +x)\right)=0,\quad\quad n\geq 2\label{V-1xi0A}\\
&&(V^{(-1)}(\xi)\,A)_i=0,\quad\quad -1\leq i\leq 1\label{V-1xiA0}
\ee
In (\ref{V-1xi0A}) the limits have to be taken in such a way that
the result be 0, so, as above, we introduce a regulator $\epsilon$
to represent the eigenvalue $\xi=0$.
Notice that $\lim_{x\to 0} V^{(-1)}_n(\xi +x)=0$ for $n\geq 2$,
while $\lim_{x\to 0} V^{(-1)}_i(\xi +x)=V^{(-1)}_i(\xi)$ and that
$V^{(-1)}_n(\xi +x)$ for $n\geq -1$ is bi-orthogonal to both
$V^{(2)}_n(\xi)$ and $V^{(2)}_n(\k)$ in the limit $x\to 0$.

\noindent Apart from the eigenvalue equation (\ref{V-1xi0C}),
the above equations (\ref{V-1xiC},\ref{V-1Cj},\ref{V-1xiA},\ref{V-1xi0A},
\ref{V-1xiA0}) represent `almost' eigenvalue equations. They will be used
to compute the discrete eigenvalues of the wedge state Neumann matrices.

\noindent In I we proved, eq.(5.19), that $\Delta=C^2-AB$ is diagonal in the
continuous basis ${V}^{(-1)}(\k)$. It is easy to prove that
it is diagonal also in the discrete basis, using (4.21) of I.
For instance it is elementary to prove that
\be
&&\sum_{n=-1} V_n^{(-1)}(0)\Delta_{n,0} =0 \0\\
&&\sum_{n=-1} V_n^{(-1)}(0)\Delta_{n,\pm 1}=0\0\\
&&\sum_{n=-1} V_n^{(-1)}(2i)\Delta_{n,\pm 1}=-\pi^2 V_{\pm 1}^{(-1)}(2i)\0\\
&&\sum_{n=-1} V_n^{(-1)}(2i)\Delta_{n,0}=-\pi^2 V_{0}^{(-1)}(2i)\0
\ee
So we can write
\be
\sum_{n=-1} V_n^{(-1)}(\xi)\Delta_{n,m}=\frac {\pi^2}4 \xi^2\,
V_{m}^{(-1)}(\xi)\label{V-1Delta}
\ee
Thanks to these results, we set out to compute the discrete eigenvalues
of the wedge state Neumann matrices.

\noindent We make a new ansatz for the solution to the KP equation
\be
\alpha_2(t) = Q_2(t) A\label{ansatz2}
\ee
Using (\ref{ADT=CAnat}) we get
\be
\dot Q_2A= (1+ Q_2C+C\,Q_2 +Q_2AB\,Q_2)A\label{eq2}
\ee
It is obvious that, if $Q_2$ satisfies
\be
\dot Q_2= 1+ Q_2 C+C\,Q_2 + Q_2 AB\,Q_2\label{intermediate2}
\ee
with $Q_2(0)=0$, $\alpha_2$ will satisfy (\ref{KP}).
Now, let us remember that $C$ and $C^2-AB$
are diagonal in the $V^{(-1)}(\xi)$ basis as far
as the $\xi=\epsilon \to 0$ eigenvalue is concerned. Arguing
as before we can conclude
that also $AB$ is diagonal. Proceeding from now on for this single
eigenvalue, we can replace the
matrices in (\ref{intermediate2}) with their discrete eigenvalues. The solution to
(\ref{intermediate2}) is
\be
Q_2(t) = \frac {{\rm sinh}\left(\sqrt{C^2-AB}\,t\right)}
{ \sqrt{C^2-AB}\, {\rm cosh} \left(\sqrt{C^2-AB}\,t\right)
-C\, {\rm sinh}\left(\sqrt{C^2-AB}\,t\right)}\label{Q1}
\ee
where the matrix symbols have to be understood as representing the corresponding
discrete eigenvalues. Now remember that $A$ is also diagonal in the same basis.
Therefore
\be
\alpha_2(t) = \frac {{\rm sinh}\left(\sqrt{C^2-AB}\,t\right)}
{ \sqrt{C^2-AB}\, {\rm cosh} \left(\sqrt{C^2-AB}\,t\right)
-C\, {\rm sinh}\left(\sqrt{C^2-AB}\,t\right)}A\label{4.16}
\ee
The multiplication by (the eigenvalue of) $A$ requires a specification.
In fact applying the equation (\ref{eq2}) from the right to $V^{(-1)}$,
everything is all right as far as the
last $A$ factor on the right. When applying this to $V^{(-1)}$
the first three entries on the RHS of (\ref{eq2}) get cut out.

At this point however we can apply a remark similar to the one made
in sec. 2.4. An expression like
$e^{c_N^\dagger \tilde\alpha_{Nm}(t) b_m^\dagger}$ in (\ref{ghostwedge}) manifests
an ambiguity when applied to $|0\rangle$. We could add any term
that is killed by $|0\rangle$. This is the case if
we consider $:e^{c_i^\dagger \tilde\tau_{ij} b_j^\dagger+
c_N^\dagger \tilde\alpha_{Nm}(t) b_m^\dagger}:|0\rangle$, for
\be
:e^{c_i^\dagger \tau_{ij} b_j^\dagger+
c_N^\dagger \tilde\alpha_{Nm}(t) b_m^\dagger}:|0\rangle
= e^{c_N^\dagger \tilde\alpha_{Nm}(t) b_m^\dagger}|0\rangle\0
\ee
for any $3\times 3$ matrix $\tau$.
We therefore take advantage of
this ambiguity by adding to $\alpha_2(t)$ an upper left $3\times 3$
non zero matrix that solves the problem. The latter is constructed
as follows. Let us denote by $ {\mathfrak t}_t(\xi) $ the discrete
eigenvalues of $\tilde\alpha_2(t)$, then the $3\times 3$ matrix we are looking for
will be
\be
\tilde\alpha_{3\times 3}(t) = \sum_\xi \tilde {\bar V}^{(2)}(\xi)\,
{\mathfrak t}_t(\xi) \,
\tilde V^{(-1)}(\xi)\label{3x3}
\ee
where $\bar V^{(2)}(\xi)$ is $V^{(2)}(\xi)$ limited to the first three
entries. By adding this matrix to $\alpha_2(t)$ now we recreate the
missing entries in the RHS of (\ref{alphadiag-1}). Setting
$\alpha'_2(t)= \alpha_2(t)+ \alpha_{3\times 3}(t)$ we can now write
\be
V^{(-1)}\tilde \alpha'_2(t) = {\mathfrak t}_t(\xi) V^{(-1)}\label{alphadiag-1'}
\ee
We are now in the condition to compute the discrete eigenvalue of $\alpha_2(t)$
for $\xi=0$. We insert the corresponding discrete eigenvalue
of $A,C, \Delta$ calculated above. Eq.(\ref{4.16})
become rather singular and some care has to be used: one must take
$x \to 0$ with $\xi=\epsilon$ first. As one can see the eigenvalue of
$A$ and $C$ explode and the first term in the denominator
of (\ref{4.16}) becomes irrelevant. Since $A$ and $C$ have the same eigenvalue
the result is
\be
{\mathfrak t}_n(\xi=0)=-1\label{discretetn0}
\ee
We notice that this result is not affected by the limit $\epsilon\to 0$.

As for $\xi\neq 0$, try as we may, we cannot repeat the same derivation.
The reason for these failures will be understood in the next paper, when it
will become clear that the Neumann matrices for ghost number 0 wedge states
are not diagonal in the $V^{(-1)}(\xi)$ bases with $\xi\neq 0$.
However these eigenvalues are important for the ghost number 3 states.
Since the continuous eigenvalue formula (\ref{tt}) evaluated
at $\xi=\pm 2i$ gives an unambiguous result (keep $t$ generic and use
standard trigonometric identities):
\be
{\mathfrak t}_n(\pm 2i)=1\label{discretetn}
\ee
it logical to try this. As we have shown in section 5,6 and 7 this turns
out to be the correct value.

\noindent Finally, let us remark that, inserting these values in (\ref{3x3}),
we find once again the matrix
\be
\tilde\alpha_{3\times 3}(n)= \left(\begin{matrix}
0 &0& -1\\
0&1&0\\
-1&0&0
\label{3x3n}
\end{matrix}\right)
\ee
Let us also remark an important feature of the addition of the matrix $z$.
The way we implement it is by adjoining it to the matrix $A$, i.e.
$A\to A'=A + z$.
Looking at eq.(\ref{eq2}) we notice that $A'B= AB$, therefore the matrix
$z$ disappears from the equation for $Q_2$. It appears only in the last
step when we reconstruct the solution $\alpha_2= Q_2 A'$.

\section{Other reconstructions}

\subsection{Reconstruction of $X^{\pm}_{(i)}$}

In this section, we deal with the reconstruction of the matrices
$X^{+}_{(i)}={\cal C}{\hat V}^{12}_{(i)}$ and
$X^{-}_{(i)}={\cal C}{\hat V}^{21}_{(i)}$. The process of reconstructing these
matrices runs analogously to the one of reconstructing $X_{(i)}$, just with some slight
differences. The first of them is that, in this case, the discrete spectrum does
not contribute, hence we have just to consider the continuous part of the spectrum
\begin{equation}
X^{\pm}_{(i)nm} = \int_{C_{1}} d\kappa ~ \mu^{\pm}(\kappa) V^{(2)}_{n}(\k)
V^{(-1)}_{m}(\k), ~~ \mu^{\pm}(\k)=-(1+{\rm e}^{\pm \frac{\pi \k}{2}}) \mu(\k)\label{D1}
\end{equation}
where $C_1$ is a straight contour from $-\infty$ to $+\infty$ with
$\frac 43 < \Im (\k) < 2$. Note that for off--diagonal Neumann coefficients there is
no need to correct the contour shifting with the
universal nilpotent $P_{nm}$ discussed in section 7.1. If we move the upper contour toward the
real axis we are bound to meet two poles of the measure, one at
$\k=\k_1= \frac {4 i }3$
and another at $\k=\k_0=0$. So finally we obtain the usual integral along
the real axis (which corresponds to $X^{\pm}$) plus two contributions
from the two poles that remain trapped inside the contour. The latter
are clockwise oriented, so we have to change the sign when calculating the
residues. We have then
\be
X^{\pm}_{(i)nm} &=& X^{\pm}_{nm} - \oint_{\k_{1}=\frac{4 i}{3}} d\k ~ \mu^{\pm}(\k)
V^{(2)}_{n}(\k) V^{(-1)}_{m}(\k) \nonumber \\
&-& \frac{1}{2} \oint_{\k_{0}=0} d\k ~ \mu^{\pm}(\k) V^{(2)}_{n}(\k) V^{(-1)}_{m}(\k)
\label{D2}
\ee
This gives us finally

\begin{equation}
X^{\pm}_{(i)nm} = X^{\pm}_{nm} - \frac{2 i}{3} V^{(2)}_{n}(0) V^{(-1)}_{m}(0) +
\frac{2 i}{3}(1 \pm i \sqrt{3}) V^{(2)}_{n} \left( \frac{4 i}{3}
\right) V^{(-1)}_{m} \left( \frac{4 i}{3} \right)\label{D3}
\end{equation}
Now, let us see some examples:
\begin{eqnarray}
X^{+}_{(i)2,0} &=& \frac{16}{27} - 0 -\frac{8}{9}(1+i \sqrt{3}) =
-\frac{8}{27}(1+3 i \sqrt{3}) \nonumber \\
X^{+}_{(i)2,-1} &=& \frac{2}{3\sqrt{3}} - \frac{2 i}{3} -\frac{2}{3}(-i+
\sqrt{3}) = -\frac{4}{3\sqrt{3}} \nonumber \\
X^{+}_{(i)3,0} &=& \frac{64}{81 \sqrt{3}} - 0 + \frac{32}{27}(-i+ \sqrt{3}) =
\frac{32}{243}(-9 i+ 11 \sqrt{3}) \nonumber \\
X^{+}_{(i)3,3} &=& \frac{10112}{19683} - 0 +\frac{320}{2187}(1+i \sqrt{3}) =
\frac{64}{19683}(203+45 i \sqrt{3}) \nonumber
\end{eqnarray}
and

\begin{eqnarray}
X^{-}_{(i)2,0} &=& \frac{16}{27} - 0 +\frac{8}{9}i(i+ \sqrt{3}) =
\frac{8}{27}(-1+3 i \sqrt{3}) \nonumber \\
X^{-}_{(i)2,-1} &=& -\frac{2}{3\sqrt{3}} - \frac{2 i}{3} +\frac{2}{3}(i+
\sqrt{3}) = \frac{4}{3\sqrt{3}} \nonumber \\
X^{-}_{(i)3,0} &=& -\frac{64}{81 \sqrt{3}} - 0 - \frac{32}{27}(i+ \sqrt{3}) =
-\frac{32}{243}(9 i+ 11 \sqrt{3}) \nonumber \\
X^{-}_{(i)3,3} &=& \frac{10112}{19683} - 0 +\frac{320}{2187}(1-i \sqrt{3}) =
\frac{64}{19683}(203-45 i \sqrt{3}) \nonumber
\end{eqnarray}
These results agree perfectly with the ones computed directly using (\ref{Vnmrsi})
(multiplied by ${\cal C}$ to give the corresponding X's).

\subsection{Reconstruction of $X^{\pm}_{(-i)}$}

In order to reconstruct $X^{+}_{(-i)}={\cal C}{\hat V}^{12}_{(-i)}$ and
$X^{-}_{(-i)}={\cal C}{\hat V}^{21}_{(-i)}$, it is clear that we should use
a different
prescription than the one used above. This prescription can be inferred from the
fact that $X^{+}_{(-i)}={\cal C} X^{-}_{(i)} {\cal C}$ and
$X^{-}_{(-i)}={\cal C} X^{+}_{(i)} {\cal C}$. We should use then
\begin{equation}
X^{\pm}_{(-i)nm} = \int_{C_{2}} d\kappa ~ \mu^{\mp}(\kappa) V^{(2)}_{n}(\k)
V^{(-1)}_{m}(\k), ~~ \mu^{\mp}(\k)=-(1+{\rm e}^{\mp \frac{\pi \k}{2}}) \mu(\k)\label{D4}
\end{equation}
where $C_2$ is a straight contour from $-\infty$ to $+\infty$ with
$-2 < \Im (\k) < -\frac{4 i}{3}$. If we move the lower contour toward the
real axis we are bound to meet two poles of the measure,
one at $\k=\k_1= -\frac {4 i }3$
and another at $\k=\k_0=0$. So finally we obtain the usual integral along
the real axis (which corresponds to $X^{\pm}$) plus two contributions
from the two poles that remain trapped inside the contour. The latter
are anticlockwise oriented this time, so we do not have to change the sign
when calculating the residues. We then have
\begin{eqnarray}
X^{\pm}_{(-i)nm} &=& X^{\pm}_{nm} + \oint_{\k_{1}=-\frac{4 i}{3}} d\k ~
\mu^{\mp}(\k) V^{(2)}_{n}(\k) V^{(-1)}_{m}(\k)\nonumber \\
&+& \frac{1}{2} \oint_{\k_{0}=0} d\k ~ \mu^{\mp}(\k) V^{(2)}_{n}(\k) V^{(-1)}_{m}(\k)
\label{D5}
\end{eqnarray}
This gives us finally
\begin{equation}
X^{\pm}_{(-i)nm} = X^{\pm}_{nm} + \frac{2 i}{3} V^{(2)}_{n}(0) V^{(-1)}_{m}(0) -
\frac{2 i}{3}(1 \mp i \sqrt{3}) V^{(2)}_{n}
\left( -\frac{4 i}{3} \right) V^{(-1)}_{m} \left( -\frac{4 i}{3} \right) \label{D6}
\end{equation}
Let us check some components:
\begin{eqnarray}
X^{+}_{(-i)2,0} &=& \frac{16}{27} + 0 +\frac{8}{9}i(i+ \sqrt{3}) =
\frac{8}{27}(-1+3 i \sqrt{3}) \nonumber \\
X^{+}_{(-i)2,-1} &=& \frac{2}{3\sqrt{3}} + \frac{2 i}{3} -\frac{2}{3}(i+
\sqrt{3}) = -\frac{4}{3\sqrt{3}} \nonumber \\
X^{+}_{(-i)3,0} &=& \frac{64}{81 \sqrt{3}} + 0 + \frac{32}{27}(i+ \sqrt{3}) =
\frac{32}{243}(9 i+ 11 \sqrt{3}) \nonumber \\
X^{+}_{(-i)3,3} &=& \frac{10112}{19683} + 0 +\frac{320}{2187}(1-i \sqrt{3}) =
\frac{64}{19683}(203-45 i \sqrt{3}) \nonumber
\end{eqnarray}
and
\begin{eqnarray}
X^{-}_{(-i)2,0} &=& \frac{16}{27} + 0 -\frac{8}{9}(1+i \sqrt{3}) =
-\frac{8}{27}(1+3 i \sqrt{3}) \nonumber \\
X^{-}_{(-i)2,-1} &=& -\frac{2}{3\sqrt{3}} + \frac{2 i}{3} +\frac{2}{3}(-i+
\sqrt{3}) = \frac{4}{3\sqrt{3}} \nonumber \\
X^{-}_{(-i)3,0} &=& -\frac{64}{81 \sqrt{3}} + 0 - \frac{32}{27}(-i+ \sqrt{3}) =
 -\frac{32}{243}(-9 i+ 11 \sqrt{3}) \nonumber \\
X^{-}_{(-i)3,3} &=& \frac{10112}{19683} + 0 +\frac{320}{2187}(1+i \sqrt{3}) =
 \frac{64}{19683}(203+45 i \sqrt{3}) \nonumber
\end{eqnarray}
These results perfectly agree with the ones computed directly using
(\ref{Vnmrsmi}) (multiplied by ${\cal C}$ to get the corresponding X's).

\subsection{Reconstruction of $T^{(N)}_{(-i)}$}

Analogously to the reconstruction of $T^{(N)}_{(i)}$, we must choose a contour
to compute the contribution of the continuous spectrum. In this case, we should
choose a contour $C^{(-)}_{N}$ which stretches from $-\infty$ to $\infty$ with
$\Im (\k)$ just below $-\frac{4}{N}$. When we move the contour up toward the
real axis, we get two contributions coming from the poles at $\k_{0}=0$
(actually, half of it) and $\k_{1}=-\frac{4 i}{N}$, that is

\begin{eqnarray}
T^{(N)}_{(-i)nm} &=& \int_{C^{(-)}_{N}} d\k ~ \mu(N,\k) V^{(2)}_{n}(\k)
V^{(-1)}_{m}(\k)+iP_{nm} \nonumber \\
&=& \int_{- \infty}^{\infty} d\k ~ \mu(N,\k) V^{(2)}_{n}(\k)V^{(-1)}_{m}(\k) +
 \oint_{\k_{1}} d\k ~ \mu(N,\k) V^{(2)}_{n}(\k)V^{(-1)}_{m}(\k) \nonumber \\
&+& \frac{1}{2} \oint_{\k_{0}} d\k ~ \mu(N,\k) V^{(2)}_{n}(\k)V^{(-1)}_{m}(\k)
+iP_{nm}\label{D7}
\end{eqnarray}
where $\mu (N, \k)$ was defined in (\ref{meas}). The first term is just $T^{(N)}_{nm}$,
and computing the residues we get
\begin{equation}
T^{(N)}_{(-i)nm} = T^{(N)}_{nm} + \frac{2 i}{N} V^{(2)}_{n}(0)V^{(-1)}_{m}(0) +
\frac{4 i}{N} V^{(2)}_{n}(-\frac{4 i}{N})V^{(-1)}_{m}(-\frac{4 i}{N})\label{D8}
\end{equation}
Here are some examples for $N=4$
\begin{eqnarray}
T^{(4)}_{(-i)2,-1} &=& 0 + \frac{i}{2} + i = \frac{3 i}{2} \nonumber \\
T^{(4)}_{(-i)3,1} &=& \frac{11}{16} + 0 + \frac{1}{2} = \frac{19}{16} \nonumber \\
T^{(4)}_{(-i)4,2} &=& \frac{5}{16} + 0 - \frac{5}{4} = - \frac{15}{16} \nonumber \\
T^{(4)}_{(-i)5,4} &=& 0 + 0 - \frac{5 i}{16} = - \frac{5 i}{16} \nonumber
\end{eqnarray}
This agrees precisely with the components of the dual wedge state with Neumann
coefficients (one should multiply it by ${\cal C}$, since
$T^{(N)}_{(-i)}={\cal C} S^{(N)}_{(-i)}$)
\begin{equation}
{S_{(-i)}}^{(N)}_{pM}=\oint_0\frac{dz}{2\pi i}\oint_0\frac{dw}{2\pi i}
\frac{1}{z^{p-1}}\frac{1}{w^{M+2}} \cdot\left[\frac {f'_N(z)^2}{f'_N(w)}
\frac 1{f_N(z)-f_N(w)}
-\frac 1{z-w}\right] \nonumber
\end{equation}
From (\ref{D8}) we see that $T^{(N)}_{(-i)}={\cal C} T^{(N)}_{(i)} {\cal C}$, as expected.

\section{A simple proof of $U_{(i)}U_{(-i)}=1$}

From the argument of the path shifting it is obvious that the twisted
matrices of the vertices will commute between themselves.
But it is instructive to see this directly using our usual $U^2=1$ argument.

\noindent Let us start by tracing back the $U$ matrices.  The relevant Neumann coefficients are
defined by
\be
\bra{\hat V_{(i)}}&=&\bra {V_3}Y(i)=\bra{\hat0}\, e^{-c_n^r\,V^{rs}_{nM}
\,b_M^s}\label{E0}\\
V_i^{rs}(z,w)&=&\left[\frac{\partial f_3^r (z)^2}{\partial f_3^s(w)}
\frac{1}{f_3^r(z)-f_3^s(w)}\left(\frac{f_3^s(w)}{f_3^r(z)}\right)^3-
\frac{\delta^{rs}}{z-w}\right]\0\\
&=&\hat V^{rs}(z,w)-\frac{4i}{3}\left(\alpha^{r-s}f^{(2)}_{-\frac{4i}{3}}(z)
f^{(-1)}_{\frac{4i}{3}}(w)+\frac12f^{(2)}_{k=0}(z)f^{(-1)}_{k=0}(w)\right)\0\\
&=&\int_{\Im{\k}>\frac 43}\,\frac{v^{rs}(\k)}{2\sinh\frac{\pi \k}{2}}
f^{(2)}_{-\k}(z)f^{(-1)}_{\k}(w),\quad\texttt{Only in the bulk}\label{E2}
\ee
where $\xi_3=\frac {4i}3$.

\noindent From now on, for simplicity, we focus on the bulk.
This allows to discard
the $\k=0$ contribution and concentrate on the rest. The vertex is not
twist invariant but still it is cyclic, so we can write (only bulk)
\be
V_{(i)}^{rs}=\frac13(E_{(i)}+\alpha^{s-r}
U_{(i)}+\alpha^{r-s}\bar U_{(i)})\label{E3}
\ee
From the explicit reconstruction formula presented in the main text, we can see
that (in the bulk) the only difference wrt the average vertex is in the
matrix $\bar U_{(i)}$ (the bulk--residual contribution is proportional to
$\alpha^{r-s}$).
So, using a subscript "$_0$" for principal parts, we have
(we inaugurate a `bra--ket' notation

$\ket{x}\to V^{(2)}(x)$, $\bra{x}\to V^{(-1)}(x)$)
\be
E_{(i)}&=&{\cal C}\0\\
U_{(i)}&=&U_0=U\0\\
\bar U_{(i)}&=&\bar U_0-\frac{4i}{3}\ket{-\xi_3}\bra{\xi_3}
\equiv {\cal C}U{\cal C}-\beta.\label{E4}
\ee
It is easy to see that the commutation of twisted bulk matrices
(${\cal C}V_{(i)}^{rs}$)
will work iff
\be
{\cal C}U_{(i)}{\cal C}\bar U_{(i)}-{\cal C}\bar U_{(i)}{\cal C}U_{(i)}
=({\cal C}U{\cal C}\bar U-{\cal C}\bar U{\cal C}U)-({\cal C}U{\cal C}\beta-
{\cal C}\beta {\cal C}U)=0.\0
\ee
Since we know that $U^2=\bar U^2=1,$ we only have to prove that
${\cal C}U{\cal C}\beta-{\cal C}\beta {\cal C}U.$ We can actually
do better: we can prove that
\be
{\cal C}U{\cal C}\beta={\cal C}\beta {\cal C}U=0,\label{E5}
\ee
which means
\be
{\cal C}U\,\ket{\xi_3}=0,\quad\quad
\bra{\xi_3}\,{\cal C}U=0\label{E6}
\ee
It takes a line to compute this quantities using reconstruction formulas.
We have indeed
\be
{\cal C}U=\int_R \frac{d\k}{2\sinh\frac {\pi \k}{2}}\;u(\k)\ket \k\bra \k\label{E7}
\ee
The ${\cal C}U$ eigenvalue can be computed from the known eigenvalues of $X^{rs}$
and it turns out to be
\be
u(\k)=x^{11}(k)+\bar\alpha   x^{12}(\k)+\alpha x^{21}(\k)=
2\left(\cosh\frac{\pi(\k-\xi_3)}{2}-1\right)\,
\frac{\sinh\frac {\pi \k}{4}}{\sinh\frac{3\pi \k}{4}}\label{E8}
\ee
Here we come to the point: $u$ eigenvalues are {\it vanishing} at $\k=\frac{4i}{3}$,
while they are divergent at $\k=-\frac{4i}{3}$
\be
u(\xi_3)=0,\quad\quad u(-\xi_3)&=&\infty.\0
\ee
So, in the $\k$-UHP, there are no poles and the delta functions will work
without producing any divergence
\be
{\cal C}U\ket{\xi_3}=u(\xi_3)\ket{\xi_3}=0,\quad\quad
\bra{\xi_3}{\cal C}U=u(\xi_3)\bra{\xi_3}=0\label{E9}
\ee


\begin{thebibliography}{99}

\bibitem{Rastelli:2000iu}
  L.~Rastelli and B.~Zwiebach,
  {\it Tachyon potentials, star products and universality,}
  JHEP {\bf 0109} (2001) 038
  [arXiv:hep-th/0006240].




\bibitem{Schnabl:2002gg}
  M.~Schnabl,
  {\it Wedge states in string field theory,}
  JHEP {\bf 0301} (2003) 004
  [arXiv:hep-th/0201095].

\bibitem{BMST} L.~Bonora, C.~Maccaferri, R.~J.~Scherer Santos
and D.~D.~Tolla,
{\it Ghost story. I. Wedge states in the oscillator formalism,}
arXiv:0706.1025 [hep-th].


\bibitem{Schnabl05}
  M.~Schnabl,
  {\it Analytic solution for tachyon condensation in open string field theory,}
  Adv.\ Theor.\ Math.\ Phys.\  {\bf 10} (2006) 433
  [arXiv:hep-th/0511286].

\bibitem{Okawa1}
  Y.~Okawa,
 {\it Comments on Schnabl's analytic solution for tachyon condensation in
  Witten's open string field theory,}
  JHEP {\bf 0604} (2006) 055
  [arXiv:hep-th/0603159].

\bibitem{Fuchs0}
  E.~Fuchs and M.~Kroyter,
 {\it On the validity of the solution of string field theory,}
  JHEP {\bf 0605} (2006) 006
  [arXiv:hep-th/0603195].

\bibitem{Ellwood}
  I.~Ellwood and M.~Schnabl,
 {\it Proof of vanishing cohomology at the tachyon vacuum,}
  JHEP {\bf 0702} (2007) 096
  [arXiv:hep-th/0606142].

\bibitem{RZ06}
  L.~Rastelli and B.~Zwiebach,
  {\it Solving open string field theory with special projectors,}
arXiv:hep-th/0606131.

\bibitem{ORZ}
  Y.~Okawa, L.~Rastelli and B.~Zwiebach,
 {\it Analytic solutions for tachyon condensation with general projectors,}
  arXiv:hep-th/0611110.

\bibitem{Erler:2009uj}
  T.~Erler and M.~Schnabl,
  {\it A Simple Analytic Solution for Tachyon Condensation,}
  arXiv:0906.0979 [hep-th].

\bibitem{Erler:2007xt}
  T.~Erler,
  {\it Tachyon Vacuum in Cubic Superstring Field Theory,}
  JHEP {\bf 0801} (2008) 013
  [arXiv:0707.4591 [hep-th]].

\bibitem{Aref'eva:2008ad}
  I.~Y.~Aref'eva, R.~V.~Gorbachev and P.~B.~Medvedev,
  {\it Tachyon Solution in Cubic Neveu-Schwarz String Field Theory,}
  Theor.\ Math.\ Phys.\  {\bf 158} (2009) 320
  [arXiv:0804.2017 [hep-th]].

\bibitem{Okawa2}
  Y.~Okawa,
{\it Analytic solutions for marginal deformations in open superstring field
  theory,}
  arXiv:0704.0936 [hep-th].

\bibitem{Okawa3}
  Y.~Okawa,
{\it Real analytic solutions for marginal deformations in open
superstring field theory,}
  arXiv:0704.3612 [hep-th].

\bibitem{Schnabl:2007az}
  M.~Schnabl,
 {\it Comments on marginal deformations in open string field theory,}
  arXiv:hep-th/0701248.

\bibitem{KORZ}
  M.~Kiermaier, Y.~Okawa, L.~Rastelli and B.~Zwiebach,
 {\it Analytic solutions for marginal deformations in open string field theory,}
  arXiv:hep-th/0701249.


\bibitem{Fuchs3}
  E.~Fuchs, M.~Kroyter and R.~Potting,
  {\it Marginal deformations in string field theory,}
   arXiv:0704.2222 [hep-th].

\bibitem{Kwon:2008ap}
  O.~K.~Kwon,
  {\it ``Marginally Deformed Rolling Tachyon around the Tachyon Vacuum in Open
  String Field Theory,''}
  Nucl.\ Phys.\  B {\bf 804} (2008) 1
  [arXiv:0801.0573 [hep-th]].




\bibitem{Lee:2007ns}
  B.~H.~Lee, C.~Park and D.~D.~Tolla,
  {\it  Marginal Deformations as Lower Dimensional D-brane Solutions in Open String
  Field theory,}
  arXiv:0710.1342 [hep-th].

\bibitem{Kiermaier:2007ki}
  M.~Kiermaier and Y.~Okawa,
  {\it General marginal deformations in open superstring field theory,}
  arXiv:0708.3394 [hep-th].



\bibitem{Kiermaier:2007vu}
  M.~Kiermaier and Y.~Okawa,
  {\it Exact marginality in open string field theory: a general framework,}
  arXiv:0707.4472 [hep-th].

\bibitem{Erler:2007rh}
  T.~Erler,
  {\it  Marginal Solutions for the Superstring,}
  JHEP {\bf 0707} (2007) 050
  [arXiv:0704.0930 [hep-th]].



\bibitem{Kroyter:2009zj}
  M.~Kroyter,
  {\it On string fields and superstring field theories,}
  arXiv:0905.1170 [hep-th].

\bibitem{Kroyter:2009bg}
  M.~Kroyter,
  {\it Comments on superstring field theory and its vacuum solution,}
  arXiv:0905.3501 [hep-th].

\bibitem{Ellwood:2009zf}
  I.~Ellwood,
  {\it Singular gauge transformations in string field theory,}
  arXiv:0903.0390 [hep-th].

\bibitem{Kiermaier:2008qu}
  M.~Kiermaier, Y.~Okawa and B.~Zwiebach,
  {\it The boundary state from open string fields, }
  arXiv:0810.1737 [hep-th].


\bibitem{Ellwood:2008jh}
  I.~Ellwood,
  {\it The closed string tadpole in open string field theory, }
  JHEP {\bf 0808} (2008) 063
  [arXiv:0804.1131 [hep-th]].

\bibitem{Fuchs4}
E.~Fuchs and M.~Kroyter,
{\it Analytical Solutions of Open String Field Theory,}
  arXiv:0807.4722 [hep-th].



\bibitem{BST}
  L.~Bonora, R.~J.~Scherer Santos and D.~D.~Tolla,
{\it Spectral properties of ghost Neumann matrices,}
  Phys.\ Rev.\  D {\bf 77} (2008) 106001
  [arXiv:0801.2099 [hep-th]].

\bibitem{BMT3} L.~Bonora, C.~Maccaferri
and D.~D.~Tolla,
{\it Ghost story. III. Back to ghost number zero,}
arXiv:0908.0056 [hep-th].




\bibitem{Belov1}
  D.~M.~Belov,
  {\it Witten's ghost vertex made simple (bc and bosonized ghosts),}
  Phys.\ Rev.\  D {\bf 69} (2004) 126001
  [arXiv:hep-th/0308147].

\bibitem{Belov2}
  D.~M.~Belov and C.~Lovelace,
  {\it Star products made easy,}
  Phys.\ Rev.\  D {\bf 68} (2003) 066003
  [arXiv:hep-th/0304158].


\bibitem{BeLove}
  D.~M.~Belov and C.~Lovelace, {\it Unpublished}







\bibitem{Samu}
S.~Samuel,
{\it The Ghost Vertex In E. Witten's String Field Theory},
Phys.\ Lett.\ B {\bf 181} (1986) 255.


\bibitem{CST}
E.Cremmer,A.Schwimmer, C.Thorn, {\it "The vertex function in Witten's
formulation of string field theory}, Phys.Lett. {\bf 179B} (1986) 57.

\bibitem{GJ1} D.J.Gross and A.Jevicki, {\it Operator Formulation
of Interacting String Field Theory}, Nucl.Phys. {\bf B283} (1987) 1.

\bibitem{GJ2} D.J.Gross and A.Jevicki, {\it Operator Formulation
of Interacting String Field Theory, 2}, Nucl.Phys. {\bf B287} (1987) 225.

\bibitem{Ohta}
N.~Ohta,
{\it  Covariant Interacting String Field Theory In The Fock Space Representation, }
Phys.\ Rev.\ D {\bf 34} (1986) 3785
[Erratum-ibid.\ D {\bf 35} (1987) 2627].

\bibitem{RSZ3} L.Rastelli, A.Sen and B.Zwiebach, {\it Half-strings,
Projectors, and Multiple D-branes in Vacuum String Field Theory},
JHEP {\bf 0111} (2001) 035 [hep-th/{0105058}].

\bibitem{GRSZ2} D.Gaiotto, L.Rastelli, A.Sen and B.Zwiebach,
{\it Ghost Structure and Closed Strings in Vacuum String Field
Theory}, [hep-th/{0111129}].

\bibitem{MM} C.~Maccaferri and D.~Mamone,
 {\it Star democracy in open string field theory,}
  JHEP {\bf 0309} (2003) 049
  [arXiv:hep-th/0306252].

\bibitem{Oku2} K.Okuyama, {\it Ghost Kinetic Operator of Vacuum
String Field Theory}, JHEP {\bf 0201} (2002) 027 [hep-th/{0201015}].

\bibitem{Furu}
  K.~Furuuchi and K.~Okuyama,
  {\it  Comma vertex and string field algebra, }
  JHEP {\bf 0109} (2001) 035
  [arXiv:hep-th/0107101].

\bibitem{Kishimoto}
  I.~Kishimoto,
  {\it Some properties of string field algebra,}
  JHEP {\bf 0112} (2001) 007
  [arXiv:hep-th/0110124].



\bibitem{FKM} E.Fuchs, M.Kroyter and A.Marcus, {\it Squeezed States Projectors
in String Field Theory}, JHEP {\bf 0209} (2002) 022 [hep-th/{0207001}].


\bibitem{Fuchs1}
E.~Fuchs and M.~Kroyter,
{\it Schnabl's L(0) operator in the continuous basis,}
JHEP {\bf 0610} (2006) 067
[arXiv:hep-th/0605254].





\bibitem{RSZ2} L.Rastelli, A.Sen and B.Zwiebach,
{\it Classical solutions in string field theory around the tachyon vacuum,}
  Adv.\ Theor.\ Math.\ Phys.\  {\bf 5} (2002) 393
  [arXiv:hep-th/0102112].




\bibitem{tope}
L.~Bonora, C.~Maccaferri, D.~Mamone and M.~Salizzoni,
{\it  Topics in string field theory, }
arXiv:hep-th/0304270.

\bibitem{KPot} V.A.Kostelecky and R.Potting, {\it Analytical construction
of a nonperturbative vacuum for the open bosonic string},
Phys.\ Rev.\ D {\bf 63} (2001) 046007
[hep-th/{0008252}].



\bibitem{Fuchs2}
  E.~Fuchs and M.~Kroyter,
  {\it  Universal regularization for string field theory, }
  JHEP {\bf 0702} (2007) 038
  [arXiv:hep-th/0610298].









\bibitem{RSZ1} L.Rastelli, A.Sen and B.Zwiebach, {\it Star Algebra
Spectroscopy}, [\hepth{0111281}].



\bibitem{leclair} A.Leclair, M.E.Peskin, C.R.Preitschopf,
{\it String Field
Theory on the Conformal Plane. (I) Kinematical Principles},
Nucl.Phys. {\bf B317} (1989) 411.

\bibitem{LeClair:1988sj}
  A.~LeClair, M.~E.~Peskin and C.~R.~Preitschopf,
  {\it  String Field Theory on the Conformal Plane. 2. Generalized Gluing, }
  Nucl.\ Phys.\  B {\bf 317} (1989) 464.








\end{thebibliography}
\end{document}